\documentclass[11pt,a4paper]{article}

\usepackage{cite}
\usepackage{amsmath, amssymb}
\usepackage{graphicx}      
\usepackage{caption}
\usepackage{xcolor}

\usepackage[
    colorlinks=true,
    citecolor=blue,
    linkcolor=blue,
    urlcolor=blue
]{hyperref}

\usepackage{tikz}
\usepackage{mathtools}

\numberwithin{equation}{section}

\newlength{\textlength}
\newlength{\margin}
\newcommand\blfootnote[1]{%
  \begingroup
  \renewcommand\thefootnote{}\footnote{#1}%
  \addtocounter{footnote}{-1}%
  \endgroup
}
\newcommand{\diagram}[2]{%
    \vcenter{\hbox{\includegraphics[height=#1]{#2}}}%
}

\begin{document}
\thispagestyle{empty}

\begin{flushright}
{\small
MITP-26-065}
\end{flushright}

\vspace{-0.5cm}

\begin{center}
\Large\bf\boldmath
Superdaisy Resummation in Theories with Broken Gauge Symmetries
\unboldmath
\end{center}

\vspace{-0.2cm}

\begin{center}
Martin M\"unzenberg$^{*}$\blfootnote{$^*$ mamuenze@uni-mainz.de} and Carlos Tamarit$^\dagger$\blfootnote{$^\dagger$ ctamarit@uni-mainz.de}\\
\vskip0.4cm

{\it PRISMA++ Cluster of Excellence \& Mainz Institute for Theoretical Physics,\\
Johannes Gutenberg-Universität Mainz, 55099 Mainz, Germany}

\vskip1.cm
\end{center}

\begin{abstract}
We extend the partial dressing resummation scheme for the finite-temperature effective potential to the gauge and ghost sectors of spontaneously broken gauge theories and apply the resulting framework to the Standard Model. Going beyond the usual high-temperature treatment, we retain the full mass dependence of the thermal self-energies and determine the scalar, gauge-boson and ghost masses self-consistently from a coupled system of gap equations. The solutions show that corrections beyond the leading $\mathcal{O}(T^2)$ approximation are particularly relevant for the gauge-boson thermal masses, while they induce a comparatively small shift in the scalar spectrum for the benchmark studied. The self-consistent gauge-boson dressing in the longitudinal sector reduces the barrier in the finite-temperature Higgs effective potential. Implementing the resummation in the transverse sector removes the barrier and would seem to exclude a first-order phase transition in accordance with lattice results. While such resummation is not quantitatively reliable, due to the transverse excitations becoming strongly coupled, the results showcase the crucial role played by transverse modes in determining the nature of the electroweak phase transition.

\end{abstract}

\section{Introduction}
\label{sec:introduction}

The thermal history of the early Universe provides a setting in which physics at energy scales far above those directly accessible today can leave observable imprints. A particularly well-motivated example is the electroweak epoch, where the Higgs expectation value evolves from the symmetric high-temperature phase to the electroweak broken phase. While this change is a smooth crossover in the Standard Model (SM) for the observed Higgs mass \cite{Kajantie:1996mn,Kajantie:1996qd,DOnofrio:2014rug,DOnofrio:2015gop}, it can become a first-order phase transition in a wide range of extensions, including singlet extensions \cite{Anderson:1991zb,Espinosa:1993bs,Profumo:2007wc,Espinosa:2011ax,Curtin:2014jma}, two-Higgs-doublet models \cite{Turok:1990zg,Cline:1996mga,Fromme:2006cm,Cline:2011mm,Dorsch:2013wja,Laine:2017hdk} and scalar triplet extensions \cite{Patel:2012pi,Chiang:2014hia,Niemi:2018asa,Niemi:2020hto}. These examples are by no means exhaustive; broader discussions of the underlying mechanisms and model realisations can be found in Refs.~\cite{Chung:2012vg,Ramsey-Musolf:2019lsf}.
This makes the electroweak transition an interesting probe of physics beyond the SM, since its dynamics --- from bubble nucleation and wall propagation to the completion of the transition --- can have observable consequences. In electroweak baryogenesis, a first-order transition provides the departure from thermal equilibrium required by the Sakharov conditions \cite{Sakharov:1967dj}, while electroweak sphaleron processes violate baryon number and CP-violating interactions at the bubble walls generate charge asymmetries in the plasma \cite{Kuzmin:1985mm,Cohen:1993nk,Morrissey:2012db}. A first-order transition can also source a stochastic gravitational-wave background through the dynamics of expanding bubbles and the surrounding plasma \cite{Witten:1984rs,Kosowsky:1991ua,Kamionkowski:1993fg,Hindmarsh:2013xza,Hindmarsh:2015qta}, which for sufficiently strong transitions may be observable by future space-based interferometers such as LISA \cite{Caprini:2015zlo,Caprini:2019egz}.

These possibilities make a reliable description of finite-temperature electroweak physics essential for quantitative predictions. Although the SM transition itself is a crossover, it provides a benchmark containing the full non-Abelian electroweak gauge structure and offers a non-trivial test of the theoretical description of finite-temperature effects. In this sense, it serves as a starting point for extending the framework to theories beyond the SM with first-order transitions and associated bubble and gravitational-wave dynamics. Taking a broader view that goes beyond the classification of phase transitions, the finite-temperature free energy determines the pressure, entropy and energy density of the thermal plasma, and therefore enters the cosmological equation of state and the effective numbers of relativistic degrees of freedom. A reliable description of thermal field theory is thus useful beyond the question of whether a particular transition is first order.

\begin{figure}[t!]
    \centering
    \includegraphics[width=\textwidth]{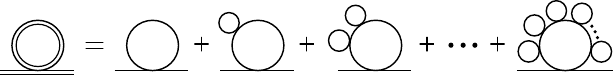}
\caption{
Schematic representation of the daisy series generated by diagrams with repeated thermal self-energy insertions into a soft bosonic propagator. Each additional insertion introduces the parametric factor $\alpha=\lambda T^2/m^2$, arising from a hard thermal loop $\sim\lambda T^2$ and an additional soft propagator $\sim1/m^2$. The resummed series is represented by the dressed propagator on the left.}
\label{fig:daisy_series}
\end{figure}

The main perturbative tool for studying this physics is the finite-temperature effective potential. Its straightforward loop expansion, however, becomes problematic precisely in the regime relevant for thermal phase transitions. This manifests itself for example in self-energy insertions not being straightforwardly suppressed by additional powers of the coupling constants: the existence of an additional dimensionful scale, the temperature $T$, can lead to effective couplings involving ratios of scales that can become large. In a scalar theory, for example, successive one-loop self-energy insertions, leading to so-called ``daisy'' diagrams as illustrated in Fig.~\ref{fig:daisy_series}, generate an effective expansion parameter of the form
\begin{equation}
    \alpha \sim \frac{\lambda T^2}{m^2},
\end{equation}
where $m^2$ denotes the unresummed field-dependent mass squared. The magnitude of this ratio can become of order one or larger near a thermally induced transition. The conventional loop expansion must therefore be reorganised by resumming the infrared-sensitive contributions
\cite{Dolan:1973qd,Weinberg:1974hy,Fendley:1987ef}. At leading order this amounts to including thermal mass-squared corrections proportional to $T^2$ in the propagators of the relevant bosonic zero modes, corresponding to the familiar ring or daisy resummation \cite{Carrington:1991hz,Arnold:1992rz,Parwani:1991gq}.
Different prescriptions for carrying out this reorganisation have been developed. The widely used Parwani prescription \cite{Parwani:1991gq} incorporates the leading thermal mass corrections throughout the bosonic one-loop potential, yet it has been shown that this procedure miscounts daisy and superdaisy diagrams \cite{Dine:1992wr,Espinosa:1992gq,Boyd:1993tz}. Alternatives that avoid this problem include the Arnold--Espinosa prescription \cite{Arnold:1992rz}, which restricts the resummation to the Matsubara zero-mode contributions, and the partial dressing (PD) method --- also known as tadpole resummation --- which is constructed from resumming thermal masses in the expression for the derivative of the effective potential with respect to the field expectation value. When the thermal masses are computed by solving gap equations, the resulting effective potential correctly resums superdaisy diagrams \cite{Espinosa:1992gq,Boyd:1992xn,Espinosa:1992kf,Curtin:2016urg}.

Despite enabling a correct counting of diagrams, PD does not constitute a complete higher-loop calculation. So-called two-loop ``lollipop'' contributions are not generated by the resummed one-loop tadpole, while ``sunset''-type contributions require additional treatment: their diagrammatic weights are not in general reproduced correctly, and evaluating the self-energies in the limit of zero external momentum cannot be justified in terms of a separation of scales in the momentum integrals \cite{Arnold:1992rz,Boyd:1993tz,Lofgren:2023sep}. A way to avoid this overlapping-momentum problem is the hybrid construction proposed in Ref.~\cite{Boyd:1993tz}, which resums momentum-independent self-energy insertions through a gap equation and adds the sunset contribution explicitly to the potential. Ref.~\cite{Berthiaume:2026dkt} recently developed this construction into hybrid partial dressing (HPD) for scalar theories.

A distinct approximation concerns the evaluation of the thermal self-energies themselves. In many practical resummation calculations these are obtained in the high-temperature expansion. At leading order, the resulting thermal self-energies are proportional to $T^2$ and independent of the masses circulating in the loop. This is a good approximation for $m/T\ll1$, but phase transitions can probe regions in which relevant masses become comparable to the temperature and vary substantially with the background field. In this regime, retaining the full thermal functions is important in order to capture the mass dependence and eventual Boltzmann suppression of heavy modes. Ref.~\cite{Curtin:2016urg} revisited PD in this context and developed practical PD and optimised partial dressing (OPD) prescriptions for phenomenological phase-transition calculations. The numerical solution and convergence properties of tadpole resummation were subsequently refined in Ref.~\cite{Curtin:2022ovx}, which also demonstrated sizeable differences relative to conventional thermal-mass prescriptions in phase-transition and gravitational-wave observables. PD was recently generalised to multiple mixing scalar fields in Ref.~\cite{Bahl:2024bcg} and applied to the two-Higgs-doublet model in Refs.~\cite{Bittar:2025lcr,Zhao:2026tob}, where different resummation prescriptions were shown to alter even qualitative features of the thermal phase structure.

The previous developments have so far concentrated mainly on scalar thermal masses. In phenomenological implementations of PD, the electroweak gauge sector is generally still treated with its leading high-temperature screening masses. Gauge-boson gap equations themselves are not new: coupled scalar and longitudinal and transverse gauge-boson masses were already studied in the SM in Refs.~\cite{Espinosa:1992kf,Buchmuller:1993bq}. Those analyses, however, relied on a high-temperature expansion of the thermal self-energies and the simplifying limit of a zero hypercharge gauge coupling, which renders the electroweak gauge bosons degenerate and removes the neutral \(Z\)--photon mixing present in the physical SM. It therefore remains unclear how the gauge sector behaves once the full mass dependence of the thermal loops and the complete electroweak gauge structure are retained. A modern PD treatment can address this question self-consistently: how strongly does the electroweak gauge sector feed back into the coupled thermal system, and how does this propagate to the resulting finite-temperature effective potential?

Alongside the thermal-mass resummation prescriptions, alternative frameworks have been developed for organising finite-temperature calculations. In particular, high-temperature dimensional reduction (DR) integrates out the non-zero Matsubara modes to obtain a systematically improvable three-dimensional effective theory~\cite{Kajantie:1995dw,Braaten:1995cm,Farakos:1994kx}. Higher-order matching, together with a consistent perturbative expansion in the three-dimensional effective theory, improves control over phase-transition predictions~\cite{Croon:2020cgk,Gould:2021oba,Schicho:2022wty,Ekstedt:2024etx}, and tools such as DRalgo facilitate its implementation in general models~\cite{Ekstedt:2022bff}. Its standard high-temperature formulation nevertheless relies on a hierarchy between thermal and field-dependent mass scales. Recent studies have highlighted the limitations of the high-temperature expansion as this hierarchy weakens, including the importance of higher-dimensional operators~\cite{Chala:2024xll,Chala:2025oul,Bernardo:2025vkz,Navarrete:2025yxy,Gould:2026rqz}. Functional methods based on the Cornwall--Jackiw--Tomboulis two-particle-irreducible (2PI) effective action instead determine dressed propagators self-consistently~\cite{Cornwall:1974vz,Amelino-Camelia:1993rvt}. Recent applications to the Abelian Higgs model and to mixing scalar theories illustrate their use beyond the high-temperature expansion~\cite{Klose:2026pux,Banik:2026ljf}, although consistent truncation and renormalisation of the resummed effective action remain technically involved. PD offers a comparatively simple four-dimensional resummation framework that retains the full mass dependence of the thermal functions while remaining tractable for sizeable coupled scalar and gauge systems.

In this work, we develop a prescription for incorporating mixing gauge bosons into the coupled gap equations and tadpoles underlying PD. We treat the longitudinal and transverse sectors separately and consistently account for the misalignment between the tree-level and diagonal basis. The construction is formulated for spontaneously broken non-Abelian gauge theories in a general $R_\xi$ gauge, including the associated Faddeev--Popov ghosts.

After formulating this prescription, we apply it to the electroweak sector of the SM, computing the mass-dependent thermal self-energies of the Higgs, Goldstone, gauge and ghost fields and solving the coupled gap equations with self-consistent $Z$--photon mixing. Only thermal self-energy contributions are iterated, while zero-temperature one-loop corrections remain outside the gap iteration. By comparing the full mass-dependent treatment with the leading high-temperature prescription, we analyse changes in the gauge-boson screening masses themselves and in their feedback through gauge loops in the scalar self-energies. We also examine how these effects propagate to the reconstructed PD effective potential. In addition to the latter, we study how the resummation impacts the effective number of entropy degrees of freedom $g_{\star s}$, which governs the relation between temperature and cosmological expansion and thereby affects the evolution of cosmological relics. While standard treatments often approximate $g_{\star s}$ by its ideal gas limit, it is known to be corrected by interactions \cite{Laine:2015kra,Saikawa:2018rcs}, and our calculation provides a complementary test of our implementation of PD. In addition, we also include a study of the impact of modifying the resummation to include the leading momentum-dependent contributions to the self-energies, which are not captured in traditional implementations of partial dressing.

The remainder of this paper is organised as follows: Sec.~\ref{sec:2} introduces the finite-temperature effective potential, starting from the zero-temperature one-loop contribution and its thermal extension, before discussing the infrared breakdown of the conventional perturbative expansion and the need for thermal resummation. The standard resummation prescriptions and the diagrammatic motivation for PD are reviewed afterwards. In Sec.~\ref{sec:3}, we construct the dressing procedure of the potential with coupled gap equations for the scalar and electroweak gauge sectors, including the separation of longitudinal and transverse gauge modes and the temperature-dependent mixing in the neutral gauge sector. Sec.~\ref{sec:4} collects the self-energies beyond the leading high-temperature expansion for all relevant sectors and verifies that their high-temperature limits reproduce the known leading-order results. Our numerical implementation and results are presented in Sec.~\ref{sec:5}. There we study the convergence of the coupled gap equations and the resulting thermal masses, compare the different resummation prescriptions at the level of the effective potential, and finally investigate their consequences for the effective number of entropy degrees of freedom $g_{\star s}$. We summarise our findings and discuss future work in Sec.~\ref{sec:6}.

\section{Finite-temperature effective potential}
\label{sec:2}
To study a thermal phase transition, we need to understand how quantum and thermal corrections change the effective potential. Its minima correspond to the possible phases, and their relative depths show which phase is favoured at a given temperature. The thermal corrections also depend on the particle masses, which change with the background field. Near the transition, bosonic modes can become light and higher-loop corrections can be as important as lower-order terms. We therefore need to resum these enhanced contributions to obtain a reliable perturbative description. This is the starting point for the self-consistent treatment developed in this work.

\subsection{Zero-temperature one-loop effective potential}
We begin by reviewing the zero-temperature effective potential and fixing the notation used throughout this work. We use the electroweak SM as a reference example to fix the effective-potential notation. Along a homogeneous Higgs background $h$, the tree-level potential is given by
\begin{equation}
    V_0(h)
    =\frac{\mu^2}{2}h^2
    +\frac{\lambda}{8}h^4,
    \label{eq:tree_level_potential}
\end{equation}
with $\mu^2<0$ and $\lambda>0$. The broken tree-level minimum is determined by $V_0'(v_0)=0$, giving
\begin{equation}
    v_0^2=\frac{-2\mu^2}{\lambda}.
\end{equation}
A field-independent constant $\mu^4/(2\lambda)$ may be added to Eq.~\eqref{eq:tree_level_potential} to normalise the potential to
$V_0(v_0)=0$.

Quantum fluctuations modify the classical potential. The appropriate object for describing these corrections is the effective action $\Gamma[h]$, i.e. the generating functional of one-particle-irreducible (1PI) correlation functions \cite{Jackiw:1974cv}. For a constant background field, the derivative terms in the effective action vanish, such that
\begin{equation}
    \Gamma[h]
    =
    - \int \mathrm{d}^4x \,
    V_{\mathrm{eff}}(h) .
    \label{eq:effective_action_potential}
\end{equation}
The effective potential thus contains the zero-external-momentum limit of the quantum corrections to the theory.

At one-loop order, the quantum corrections are obtained by expanding the action to quadratic order in fluctuations around the background field. For a particle species $i$ with field-dependent tree-level mass $m_i^2(h)$, the corresponding contribution is schematically given by the functional determinant 
\begin{equation}
    V_i^{(1)}(h)
    =
    \frac{n_i}{2}
    \mu_R^{2\epsilon}
    \int
    \frac{\mathrm{d}^d k_E}{(2\pi)^d}
    \log\!\left[
        k_E^2 + m_i^2(h)
    \right],
    \label{eq:one_loop_functional_determinant}
\end{equation}
with $d = 4-2\epsilon$, $k_E$ denoting the Euclidean momentum and $\mu_R$ being the renormalisation scale. The multiplicity $n_i$ includes the number of degrees of freedom as well as the appropriate sign for fermionic and Faddeev--Popov ghost contributions. 

After dimensional regularisation and renormalisation in the $\overline{\mathrm{MS}}$ scheme, the zero-temperature one-loop contribution takes the Coleman--Weinberg form
\cite{Coleman:1973jx,Jackiw:1974cv}
\begin{equation}
    V_{\mathrm{CW}}(h)
    =
    \sum_i
    \frac{n_i}{64\pi^2}
    m_i^4(h)
    \left[
        \log\!\left(
            \frac{m_i^2(h)}{\mu_R^2}
        \right)
        - c_i
    \right].
    \label{eq:coleman_weinberg}
\end{equation}
The constants $c_i$ depend on the particle species and the renormalisation scheme; in the $\overline{\mathrm{MS}}$ scheme, $c_i=3/2$ for scalars and fermions and $c_i=5/6$ for gauge bosons. The parameters entering the tree-level potential and the field-dependent masses are understood as renormalised running parameters evaluated at the renormalisation scale $\mu_R$.

The zero-temperature effective potential at one-loop order is therefore
\begin{equation}
    V_{\mathrm{eff}}^{T=0}(h)
    =
    V_0(h) + V_{\mathrm{CW}}(h),
    \label{eq:zero_temperature_effective_potential}
\end{equation}
up to an irrelevant field-independent contribution.

\subsection{Finite-temperature one-loop effective potential}
\label{sec:finite_temperature_potential}

At finite temperature, thermal fluctuations provide an additional contribution to the effective potential. Detailed discussions of finite-temperature field theory, the finite-temperature effective potential and its application to cosmological phase transitions can be found in Refs.~\cite{Quiros:1999jp,Laine:2016hma,Kapusta:2006pm}. In the following, we only summarise the ingredients needed for our subsequent discussion.

In the imaginary-time formalism, the Euclidean time is compactified with period $1/T$, and the continuous integration over the temporal momentum is replaced by a discrete sum over Matsubara frequencies \cite{Matsubara:1955ws},
\begin{equation}
    \int \frac{\mathrm{d} k_0}{2\pi}
    \;\longrightarrow\;
    T \sum_{n=-\infty}^{\infty},
    \label{eq:matsubara_replacement}
\end{equation}
with $\omega_{n}=2\pi nT$ for bosons and $\omega_{n}=(2n+1)\pi T$ for fermions. The regulated finite-temperature one-loop contribution of a species $i$ follows directly from Eq.~\eqref{eq:one_loop_functional_determinant},
\begin{equation}
    V_i^{(1)}(h,T)
    =\frac{n_iT}{2}\mu_R^{2\epsilon}\sum_n
    \int\frac{\mathrm d^{d-1}\mathbf k}{(2\pi)^{d-1}}
    \log\left[\omega_n^2+\mathbf k^2+m_i^2(h)\right].
    \label{eq:finite_temperature_functional_determinant}
\end{equation}
This expression contains both vacuum and thermal contributions, which can be separated. After renormalising the vacuum part and summing over species, the full finite-temperature potential takes the form
\begin{equation}
    V_{\mathrm{eff}}^{(1)}(h,T)
    =
    V_0(h)
    +
    V_{\mathrm{CW}}(h)
    +
    V_T(h,T).
    \label{eq:finite_temperature_effective_potential}
\end{equation}
The thermal contribution can be written as
\begin{equation}
    V_T(h,T)
    =
    \frac{T^4}{2\pi^2}
    \sum_i
    n_i\,
    J_{\rm B / \rm F}\!\left(
        \frac{m_i^2(h)}{T^2}
    \right),
    \label{eq:thermal_potential}
\end{equation}
where $J_{\rm B}$ and $J_{\rm F}$ denote the bosonic and fermionic thermal functions, respectively,
\begin{align}
    J_{\rm B/ \rm F}(y)
    &=
    \int_0^\infty \mathrm{d}x\,x^2
    \log\!\left(
        1\mp e^{-\sqrt{x^2+y}}
    \right).
    \label{eq:JB}
\end{align}
As in the zero-temperature contribution, $n_i$ contains the number of degrees of freedom and the appropriate sign for fermionic and ghost contributions. For Faddeev--Popov ghosts the bosonic thermal function is used, reflecting their periodic boundary conditions in imaginary time. 

The thermal functions retain the full dependence on the ratio $y=m_i^2/T^2$. In the high-temperature regime, $y\ll1$, they admit the expansions~\cite{Dolan:1973qd,Quiros:1999jp}
\begin{align}
    J_{\rm B}(y)
    &=
    -\frac{\pi^4}{45}
    +\frac{\pi^2}{12}y
    -\frac{\pi}{6}y^{3/2}
    -\frac{y^2}{32}
    \log\!\left(
        \frac{y}{a_{\rm B}}
    \right)
    +\mathcal{O}\!\left(y^3\right),
    \label{eq:JB_highT}
    \\
    J_{\rm F}(y)
    &=
    \frac{7\pi^4}{360}
    -\frac{\pi^2}{24}y
    -\frac{y^2}{32}
    \log\!\left(
        \frac{y}{a_{\rm F}}
    \right)
    +\mathcal{O}\!\left(y^3\right),
    \label{eq:JF_highT}
\end{align}
where $a_{\rm B} = 16\pi^2 \exp\!\left( \tfrac{3}{2}-2\gamma_E \right)$ and $a_{\rm F} = \pi^2 \exp\!\left( \tfrac{3}{2}-2\gamma_E \right)$.
In particular, the bosonic thermal potential contains the non-analytic contribution
\begin{equation}
    V_T^{\rm B}(h,T)
    \supset
    -\frac{T}{12\pi}
    \sum_{i\in {\rm B}}
    n_i\,m_i^3(h),
    \label{eq:bosonic_cubic_term}
\end{equation}
which has no fermionic analogue. This term originates from the bosonic Matsubara zero mode and will play an important role in the infrared behaviour of the perturbative expansion discussed below.

Conversely, when $m_i/T \gg 1$, the thermal contribution of the corresponding degree of freedom is exponentially suppressed. The full thermal functions therefore interpolate between the high-temperature regime and the decoupling of particles whose masses become large compared with the temperature. This mass dependence will be particularly relevant when discussing thermal resummation beyond the high-temperature approximation.

\subsection{Breakdown of fixed-order perturbation theory}
\label{sec:perturbative_breakdown}

The bosonic Matsubara zero mode discussed above is also responsible for a well-known breakdown of the conventional perturbative expansion at finite temperature. In particular, higher-loop diagrams can become parametrically enhanced when the masses of bosonic excitations are small compared with the temperature, such that the smallness of the underlying couplings alone no longer guarantees perturbative convergence \cite{Dolan:1973qd,Carrington:1991hz,Dine:1992wr,Espinosa:1992gq,
Curtin:2016urg}.

The origin of this behaviour can be illustrated in a simple real scalar theory with the same potential as in Eq.~\eqref{eq:tree_level_potential}. Here, $h$ is to be interpreted as the expectation value of a scalar field $\phi$, $\langle\phi\rangle=h$, and we neglect other fields for the sake of illustration. Expanding around a homogeneous background $h$, the field-dependent mass of the scalar fluctuations is $m^2(h)=\mu^2+ \tfrac32\lambda h^2$. In the following, we suppress the explicit background-field dependence of $m^2$.

Consider the leading one-loop contribution to the scalar self-energy, corresponding to the first diagram on the right-hand side of Fig.~\ref{fig:daisy_series}, given by
\begin{equation}
    \Pi
    =
    \frac{3}{2}\,\lambda \,I_1(m,T),
    \label{eq:scalar_tadpole}
\end{equation}
where we defined
\begin{equation}
    I_1(m,T)
    =
    T\sum_n
    \int\frac{\mathrm{d}^3k}{(2\pi)^3}
    \frac{1}{\omega_n^2+\mathbf{k}^2+m^2}.
    \label{eq:I1_definition}
\end{equation}
For the purpose of thermal power counting, $I_1$ can be separated into the
bosonic zero mode ($n=0$) and the non-zero Matsubara modes ($n\neq 0$):
\begin{equation}
    I_1(m,T)
    =
    T\int\frac{\mathrm{d}^3k}{(2\pi)^3}
    \frac{1}{\mathbf{k}^2+m^2}
    +
    T\sum_{n\neq0}
    \int\frac{\mathrm{d}^3k}{(2\pi)^3}
    \frac{1}{\omega_n^2+\mathbf{k}^2+m^2}.
    \label{eq:I1_soft_hard}
\end{equation}
In the high-temperature regime, the zero-mode contribution is governed by the soft scale $k\sim m$ and scales as $I_1^{\mathrm{soft}}\sim mT$, whereas the non-zero modes are governed by the hard thermal scale $k\sim T$ and scale as $I_1^{\mathrm{hard}}\sim T^2$. The leading thermal self-energy therefore scales as $\Pi_{\mathrm{hard}}\sim\lambda T^2$.

The perturbative problem becomes apparent when the hard thermal self-energy is repeatedly inserted into an infrared-sensitive zero-mode loop. A daisy contribution with $n$ such insertions, illustrated in Fig.~\ref{fig:daisy_series}, scales parametrically as
\begin{equation}
    \Pi_{\mathrm{daisy}}^{(n+1)}
    \sim
         \underbrace{ \vphantom{\left(\lambda T^2\right)^n}\lambda T m}_{\text{soft loop}}
        \times
        \underbrace{  \vphantom{\left(\lambda T^2\right)^n} m^{-2n}
    }_{\text{soft propagators}}
    \times
    \underbrace{
        \vphantom{\lambda T\,m^{1-2n}}
        \left(\lambda T^2\right)^n
    }_{\text{hard loops}}
    =
    \lambda^2\frac{T^3}{m}
    \left(
        \frac{\lambda T^2}{m^2}
    \right)^{n-1},
    \qquad n\geq1,
    \label{eq:daisy_self_energy_scaling}
\end{equation}
where numerical and symmetry factors have been omitted. Diagrammatically, each additional hard thermal self-energy insertion contributes a factor of order $\lambda T^2$, while the accompanying additional soft propagator produces an inverse factor of order $1/m^2$. Successive diagrams in the daisy series are therefore controlled by the thermal expansion parameter
\begin{equation}
    \alpha
    \equiv
    \frac{\lambda T^2}{m^2}.
    \label{eq:thermal_expansion_parameter}
\end{equation}

The crucial point is that $\alpha$ need not be small, even for $\lambda\ll1$. In the region relevant for a thermal phase transition, the thermal correction is generically expected to become comparable to the characteristic mass scale of the soft bosonic mode, $m^2\sim\lambda T^2$, implying $\alpha\sim\mathcal{O}(1)$. For example, a second-order phase transition occurs when the effective mass squared including self-energy corrections crosses zero, destabilising the $h=0$ configuration. The effective mass parameter near the origin of field space goes as $m^2_{\rm eff}\sim m^2(0)+\Pi\sim\mu^2+\lambda T^2$, which crosses zero when $|m^2|\sim|\mu^2|\sim\lambda T^2$, leading to $\alpha\sim 1$. Successive daisy diagrams are then no longer parametrically suppressed, and the fixed-order loop expansion ceases to provide a consistent ordering of contributions. Eq.~\eqref{eq:daisy_self_energy_scaling} also shows that individual higher-order daisy contributions become infrared divergent in the limit $m\to0$.

This infrared problem is specific to bosonic modes: fermions possess no Matsubara zero mode and therefore do not exhibit the corresponding enhancement. A consistent treatment of the bosonic sector therefore requires a reorganisation of the finite-temperature perturbative expansion that properly accounts for these infrared-enhanced contributions.

While we concentrated on the simplified case of a single real scalar field, the previous arguments can be generalised to more general settings, e.g. when complexifying the scalar and including gauge interactions. In the latter case the power counting in the gauge coupling can be obtained from the one in terms of $\lambda$ by substituting $\lambda\rightarrow g^2$.

\subsection{Thermal resummation}
\label{sec:thermal_resummation}

The breakdown of fixed-order perturbation theory discussed above requires a reorganisation of the finite-temperature expansion, which consistently accounts for the infrared-enhanced bosonic modes. In direct thermal resummation, the enhanced self-energy corrections are incorporated into thermally dressed propagators, thereby resumming classes of higher-loop diagrams such as daisies and superdaisies \cite{Dolan:1973qd,Espinosa:1992gq}.

A conceptually different and systematically improvable framework is provided by dimensional reduction, in which the non-zero Matsubara modes are integrated out and the long-distance dynamics is described by a three-dimensional effective field theory \cite{Kajantie:1995dw,Braaten:1995cm}. This approach provides a controlled organisation of the thermal expansion and can achieve high perturbative accuracy, but its implementation becomes increasingly involved at higher orders and requires a careful separation of the relevant thermal scales. In the present work, we instead focus on four-dimensional thermal resummation and on the different prescriptions for incorporating thermal self-energies directly into the effective potential.

\subsubsection{Daisy resummation and thermal masses}
\label{sec:daisy_resummation}

The infrared-enhanced contributions identified in Sec.~\ref{sec:perturbative_breakdown} arise from repeated insertions of the thermal self-energy into soft bosonic propagators. Since successive daisy diagrams are not parametrically suppressed when $\alpha\sim\mathcal{O}(1)$, they have to be included to all orders. Daisy resummation achieves this by summing the corresponding series of self-energy insertions, as illustrated schematically in Fig.~\ref{fig:daisy_series}.

The effect of the resummation can be expressed in terms of a thermally dressed propagator, obtained by replacing the tree-level field-dependent mass by a thermal mass,
\begin{equation}
    m_i^2(h)
    \;\longrightarrow\;
     \overline{m}_i^2(h,T)
    \equiv
    m_i^2(h)
    +
    \Pi_i(0)(h,T),
    \label{eq:thermal_mass_definition}
\end{equation}
where $ m_i^2(h)$ denotes the tree-level mass of a bosonic species $i$ in the scalar background $\langle\phi\rangle=h$, while $\Pi_i(0)(h,T)$ denotes the corresponding thermal self-energy at zero external momentum. At leading order in the high-temperature expansion, $\Pi_i^{\rm HT}=c_i^{\rm HT}T^2$, where the coefficients $c_i^{\rm HT}$ are determined by the interactions of the respective field. The resulting thermal masses screen the soft bosonic modes and remove the infrared enhancement responsible for the breakdown of the fixed-order expansion. For scalar fields, the thermal self-energy directly generates a screening mass, while in the gauge sector only the electric or longitudinal zero modes acquire Debye masses at leading order in the high-temperature expansion. Fermionic fields have no Matsubara zero mode and therefore do not require thermal mass resummation.

Equation~\eqref{eq:thermal_mass_definition}, however, does not by itself specify how the dressed masses should be incorporated into the effective potential in Eq.~\eqref{eq:finite_temperature_effective_potential}. Different implementations of this replacement resum different sets of higher-order contributions and, beyond leading order, need not agree. This ambiguity underlies the various thermal resummation prescriptions discussed below.

\subsubsection{Conventional resummation prescriptions }
\label{sec:conventional_resummation}

Several prescriptions have been proposed for incorporating thermal masses into the effective potential. The two most commonly employed approaches are the Arnold--Espinosa \cite{Arnold:1992rz} and Parwani \cite{Parwani:1991gq} prescriptions. In their conventional implementations, both employ the leading high-temperature dressed masses,
\begin{equation}
    \overline{m}_{i,\mathrm{HT}}^2(h,T)
    \equiv
    m_i^2(h)
    +
    \Pi_i^{\mathrm{HT}}(T).
    \label{eq:HT_thermal_mass}
\end{equation}
The two prescriptions differ in how these masses enter
the one-loop effective potential.

The Arnold--Espinosa prescription targets directly the infrared-sensitive bosonic Matsubara zero modes. As seen from the high-temperature expansion in Eq.~\eqref{eq:JB_highT}, their contribution gives rise to the non-analytic cubic term in Eq.~\eqref{eq:bosonic_cubic_term}. Resummation is implemented by replacing the tree-level masses by thermal masses only in the non-analytic contributions. Equivalently, one adds the ring term
\begin{equation}
    V_{\mathrm{ring}}(h,T)
    =
    -\frac{T}{12\pi}
    \sum_{i}
    n_i
    \big(
        \overline{m}_{i,\mathrm{HT}}^3(h,T)-m_i^3(h)
    \big)
    \label{eq:arnold_espinosa_ring}
\end{equation}
to the one-loop potential in Eq.~\eqref{eq:finite_temperature_effective_potential}:
\begin{equation}
    V_{\mathrm{eff}}^{\rm{AE}}(h,T)
    =V_0(h)
    +V_{\mathrm{CW}}(h)
    +V_T(h,T)
    + V_\mathrm{ring}(h,T).
\end{equation}
The subtraction removes the undressed zero-mode contribution already contained in $V_T$, while the Coleman--Weinberg contribution and the non-zero Matsubara modes remain evaluated with the tree-level masses. In this way, the Arnold--Espinosa prescription modifies only those modes responsible for the leading infrared enhancement.

The Parwani prescription instead makes the mass replacement throughout the one-loop corrections in Eq.~\eqref{eq:finite_temperature_effective_potential}. For the bosonic degrees of freedom to be resummed, one therefore evaluates
\begin{equation}
    V_{\mathrm{eff}}^{\mathrm{Parwani}}(h,T)
    =
    V_0(h)
    +
    V_{\mathrm{CW}}\!\left(\overline{m}_{i,\mathrm{HT}}^2(h,T)\right)
    +
    V_T\!\left(\overline{m}_{i,\mathrm{HT}}^2(h,T),T\right).
    \label{eq:parwani_potential}
\end{equation}
Consequently, the dressed masses enter not only the infrared-sensitive zero-mode contribution, but also the non-zero thermal modes and the zero-temperature Coleman--Weinberg potential. This prescription is also commonly referred to as truncated full dressing (TFD) \cite{Curtin:2016urg}.

Both prescriptions reproduce the leading daisy resummation required to cure the infrared breakdown discussed in Sec.~\ref{sec:perturbative_breakdown}. Their difference concerns formally higher-order contributions: the Arnold--Espinosa prescription restricts the thermal-mass insertion to the infrared-sensitive zero modes, whereas the Parwani prescription generates additional higher-order terms by inserting the same thermal masses throughout the one-loop functions. The two prescriptions therefore agree at the order for which the leading daisy resummation is required, but generally differ once subleading thermal corrections become relevant
\cite{Curtin:2016urg,Bahl:2024bcg}.

A common limitation of both conventional implementations is that the self-energies entering Eq.~\eqref{eq:HT_thermal_mass} are truncated to their leading high-temperature behaviour. This approximation assumes that the masses of the particles propagating in the thermal loops are small compared with the temperature. In realistic phase-transition calculations, however, the field-dependent masses vary with the order parameter $h$ and can become comparable to, or larger than, the temperature. In this regime, the neglected mass-dependent contributions to the self-energies are no longer parametrically small. Moreover, the leading expression $\Pi_i^{\mathrm{HT}}\propto T^2$ does not account for the thermal decoupling of heavy degrees of freedom, whose contributions should become suppressed as $m/T$ increases. This can be particularly relevant when the thermally corrected mass results from a partial cancellation between the tree-level contribution and the thermal self-energy, since even moderate corrections to $\Pi_i$ may then induce sizeable changes in the resulting screening mass and, consequently, in the phase-transition dynamics \cite{Curtin:2016urg}.

A treatment valid beyond the strict high-temperature regime should therefore retain the dependence of the thermal self-energies on the masses of the particles propagating in the loops. Since these masses are themselves modified by thermal effects, this naturally leads to a self-consistent determination of the dressed masses through gap equations.

\subsubsection{Gap resummation and partial dressing}
\label{sec:partial_dressing}
\begin{figure}[t!]
    \centering
    \begin{tikzpicture}
        \node[anchor=south west, inner sep=0] (img) at (0,0)
            {\includegraphics[width=\textwidth]{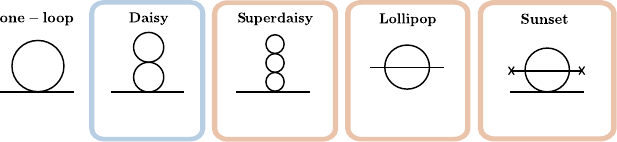}};
            \begin{scope}[
            x={(img.south east)},
            y={(img.north west)}]
            \node[font=\normalsize] at (0.060,0.17)
                {$\sim \lambda T^2$};
            \node[font=\normalsize] at (0.24,0.17)
                {$\sim \lambda^2 \frac{T^3}{M}
                  = \lambda T^2\beta$};
            \node[font=\normalsize] at (0.447,0.17)
                {$\sim \lambda^3 \frac{T^4}{M^2}
                  = \lambda T^2\beta^2$};
            \node[font=\normalsize] at (0.662,0.17)
                {$\sim \lambda^2 T^2
                  = \lambda T^2\beta^2$};
            \node[font=\normalsize] at (0.888,0.17)
                {$\sim \lambda^3 h^2 \frac{T^2}{M^2}
                  = \lambda h^2\beta^2$};
        \end{scope}
    \end{tikzpicture}
    \caption{Schematic higher-order contributions to the thermal mass \cite{Curtin:2016urg}. Besides the daisy families, superdaisy, lollipop and sunset topologies contribute at subleading order in the reorganised thermal expansion. The displayed scalings are parametric and omit numerical and symmetry factors. Diagrams contributing at $\mathcal{O}(\beta)$ are marked in blue, whereas diagrams contributing at $\mathcal{O}(\beta^2)$ are marked in brown; the powers of $\beta$ are valid in the critical region $T\sim M/\sqrt{\lambda}$. Insertions of the background field $h$ in the vertices are denoted with crosses.}
    \label{fig:higher_order_topologies}
\end{figure}

Going beyond the leading high-temperature approximation requires retaining the mass dependence of the thermal self-energies. Since the masses appearing inside the thermal loops are themselves modified by these self-energies, the dressed masses can be determined self-consistently. This leads to the coupled gap equations
\begin{equation}
    M_i^2(h,T)
    =
    m_i^2(h)
    +
    \Pi_i(0)\!\left(
        h,T;\{M_j^2(h,T)\}
    \right),
    \label{eq:gap_equation_general}
\end{equation}
where the set $\{M_j^2\}$ denotes the thermally dressed masses of the degrees of freedom propagating inside the self-energy. In general, Eq.~\eqref{eq:gap_equation_general} constitutes a non-linear system that must be solved self-consistently~\cite{Espinosa:1992gq}. If the self-energies are instead truncated to their leading high-temperature values, the mass dependence on the right-hand side disappears and one recovers the thermal masses $\overline{m}_{i,\mathrm{HT}}^2$ of Eq.~\eqref{eq:HT_thermal_mass}.

Solving the gap equations resums the thermally enhanced self-energy insertions, but it does not by itself imply that the resulting perturbative expansion is complete. It is therefore useful to ask which contributions remain relevant once the leading $\alpha$-enhanced terms have been resummed. For this purpose, we return briefly to the single-scalar theory introduced in Sec.~\ref{sec:perturbative_breakdown}. Besides the daisy diagrams discussed above, the relevant leading higher-order contributions can be organised into superdaisy, lollipop and sunset topologies, as illustrated in Fig.~\ref{fig:higher_order_topologies}.

After the leading thermal enhancement has been resummed, the hierarchy between successive contributions is controlled by a second expansion parameter,
\begin{equation}
    \beta
    \equiv
    \frac{\lambda T}{M},
    \label{eq:beta_expansion_parameter}
\end{equation}
where $M$ denotes the characteristic screened mass of the soft mode. In the region in which thermal resummation is required, i.e. $\alpha=\lambda T^2/m^2\sim 1$, one has  $T \sim M/\sqrt{\lambda}$, which, as mentioned, before is of the order of the expected critical temperature, at least for a second-order phase transition. For such temperatures one has $\beta\sim\sqrt{\lambda}$. Thus, although the original expansion in $\alpha$ has broken down, the reorganised perturbative expansion can remain controlled for weak coupling. The leading contributions to the scalar mass correction can be found in Fig.~\ref{fig:higher_order_topologies}. They scale parametrically as
\begin{align}
    \delta M_{1}^{2}
    &\sim \lambda T^2,
    &
    \delta M_{\mathrm{D}}^{2}
    &\sim \lambda^2\frac{T^3}{M},
    &
    \delta M_{\mathrm{SD}}^{2}
    &\sim \lambda^3\frac{T^4}{M^2},
    \nonumber\\
    \delta M_{\mathrm{LL}}^{2}
    &\sim \lambda^2T^2,
    &
    \delta M_{\mathrm{SS}}^{2}
    &\sim
    \lambda^3h^2\frac{T^2}{M^2},
    \label{eq:higher_order_thermal_scaling}
\end{align}
where $\delta M_{\mathrm{D}}^{2}$, $\delta M_{\mathrm{SD}}^{2}$, $\delta M_{\mathrm{LL}}^{2}$ and $ \delta M_{\mathrm{SS}}^{2}$ denote daisy, superdaisy, lollipop and sunset contributions, respectively, and numerical and symmetry factors have been omitted. In the transition region $T \sim M/\sqrt{\lambda}$, their relative scaling is therefore schematically
\begin{equation}
    \frac{\delta M_{\mathrm{D}}^2}{\delta M_1^2}
    \sim
    \frac{\delta M_{\mathrm{SD}}^2}{\delta M_{\mathrm{D}}^2}
    \sim
    \beta,
    \qquad
    \frac{\delta M_{\mathrm{LL}}^2}{\delta M_{\mathrm{D}}^2}
    \sim
    \beta,
    \qquad
    \frac{\delta M_{\mathrm{SS}}^2}{\delta M_{\mathrm{D}}^2}
    \sim
    \beta\,\frac{h^2}{T^2},
    \label{eq:beta_power_counting}
\end{equation}
up to additional numerical and symmetry suppressions. In an $O(N)$-symmetric scalar theory, the multiplicity factors provide an additional large-$N$ suppression of the non-daisy topologies, with lollipop and sunset contributions suppressed parametrically by $1/N$ and $1/N^2$, respectively \cite{Curtin:2016urg}. Consequently, after the leading daisy resummation, superdaisy, lollipop and sunset contributions enter at the next order of the improved thermal expansion. 

Importantly, solving the gap equations~\eqref{eq:gap_equation_general} while using one-loop expressions for the self-energies $\Pi_i$ amounts to resumming all possible nested insertions of one-loop daisy and sunset self-energy corrections, to arbitrary total loop order. This can be understood diagrammatically by identifying the dressed masses $M^2_i$ and the tree-level masses $m^2_i$ with dressed and tree-level 1PI correlators at zero momentum, respectively. Then Eq.~\eqref{eq:gap_equation_general} can be written as the relation between amputated 1PI diagrams:
\begin{equation}
    \raisebox{-0.7cm}{
        \includegraphics[width=0.56\textwidth]
        {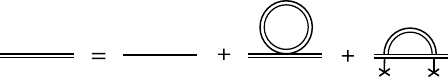}
    }
    \label{eq:gapgraphic}
\end{equation}
This relation can be solved recursively in a way that generates all possible nested insertions of one-loop self-energy corrections:
\begin{equation}
    \raisebox{-1.3cm}{
        \includegraphics[width=0.91\textwidth]{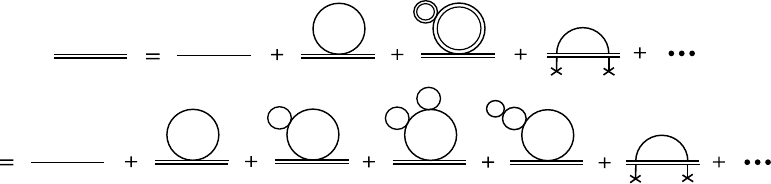}
    }
    \label{eq:gap_solution}
\end{equation}

Once the gap equations have been solved, one still has to specify how their solutions are used to construct the resummed effective potential. The most direct possibility, analogous to the Parwani prescription discussed above, is to substitute the solution of Eq.~\eqref{eq:gap_equation_general} directly into the one-loop potential of Eq.~\eqref{eq:finite_temperature_effective_potential},
\begin{equation}
    V_{\mathrm{eff}}^{\mathrm{FD}}(h,T)
    =
    V_0(h)
    +
    V_{\mathrm{CW}}\!\left(M_i^2(h,T)\right)
    +
    V_T\!\left(M_i^2(h,T),T\right),
    \label{eq:full_dressing}
\end{equation}
a procedure referred to as full dressing (FD). When the self-energies are truncated to their leading high-temperature form, this reduces at one-loop order to the TFD prescription.

Despite its intuitive form, FD already fails to reproduce the correct combinatorial factors at the level of daisy and superdaisy vacuum diagrams --- as opposed to self-energy diagrams --- beyond leading order. The effective potential can be understood as a sum over all vacuum 1PI diagrams, and while solving the gap equation captures the daisy and superdaisy corrections of the self-energy diagrams, the FD prescription does not generate the correct tower of daisy and superdaisy vacuum diagrams.

This becomes clear upon differentiating the dressed potential with respect to the background field. For a potential of the form $V_0(h)+V^{(1)}(h,T)$, substituting the dressed masses into the one-loop expression of Eq.~\eqref{eq:finite_temperature_functional_determinant} gives the FD potential in the form
\begin{align}\label{eq:V_FD}
   V^{\rm FD}_{
   \rm eff}(h,T) =V_0(h)+\sum_i\frac{n_iT}{2}\mu_R^{2\epsilon}\sum_n
    \int\frac{\mathrm d^{d-1}\mathbf k}{(2\pi)^{d-1}}
    \log\left[\omega_n^2+\mathbf k^2+M_i^2(h,T)\right],
\end{align}
implying that the tadpole corresponding to the FD potential is given by 
\begin{align}\label{eq:Vprime_FD}
\partial_h V^{\rm FD}_{
   \rm eff}(h,T)\equiv  {V^{\rm FD}}'_{
   \rm eff}(h,T) =V'_0(h)+\sum_i\frac{n_iT}{2}\mu_R^{2\epsilon}\sum_n
    \int\frac{\mathrm d^{d-1}\mathbf k}{(2\pi)^{d-1}}
    \frac{\partial_h M^2_i(h,T)}{\omega_n^2+\mathbf k^2+M_i^2(h,T)}.
\end{align}
The tadpole $V'$ is meant to be given by the sum of 1PI diagrams with one external leg, which Eq.~\eqref{eq:Vprime_FD} fails to reproduce. The culprit is the derivative of the resummed mass squared in the numerator of the last term on the right-hand side. To see this, it suffices to realise that a resummation of superdaisy diagrams and sunset insertions  can be expressed purely in terms of a resummed propagator, without any vertex corrections, as illustrated next:
\begin{equation}
\begin{aligned}
    V'_{\mathrm{eff}}(h,T)
    &= V'_0(h) \ \ 
    \raisebox{0.02cm}{%
    $\diagram{1.9cm}{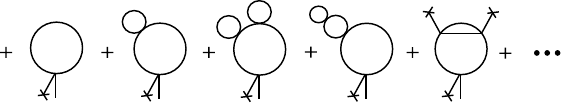}$%
    }
    \\[0.5em]
    &= V'_0(h) \ \ 
    \raisebox{-0.16cm}{%
    $\diagram{1.5cm}{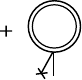}$%
    }
    \ \ 
    = V'_0(h)
    + \frac{T}{2}\,\mu_R^{2\epsilon}
      \sum_i \sum_n n_i\int \frac{\rm d^{d-1}\mathbf{k}}{(2\pi)^{d-1}}
      \frac{\partial_h m_i^2(h)}
           {\omega_n^2+\mathbf{k}^2+M_i^2(h,T)} .
\end{aligned}
\label{eq:correct_tadpole_resummation}
\end{equation}
The identification of the resummed propagator in Eq.~\eqref{eq:correct_tadpole_resummation} follows from Eq.~\eqref{eq:gap_solution}. The factor of $\partial_h m^2_i(h)$ corresponds to the Feynman rule for the tree-level vertex with an external scalar leg and a background insertion. The above result differs from Eq.~\eqref{eq:Vprime_FD} precisely in this factor, as Eq.~\eqref{eq:Vprime_FD} features $\partial_h M^2_i(h,T)$ instead of $\partial_h m^2_i(h)$. Diagrammatically,  this corresponds to  dressing not only the propagator but also the associated tadpole vertex, which leads to a miscounting of subleading daisy and superdaisy contributions, beginning at two-loop order \cite{Boyd:1993tz}.

Eq.~\eqref{eq:correct_tadpole_resummation} motivates the partial-dressing (PD), or tadpole-resummation, prescription, defined by its last equality~\cite{Boyd:1993tz,Curtin:2016urg}:
\begin{align}\label{eq:Vprime_PD_general}\begin{aligned}
{V_{\mathrm{eff}}^{\mathrm{PD}}}'(h,T)
    =&\,
    V'_0(h) +\frac{T}{2}\,\mu_R^{2\epsilon}
      \sum_i \sum_n n_i \int\frac{\mathrm d^{d-1}\mathbf k}{(2\pi)^{d-1}}
    \frac{\partial_h m^2_i(h)}{\omega_n^2+\mathbf k^2+M_i^2(h,T)}\\
    =&\,V'_0(h) +\left.\sum_i {V^{(1)}_i}'(h,T)\right|_{m^2_i(h)\rightarrow M^2_i(h,T)}\\
     =&\,V'_0(h)+\left[V_{\rm CW}'(h)+V_T'(h,T)\right]_{m^2_i(h)\rightarrow M^2_i(h,T)}.
\end{aligned}\end{align}
In the previous identities we made use of Eqs.~\eqref{eq:finite_temperature_functional_determinant} and \eqref{eq:finite_temperature_effective_potential}. In this prescription, which by construction correctly sums all superdaisy diagrams, the one-loop potential is first differentiated with respect to the background field, after which the masses in the internal propagators are replaced by the solutions of the gap equations, while the corresponding tree-level vertices are left undressed. The resummed potential is then reconstructed by integration,
\begin{equation}
    V_{\mathrm{eff}}^{\mathrm{PD}}(h,T)
    =
    V_0(h) +
    \int \mathrm{d} h \,
    \bigg[ \frac{\partial V_{\mathrm{CW}}}{\partial h}
    +
    \frac{\partial V_T}{\partial h} \bigg]_{\displaystyle m^2_i(h)\rightarrow M^2_i(h,T)}.
    \label{eq:partial_dressing}
\end{equation}
Here, $V_{\mathrm{CW}}$ and $V_T$ retain the sums over particle species defined in Eqs.~\eqref{eq:coleman_weinberg} and \eqref{eq:thermal_potential}, and the notation indicates that the derivatives are taken before the tree-level masses in the internal propagators are replaced by the corresponding solutions of the gap equations.

\subsubsection{Power counting for gauge interactions}
\label{sec:gauge_power_counting}
Before delving into the implementation of PD for non-Abelian gauge theories, it is useful to consider whether the power-counting arguments discussed so far for a simple scalar theory can be generalised to the case of the gauge interactions with a coupling constant $g$. For the gap equation of scalar self-energies, daisy and sunset-type contributions involving gauge fields enter with factors of $g^2$, so that the previous power counting arguments are expected to hold under the substitution $\lambda\rightarrow g^2$. In particular, the insertions of gauge field corrections to the scalar self-energy will be controlled by a parameter 
\begin{align}
    \beta_{g}=\frac{g^2T}{M},
\end{align}
where $M\sim |\mu|$ is again the characteristic mass of the scalar. Then for $T\sim M/\sqrt{\lambda}\sim M/g$, the effective coupling constant $\beta_{g}$ can remain perturbative, so that one can use perturbation theory with the resummed propagators. Moving on to gauge-boson self-energies, in the broken phase the role of $M$ is played by the gauge-boson masses $m_{\rm g}\sim g h$ induced by symmetry breaking. Then one can have an effective coupling in the gauge sector going as $\gamma=g^2T/m_{\rm g}\sim g\, T/h$ remaining perturbative, justifying the partial resummation implemented by the use of the gap equations. Near $h=0$, however, the tree-level mass goes to zero, and $m_{\rm g}$ should include thermal corrections. For longitudinal gauge bosons, it is known that thermal corrections generate a Debye mass $M_{g,L}\sim g T$ \cite{Carrington:1991hz}, leading to a perturbative value of the effective coupling, $\gamma_{L}\sim g$. Hence, in the longitudinal sector the theory with resummed propagators can remain under perturbative control. In the transverse sector, on the other hand, thermal corrections to the masses are expected to go as $g^2T$ \cite{Linde:1980ts,Buchmuller:1993bq, Buchmuller:1994qy}, leading to a strong effective coupling $\gamma_{T}\sim 1$. In contrast to the longitudinal sector, the resummation by means of the gap equation is not reliable because the resulting theory with dressed propagators remains nonperturbative.

In conclusion, when extending the PD method to gauge fields, the resummation is well motivated for longitudinal excitations regardless of the scalar background $h$. Transverse excitations, however, are expected to become strongly coupled for $h\rightarrow0$, so that the resummation will become unreliable in this regime. Unfortunately, the $h\rightarrow0$ region is particularly crucial in the case of a second-order or crossover phase transition. Given the above, in this paper we will put our main focus on results in which the resummation is restricted to the longitudinal gauge-boson sector. We will however study the impact of carrying out the resummation in the transverse sector, bearing in mind that the results near $h=0$ may only be interpreted qualitatively, as they are expected to have an ${\cal O}(1)$ uncertainty from the large effective coupling $\gamma_T$.

\subsubsection{Limitations of the PD approach}
\label{sec:pd_limitations}
\begin{figure}[t!]
    \centering
    \includegraphics[width=0.6\textwidth]{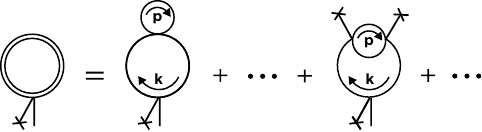}
    \caption{ Schematic self-energy insertions appearing in the tadpole expansion underlying PD. The momentum $k$ flows through the surrounding tadpole loop, while $p$ denotes the characteristic momentum inside the inserted self-energy. Momentum-independent insertions (first diagram on the right) can be evaluated directly in the static limit. For momentum-dependent insertions, the replacement $\Pi(k)\to\Pi(0)$ is justified when $k\ll p$, whereas for soft contributions with $k\sim p$ this hierarchy is lost and the loop momenta overlap, so that in the static approximation the gap equation does not reproduce the complete sunset contribution.}
\label{fig:momentum_overlap}
\end{figure}

While it was argued that PD correctly resums nested daisy and sunset-type self-energy corrections, there is a caveat.  The thermal masses entering the gap equations are defined from the self-energies in the infrared limit, 
\begin{equation}
\Pi(0) \equiv \Pi(k_0=0,\mathbf{k}\rightarrow0),
\end{equation}
Using such self-energies inside nested loops would be justified as long as one can motivate a hierarchy among the loop momenta. This, however, is not always the case, as illustrated in Fig.~\ref{fig:momentum_overlap}. Consider a momentum-dependent self-energy $\Pi(k)$, with internal loop momentum $p$, inserted into a surrounding tadpole loop with momentum $k$. If the characteristic momentum $p$ is parametrically larger than $k$, the self-energy can be expanded in $k/p$ and its leading contribution is obtained by setting $k=0$. For a hard thermal insertion one has $k\sim M\ll p\sim T$, and the infrared   approximation is therefore justified. For subleading soft contributions, however, both loop momenta can be of the same order, $k\sim p\sim M$. The required hierarchy is then lost, and replacing $\Pi(k)$ by $\Pi(0)$ factorises loop integrations whose momenta are in fact correlated. This is only a problem for momentum-dependent self-energy corrections, such as sunset-type contributions \cite{Arnold:1992rz,Boyd:1993tz}, as illustrated in the right diagram of Fig.~\ref{fig:momentum_overlap}. Note, however, that in the scalar case such contributions involve insertions of the background field $h$, and thus become suppressed near $h=0$. This argument does not extend to all momentum-dependent contributions in the electroweak theory: mixed gauge--scalar and fermion bubbles have no such vertex suppression, as discussed in Appendix~\ref{app:momentum_dependence}.

The numerical importance of these corrections has recently been investigated in scalar theories. Ref.~\cite{Curtin:2016urg} finds the sunset miscounting to be subleading compared with the dominant neglected corrections, while Ref.~\cite{Bahl:2024bcg} explicitly corrects the overlapping-momentum contribution and finds only a minor numerical effect, significantly smaller than the differences between PD and the conventional resummation schemes. Although these results cannot be transferred directly to the electroweak theory, they suggest that the corresponding correction can remain numerically small in perturbative regimes.

An additional caveat related to the PD prescription is that the superdaisy resummation does not generate the lollipop diagrams illustrated in Fig.~\ref{fig:higher_order_topologies}. These diagrams are of the same order $\beta^2$ as superdaisy diagrams, so that they should be included to achieve full $\beta^2$ accuracy. In theories with field multiplets with $N_f$ components, however, the lollipop diagrams can be suppressed by powers of $1/N_f$ with respect to superdaisy diagrams.

In the present work, we do not include explicit lollipop or momentum-overlap corrections. It is worth stressing, however, that these omissions do not represent a fundamental limitation of the resummation framework. In scalar $\phi^4$ theory, the hybrid construction proposed in Ref.~\cite{Boyd:1993tz} incorporates these contributions by resumming momentum-independent self-energy insertions through a gap equation and evaluating the remaining overlapping-momentum vacuum diagrams explicitly. Recently, Ref.~\cite{Berthiaume:2026dkt} generalised this construction to non-mixing multi-scalar theories. Their hybrid partial dressing (HPD) prescription includes the full vacuum and thermal sunset contributions, reproducing the complete two-loop effective potential at arbitrary temperature
without the momentum-overlap problem. Extending this construction to the electroweak theory requires a separate diagrammatic analysis. Derivative gauge interactions can generate leading $\mathcal O(g^2T^2)$ contributions through bubble self-energies, so restricting the gap equations to tadpole topologies would omit leading thermal terms. We leave the inclusion of explicit higher-loop contributions, with the subtractions needed to avoid double counting, for future work.

The numerical evaluation of the full PD prescription can become computationally demanding, since the mass-dependent gap equations have to be solved for each choice of background field and temperature. Ref.~\cite{Curtin:2016urg} therefore introduced optimised partial dressing
(OPD) as a computationally more efficient implementation of the same
tadpole-resummation principle. The essential observation is that the thermal functions entering the effective potential and those entering the gap equations need not be treated at the same level of approximation. In its original formulation, the gap equations were simplified using the controlled high-temperature expansions in Eqs.~\eqref{eq:JB_highT} and \eqref{eq:JF_highT}, while the full thermal functions were retained when evaluating the resummed tadpole and reconstructing the effective potential. This was found to reproduce the full PD result to good accuracy at a substantially reduced numerical cost. More recently, Ref.~\cite{Curtin:2022ovx} revisited and refined the numerical implementation of OPD, improving the practical solution of the gap equations and further assessing the validity of the approximations entering the original construction.

The phenomenological relevance of the improved resummation has become increasingly apparent. Already in singlet extensions, (O)PD was found to modify the thermal masses and, in parts of parameter space, the predicted thermal history compared with conventional truncated full dressing \cite{Curtin:2016urg}. More recent studies have found even larger effects. Ref.~\cite{Curtin:2022ovx} demonstrated in scalar benchmark theories that (O)PD substantially improves the perturbative behaviour and renormalisation-scale dependence relative to Parwani resummation, with differences in the predicted gravitational-wave signal reaching orders of magnitude when the high-temperature approximation becomes inaccurate. PD has subsequently been generalised to systems of mixing scalar fields \cite{Bahl:2024bcg} and applied to the two-Higgs-doublet model \cite{Bittar:2025lcr,Zhao:2026tob}, where the choice of resummation prescription was shown
to affect the phase structure and gravitational-wave predictions. These results demonstrate that the treatment of thermal masses can influence not only precision observables but also qualitative conclusions about the thermal history.

Despite this progress, the self-consistent gap resummation has so far been developed primarily for the scalar  sector. In the original PD implementation, gauge-boson thermal masses were retained at their conventional leading high-temperature values, while the coupled gap equations were solved for the scalar masses \cite{Curtin:2016urg}. Subsequent developments have extended PD to full thermal functions and to mixing scalar systems \cite{Curtin:2022ovx,Bahl:2024bcg}, but have not incorporated the electroweak gauge-boson screening masses into the same self-consistent gap system. This remains true in the recent two-Higgs-doublet-model application of Ref.~\cite{Bittar:2025lcr}, where the full gap equations are solved for the scalar degrees of freedom while the longitudinal gauge bosons are assigned their leading high-temperature Debye masses.

These considerations motivate a general construction of self-consistent PD for non-Abelian gauge theories, in which the scalar, gauge-boson and ghost sectors are coupled through their thermal gap equations. Retaining the complete thermal functions in the self-energies allows the contributions of heavy degrees of freedom to be continuously suppressed as their masses become large compared with the temperature, rather than retaining the unsuppressed behaviour implied by the strict high-temperature approximation. The electroweak SM serves as a concrete application of this general construction.

\section{Partial dressing beyond leading order}
\label{sec:3}

Thermal self-energies can change both the masses and the mixing of the fields running in a loop. We formulate PD in terms of general mass matrices, keeping the tadpole vertices at tree level while dressing the propagators. When mixed fields are transformed to a mass basis, the same transformations must enter their contractions with the vertices. This provides a common formulation for the scalar, gauge-boson and ghost sectors, with the dressed masses and mixing determined by coupled thermal gap equations.

We give the general expressions in each sector and then specialise them to the electroweak SM. In this application, we treat longitudinal and transverse gauge modes separately and determine the $Z$--photon mixing self-consistently. The scalar, gauge-boson and ghost contributions form a coupled system, allowing us to assess how gauge-sector corrections feed back into the scalar masses and the effective potential. We work in a general $R_\xi$ gauge and keep the gauge-fixing parameter $\xi$ arbitrary throughout this section.
\subsection{Gauge-theory setup}
\label{subsec:gauge_theory_setup}

For the gauge-theory setup, we follow a general background-field formulation (see e.g. Refs.~\cite{Patel:2011th,Martin:2018emo}). We consider a renormalisable gauge theory with canonical kinetic terms and gauge group $G=\prod_r G_r$, allowing both non-Abelian and Abelian factors. The scalars are collected in a vector of real components $\Phi_i$. We use a single index $a$ for the generators of all gauge-group factors, with $g_a$ denoting the coupling of the corresponding factor. Scalar indices are denoted by $i,j$, gauge indices by $a,b$. Repeated indices are summed unless they are free or explicitly excluded from summation. 
The Lagrangian is
\begin{equation}
    \mathcal L
    =-\frac14 F_{\mu\nu}^aF^{a\mu\nu}
    +\frac12(D_\mu\Phi)^T(D^\mu\Phi)-V_0(\Phi)
    +\mathcal L_{\rm f}+\mathcal L_{\rm gf}+\mathcal L_{\rm FP}.
    \label{eq:general_gauge_lagrangian}
\end{equation}
Here $\mathcal L_{\rm f}$ contains the fermion kinetic, mass and Yukawa terms. The gauge-fixing term $\mathcal L_{\rm gf}$ and the associated Faddeev--Popov ghost term $\mathcal L_{\rm FP}$ are specified below. We use the Minkowski metric $(+,-,-,-)$ and standard gauge-theory conventions~\cite{Peskin:1995ev},
\begin{align}
    D_\mu\Phi
    &=\partial_\mu\Phi-\mathrm i g_aA_\mu^at^a\Phi,
    \nonumber\\
    F_{\mu\nu}^a
    &=\partial_\mu A_\nu^a-\partial_\nu A_\mu^a
    +g_af^{abc}A_\mu^bA_\nu^c.
    \label{eq:general_gauge_derivatives}
\end{align}
Here $t^a$ are Hermitian, purely imaginary scalar generators and $f^{abc}$ are the structure constants.

For the effective potential, we choose a one-parameter path of homogeneous scalar backgrounds $\bar\Phi(h)$ and write
\begin{equation}
    \Phi_i(x)=\bar\Phi_i(h)+\chi_i(x).
    \label{eq:general_scalar_background}
\end{equation}
Here $h$ parametrises the background path. The background $\bar\Phi(h)$ need not minimise $V_0$. For convenience, we also define
\begin{equation}
K_{ia}(h)=-\mathrm i g_a(t_a)_{ij}\bar\Phi_j(h),
\label{eq:K_matrix}
\end{equation}
where $t_a\equiv\delta_{ab}t^b$ and there is no sum over $a$. $K$ is a real matrix and determines the gauge--scalar mixing induced by the background. We choose a common-parameter background $R_\xi$ gauge~\cite{tHooft:1971qjg,Fujikawa:1972fe},
\begin{equation}
    \mathcal L_{\rm gf}=-\frac{1}{2\xi}\mathcal F^a\mathcal F^a,
    \qquad
    \mathcal F^a=\partial^\mu A_\mu^a-\xi K_{ia}\chi_i,
    \label{eq:general_Rxi_gauge_fixing}
\end{equation}
which cancels the tree-level mixing between gauge and scalar fluctuations. The associated ghost action follows from the Faddeev--Popov construction~\cite{Faddeev:1967fc} applied to the same gauge-fixing functions,
\begin{equation}
    \mathcal L_{\rm FP}
    =-\bar\eta^a\left.
    \frac{\delta\mathcal F^a}{\delta\alpha^b}
    \right|_{\alpha=0}\eta^b,
    \label{eq:general_FP_lagrangian}
\end{equation}
where $\alpha^b$ parametrises gauge transformations of the fluctuating fields at fixed background, and $\eta^a$, $\bar\eta^a$ are the ghost and antighost fields. The resulting tree-level mass matrices are
\begin{align}
    m_A^2(h)_{ab}&=(K^TK)_{ab},
    \label{eq:general_tree_mass_matrices_A}
    \\[0.2em]
    m_\chi^2(h,\xi)_{ij}
    &=\left.\frac{\partial^2V_0}
    {\partial\Phi_i\partial\Phi_j}\right|_{\Phi=\bar\Phi(h)}
    +\xi(KK^T)_{ij},
    \label{eq:general_tree_mass_matrices_chi}
    \\[0.2em]
    m_\eta^2(h,\xi)_{ab}&=\xi(K^TK)_{ab}=\xi m_A^2(h)_{ab}.
    \label{eq:general_tree_mass_matrices_eta}
\end{align}
Up to total derivatives, the part quadratic in the fluctuations is
\begin{equation}
\begin{aligned}
    \mathcal L^{(2)}
    ={}& \frac12 A_\mu^a
    \Big[ g^{\mu\nu}\big(\partial^2 \delta_{ab}+
    m_A^2(h)_{ab}\big) -\big(1-\frac1\xi\big)
    \partial^\mu \partial^\nu \delta_{ab}
    \Big] A_\nu^b 
    \\[0.3em]
    &-\frac12\chi_i \Big[\partial^2 \delta_{ij}+m_\chi^2(h,
    \xi)_{ij}\Big]\chi_j
    -\bar\eta^a
    \Big[\partial^2 \delta_{ab}+m_\eta^2(h,\xi)_{ab}\Big]\eta^b
    +\mathcal L_{\rm f}^{(2)},
\end{aligned}
\label{eq:general_quadratic_lagrangian}
\end{equation}
where $\partial^2=\partial_\mu\partial^\mu$ and $\mathcal L_{\rm f}^{(2)}$ contains the fermions with their mass matrices evaluated at $\bar\Phi(h)$. The scalar, gauge and ghost operators determine the tree-level propagators. After Wick rotation, the scalar, gauge and ghost trace logarithms enter the one-loop effective action with coefficients $1/2$, $1/2$ and $-1$, respectively. When expressed in the form of Eq.~\eqref{eq:finite_temperature_functional_determinant}, these factors are incorporated through the signed multiplicities $n_i$. The bosonic matrices depend on $h$, and the scalar and ghost matrices also depend on $\xi$. The scalar and gauge matrices act in different field spaces and need not be diagonal in the chosen bases. In the rest of the paper we will use lowercase $m^2$ to denote a tree-level mass-squared matrix, and $M^2$ to designate its dressed counterpart,  as in Eq.~\eqref{eq:gap_equation_general}. All mass matrices are evaluated on $\bar\Phi(h)$, with their background dependence denoted simply by $h$. Background derivatives in the following subsections are taken in a fixed field basis along this path, so that $\partial_h=(\mathrm d\bar\Phi_i/\mathrm dh)\,\partial/\partial\bar\Phi_i$.

For the electroweak SM, the gauge group is $SU(2)_L\times U(1)_Y$, with couplings $g$ and $g'$. We write the Higgs doublet as
\begin{equation}
    H(x)=\frac{1}{\sqrt{2}}
    \begin{pmatrix}
        G^1(x)+\mathrm iG^2(x)\\
        h+h(x)+\mathrm iG^0(x)
    \end{pmatrix},
    \qquad
    \bar\Phi(h)=(0,0,h,0)^T,
    \label{eq:SM_background_setup}
\end{equation}
where $h(x)$ denotes the radial fluctuation about the homogeneous background $h$. We also use $h$ as the corresponding scalar mode label and define $G^\pm=(G^1\pm\mathrm iG^2)/\sqrt{2}$. The scalar potential is
\begin{equation}
    V_0(H)=
    \frac{\mu^4}{2\lambda}
    +\mu^2H^\dagger H
    +\frac{\lambda}{2}(H^\dagger H)^2.
    \label{eq:SM_tree_potential}
\end{equation}
On the chosen background this reproduces Eq.~\eqref{eq:tree_level_potential} up to the field-independent term. We retain this term here to set $V_0(v_0)=0$ at the broken tree-level minimum, where $v_0^2=-2\mu^2/\lambda$.

Using the $SU(2)_L\times U(1)_Y$ generators in the Higgs representation, Eqs.~\eqref{eq:general_tree_mass_matrices_A}-\eqref{eq:general_tree_mass_matrices_eta} give the tree-level mass matrices in the real scalar basis $(G^1,G^2,h,G^0)$ and gauge basis $(W^1,W^2,W^3,B)$,
\begin{align}
    m_A^2(h)
    &=\frac{h^2}{4}
    {\setlength{\arraycolsep}{4pt}
    \begin{pmatrix}
        g^2&0&0&0\\
        0&g^2&0&0\\
        0&0&g^2&-gg'\\
        0&0&-gg'&g'^2
    \end{pmatrix}},
    \label{eq:SM_tree_mass_spectrum_A}
    \\[0.7em]
    m_\chi^2(h,\xi)
    &=
    \left(\mu^2+\frac{\lambda h^2}{2}\right)\mathbf1_4
    +\frac{h^2}{4}
    {\setlength{\arraycolsep}{3pt}
    \begin{pmatrix}
        \xi g^2 & 0 & 0 & 0\\
        0 & \xi g^2 & 0 & 0\\
        0 & 0 & 4\lambda & 0\\
        0 & 0 & 0 & \xi(g^2+g'^2)
    \end{pmatrix}}.
    \label{eq:SM_tree_mass_spectrum_chi}
\end{align}
The ghost matrix follows from Eq.~\eqref{eq:general_tree_mass_matrices_eta} and uses the gauge ordering $(\eta^1,\eta^2,\eta^3,\eta^B)$. For the following part, we retain the top and bottom Yukawa couplings, with $m_q^2(h)=y_q^2h^2/2$ for $q=t,b$, and set $V_{\rm CKM}=\mathbf 1$. All remaining SM fermions are treated as massless and retained in gauge-boson loops. The colour multiplicities are $N_c=3$ for quarks and $N_c=1$ for leptons.

For the resummation, we use the tree-level neutral gauge basis $(Z,A^\gamma)$, related to $(W^3,B)$ by
\begin{equation}
    \begin{pmatrix}Z_\mu\\A_\mu^\gamma\end{pmatrix}
    =
    {\setlength{\arraycolsep}{1pt}
    \begin{pmatrix}
        c_w & -s_w\\[1pt]
        s_w &  c_w
    \end{pmatrix}}
    \begin{pmatrix}W_\mu^3\\B_\mu\end{pmatrix},
    \label{eq:SM_tree_neutral_rotation}
\end{equation}
where
\begin{equation}
c_w=\frac{g}{\sqrt{g^2+g'^2}}, \qquad s_w=\frac{g'}{\sqrt{g^2+g'^2}}.
\end{equation}
We also define $W_\mu^\pm=(W_\mu^1\mp\mathrm iW_\mu^2)/\sqrt{2}$ and $e=gs_w=g'c_w$. This rotation diagonalises the tree-level mass matrix in Eq.~\eqref{eq:SM_tree_mass_spectrum_A}. The charged pairs are represented by two real field components. The gauge-direction part of the vector propagator contains $\xi m_A^2$ and is distinguished from the spatially longitudinal thermal sector below.

For simplicity, we work in Euclidean space with metric $(+,+,+,+)$ from this point onwards. We write $p=(\omega_n,\mathbf p)$ as the Euclidean momentum, with $p^2=\omega_n^2+\mathbf p^2$, and contract spacetime indices with $\delta_{\mu\nu}$. The frequencies $\omega_n$ are bosonic for scalars, gauge fields and ghosts, and fermionic for fermions. The tree-level mass matrices introduced above are unchanged by the Wick rotation.

\subsection{Summary and collected general formulae for partial dressing}

Before providing more details in the subsequent sections, we collect the finite-temperature PD expressions for the resummed effective potential $V^{\rm PD}_{\rm eff}(h,T)$, assuming an $h$-independent tree-level mass basis. The potential can be recovered from integrating the tadpole $V_{\rm eff}^{\prime\,{\rm PD}}(h,T)$, which we write as
\begin{align}\label{eq:Vprime_sectors}
    V_{\rm eff}^{\prime\,{\rm PD}}(h,T)
    =V_0'(h)
    +V_f^{\prime\,{\rm PD}}(h,T)
    +V_{\rm sc}^{\prime\,{\rm PD}}(h,T)
    +V_{\rm g}^{\prime\,{\rm PD}}(h,T)
    +V_{\rm gh}^{\prime\,{\rm PD}}(h,T),
\end{align}
where we have separated the fermionic, scalar, gauge and ghost pieces. For the total fermionic contribution $V_f^{\prime\,{\rm PD}}(h,T)$ no resummation is needed and we follow the standard treatment at finite temperature \cite{Quiros:1999jp}. 

The full potential is obtained by integrating the tadpole, which leaves a field-independent function of temperature undetermined. The latter is however relevant, as it plays a role in determining the free energy and its associated thermodynamic quantities. By comparing the FD and PD prescriptions for the tadpole, Eqs.~\eqref{eq:Vprime_FD} and \eqref{eq:correct_tadpole_resummation}, it is clear that they differ in using  dressed vertices (involving $\partial_h M^2_k$) versus tree-level vertices (involving $\partial_h m^2_k$). With the field dependence of the masses squared being proportional to $h^2$, the difference between the two prescriptions disappears for $h\rightarrow0$. Moreover, the FD potential is known beyond its tadpole formula (see Eq.~\eqref{eq:V_FD}), so that there is no undetermined integration constant at $h=0$. This implies we can fix the integration constant in the PD prescription by matching it with the $h=0$ potential in the FD prescription. 
\begin{equation}
    V^{\rm PD}_{\rm eff}(h,T)
    =
    V^{\rm FD}_{\rm eff}(0,T)
    +
    \int_{0}^{h} {\rm d}h'\,
    \frac{\partial V^{\rm PD}_{\rm eff}(h',T)}{\partial h'} \, .
    \label{eq:PD_Normalisation}
\end{equation}
Notice that the full-dressing result enters only through the field-independent normalisation; the field dependence of the potential is entirely determined from the partially dressed tadpole.

In the following we give the expressions for the contributions in Eq.~\eqref{eq:Vprime_sectors} involving a resummation. First, for the scalar fluctuations $\chi$ (cf. Eq.~\eqref{eq:general_scalar_background}) one has
\begin{align}
\begin{aligned}
    V_{\rm sc}^{\prime\,{\rm PD}}(h,T)=
    \sum_{k,l}{(U_\chi)}_{lk}^2
    \frac{\mathrm d}{\mathrm dh}\bigg[
    \frac{m_{\chi,l}^4(h,\xi)}{64\pi^2}
    \Big(\log\frac{m_{\chi,l}^2(h,\xi)}{\mu_R^2}-\frac32\Big)
    +\frac{T^4}{2\pi^2}
    J_{\rm B}\!\Big(\frac{m_{\chi,l}^2(h,\xi)}{T^2}\Big)
    \bigg]_{\displaystyle m_{\chi,l}^2\to M^2_{\chi,k}}.
    \label{eq:scalar_PD_summary}
\end{aligned}
\end{align}
Above, $l$ labels the eigenstates of the tree-level mass matrix $m_\chi^2(h,\xi)_{ij}$ defined in Eq.~\eqref{eq:general_tree_mass_matrices_chi}, whose eigenvalues are denoted by $m_{\chi,l}^2(h,\xi)$. Analogously, $k$ labels the eigenstates of the resummed mass matrix defined by the gap equation $(M^2_{\chi})_{ij}=m_\chi^2(h,\xi)_{ij}+\Pi_{\chi}(0)(h,T;\{M^2_{\varphi}\})_{ij}$, whose eigenvalues are denoted by $M^2_{\chi,k}$ and where $\Pi_{\chi}(0)$ is the one-loop thermal scalar self-energy matrix at vanishing external momentum computed with resummed propagators for all species $\varphi$. The substitution $m_{\chi,l}^2\to M_{\chi,k}^2$ only applies to the masses with no derivatives acting on them, i.e. contributions of the form $\partial_h m^2_{\chi,l}$ are left unchanged. Finally, $U_\chi$ is an orthogonal mixing matrix satisfying  $U_\chi^TM_\chi^2U_\chi
    =\operatorname{diag}\!\left(M_{\chi,1}^2,M_{\chi,2}^2,\ldots\right)$.
    
In the gauge sector, one similarly has
\begin{align}\begin{aligned}
&V_\mathrm{g}^{\prime\,{\rm PD}}(h,T)
 = \\
&\sum_{X,k,l} n_X ({U_{A,X}})_{lk}^2
\frac{\mathrm d}{\mathrm dh}
\bigg[
\frac{m_{A,X,l}^4(h)}{64\pi^2}
\Big(\log\frac{m_{A,X,l}^2(h)}{\mu_R^2}-c_X\Big)
+\frac{T^4}{2\pi^2}J_{\rm B}\!\Big(\frac{m_{A,X,l}^2(h)}{T^2}\Big)
\bigg]_{\displaystyle m_{A,l}^2\to M_{A,X,k}^2}.
\label{eq:gauge_PD_Tadpole_compact}
\end{aligned}
\end{align}
Here, $X=T,L,D$ corresponds to a decomposition of the gauge boson propagators into 3D transverse, 3D longitudinal,  and 4D longitudinal contributions  \cite{Weldon:1996kb}, which can be resummed independently. Additionally, one has $n_T=2$, $n_D=n_L=1$, $c_T=1/2$ and $c_L=c_D=3/2$. 

The $m_{A,X,l}^2(h)$ are related to the tree-level
eigenvalues $m_{A,l}^2(h)$ of the mass matrix defined in Eq.~\eqref{eq:general_tree_mass_matrices_A} by the relation $m_{A,X,l}^2(h)\equiv\kappa_X(\xi)m_{A,l}^2(h)$, where $\kappa_T(\xi)=\kappa_L(\xi)=1$ and $\kappa_D(\xi)=\xi$. The $M_{A,X,k}^2$ are eigenvalues of resummed mass matrices $(M_{A,X}^2)_{ab}$ (with $a,b$ Lie algebra indices) which solve the  gap equations $ (M_{A,X}^2)_{ab}=m_{A}^2(h)_{ab}+\Pi_{{A},X}(0)(h,T;\{M^2_{\varphi}\})_{ab}$, with $\Pi_{{A},X}(0)(h,T;\{M^2_{\varphi}\})$ given by one-loop gauge-field self-energies in the $T,L,D$ sectors at vanishing external momenta and computed with resummed propagators. We find $\Pi_{A,D}(0)=\Pi_{A,T}(0)$.
Taking the zero momentum limit $(k^0=0,{\bf k}\rightarrow0)$ and assuming a finite, direction-independent limit consistent with spatial rotational symmetry, one can express the $\Pi_{A,X}(0)$ in terms of the usual Euclidean polarisation tensor $\Pi_{A}^{\mu\nu}(0)_{ab}$ as follows:
$\Pi_{A,L}(0)_{ab}= \Pi_{A}^{00}(0)_{ab},$ $
    \delta^{mn}\Pi_{{A},T/D}(0)_{ab}=\Pi_{A}^{mn}(0)_{ab},$
with $m,n$ spatial indices. Finally, the $U_{A,X}$ in Eq.~\eqref{eq:gauge_PD_Tadpole_compact} are orthogonal transformations that diagonalise the $M^2_{A,X}$ matrices: $U_{A,X}^T M_{A,X}^2U_{A,X}
=\operatorname{diag} (M_{A,X,1}^2,M_{A,X,2}^2,\ldots)$. Again, the substitution of mass eigenvalues only applies to masses on which no field derivatives are acting.
    
The last resummation concerns the ghost fields, for which one can write
\begin{align}
\label{eq:ghost_PD_Tadpole}
    \begin{aligned}
   &V_{\rm gh}^{\prime\,{\rm PD}}(h,T)
    =\\
    &-2\sum_{k,l}  (U_\eta^{-1})_{kl}(U_{\eta})_{lk}
    \frac{\mathrm d}{\mathrm dh}\bigg[
    \frac{m_{\eta,l}^4(h,\xi)}{64\pi^2}
    \Big(\log\frac{m_{\eta,l}^2(h,\xi)}{\mu_R^2}-\frac32\Big)
    +\frac{T^4}{2\pi^2}
    J_{\rm B}\!\Big(\frac{m_{\eta,l}^2(h,\xi)}{T^2}\Big)
    \bigg]_{\displaystyle m_{\eta,l}^2\to M_{\eta,k}^2}.
\end{aligned}
\end{align}
Here $l$ labels the tree-level ghost eigenstates, with
squared masses $m_{\eta,l}^2(h,\xi)$ given by the eigenvalues of the mass matrix defined in
Eq.~\eqref{eq:general_tree_mass_matrices_eta}, while $k$ labels the dressed eigenstates, with squared masses $M_{\eta,k}^2$. The dressed mass matrix satisfies the gap equation
\[
    (M_\eta^2)_{ab}
    =m_\eta^2(h,\xi)_{ab}
    +\Pi_\eta(0)(h,T;\{M_\varphi^2\})_{ab},
\]
where $\Pi_\eta(0)$ is the one-loop thermal ghost self-energy matrix at zero external momentum, computed using resummed propagators. $M_\eta^2$ is diagonalised by an invertible matrix $U_\eta$ as $U_\eta^{-1}M_\eta^2U_\eta =\operatorname{diag}\! \left(M_{\eta,1}^2,M_{\eta,2}^2,\ldots\right)$.

The computation of $V_{\rm eff}^{\prime\,{\rm PD}}(h,T)$ can then be carried out once the one-loop self-energies are known. One can start from a standard one-loop calculation in the tree-level mass basis, with the result parametrised in terms of vertex factors multiplying thermal loop integrals depending on particle masses. This yields self-energies $\Pi_\varphi(0)(h,T;\{m^2_{\varphi'}(h)\})$. From them one can obtain a first estimate of the resummed masses as $M^2_{\varphi,(1)}=m^2_{\varphi}+\Pi_\varphi(0)(h,T;\{m^2_{\varphi'}(h)\})$. Using these masses in the self-energies and proceeding iteratively, i.e. taking $M^2_{\varphi,(n)}=m^2_{\varphi}+\Pi_\varphi(0)(h,T;\{M^2_{\varphi',(n-1)}\})$ one eventually approaches a solution to the gap equations upon convergence of the results of successive iterations. As the mass eigenstates in each iteration are not necessarily aligned with the initial tree-level eigenstates, in the formulae for  $\Pi_\varphi(0)(h,T;\{M^2_{\varphi',(n-1)}\})$ one has to modify the interaction vertices  with insertions of mixing matrices $U_{\varphi',(n-1)}$ for each internal line. This ensures that the self-energies are computed with the appropriate Feynman rules corresponding to the mass basis of the matrices $M^2_{\varphi',(n-1)}$.  For a derivation of
Eqs.~\eqref{eq:scalar_PD_summary}, \eqref{eq:gauge_PD_Tadpole_compact} and \eqref{eq:ghost_PD_Tadpole}, plus additional details concerning the application to the SM, the reader is referred to
Secs.~\ref{subsec:scalar_resummation}, \ref{subsec:selfenergydecomp} and \ref{subsec:ghost_resummation}, respectively. The necessary modifications of the vertex factors in the self-energies to account for mixing effects are  discussed in Section~\ref{subsec:gap_equations} and in Appendix~\ref{app:neutral_mixing_replacements}, while the results for the self-energies in the SM are given in Sec. \ref{sec:4}.

The PD method described above only resums the zero-momentum contributions to self-energies, which does not accurately capture loops with overlapping momentum regions. In order to estimate the impact of gradient corrections, the above resummations can be improved by incorporating the leading $p^2$ corrections. The tadpole formulae can then be corrected with the substitution $\omega_n^2+\mathbf p^2+M_\varphi^2 \rightarrow Z_0\,\omega_n^2+Z_s\,\mathbf p^2+M_\varphi^2$, with appropriate wave-function factors $Z_0$ and $Z_s$ computed from the self-energies; more details are given in Appendix~\ref{app:momentum_dependence}.


\subsection{Scalar sector}
\label{subsec:scalar_resummation}
The starting point for implementing the tadpole resummation in the scalar sector is  the  derivative 
with respect to the background $h$ of the scalar functional determinant in Eq.~\eqref{eq:finite_temperature_functional_determinant}, which yields
\begin{equation}
    V_{\rm sc}^{\prime}(h,T)
    =\frac{T}{2}\mu_R^{2\epsilon}\sum_n
    \int\frac{\mathrm d^{d-1}\mathbf p}{(2\pi)^{d-1}}
    \operatorname{Tr}\!\left[
    \mathcal D_{\chi}^{(0)}(\mathbf p,\omega_n)\,
    \partial_h m_\chi^2(h,\xi) \right],
    \label{eq:tadpole_scalar}
\end{equation}
where the tree-level scalar propagator is
\begin{equation}
    \mathcal D_\chi^{(0)}(\mathbf p,\omega_n)_{ij}
    =\big( \omega_n^2+\mathbf p^2
    +m_\chi^2(h,\xi)\big)_{ij}^{-1}.
    \label{eq:scalar_tree_propagator}
\end{equation}
Here $m_\chi^2(h,\xi)$ is the scalar mass matrix defined in Eq.~\eqref{eq:general_tree_mass_matrices_chi}. The trace and the matrix inverse act on scalar field space. The background derivatives are taken at fixed $T$, $\xi$ and $\mu_R$.

Considering arbitrary insertions of one-loop scalar self-energies in the scalar propagator leads to a summable geometric series, equivalent to the replacement
\begin{equation}
\begin{aligned}
    \mathcal D_\chi^{(0)}(\mathbf p,\omega_n)_{ij}
    &\longrightarrow\mathcal D_\chi(\mathbf p,\omega_n)_{ij}
    =\big(\omega_n^2+\mathbf p^2+M_\chi^2\big)_{ij}^{-1},
\end{aligned}
\label{eq:scalar_dressed_propagator}
\end{equation}
where the effective mass is defined by
\begin{align}
    \big( M_\chi^2\big)_{ij}&=m_\chi^2(h,\xi)_{ij}+\Pi_\chi(0)(h,T;\{m^2_\varphi\})_{ij},
\end{align}
with $\Pi_\chi(0)$  the thermal scalar self-energy matrix at vanishing external momentum, and $m^2_\varphi$ the tree-level masses of the particle species contributing to the one-loop self-energies. The tree-level tadpole vertex $\partial_hm_\chi^2$ is kept unchanged. By evaluating $\Pi_\chi(0)$ with  dressed masses $M^2_\varphi$ as opposed to the tree-masses $m^2_\varphi$, one obtains a gap equation,
\begin{align}
    \big( M_\chi^2\big)_{ij}&=m_\chi^2(h,\xi)_{ij}+\Pi_\chi(0)(h,T;\{M^2_\varphi\})_{ij},
\end{align}
which effectively resums nested insertions of one-loop self-energies, as illustrated in Eq.~\eqref{eq:gap_solution}.

In general, $M_\chi^2$ need not be diagonal in the original field basis. For real scalar components, interchange of the external legs gives $\Pi_\chi(0)_{ij}=\Pi_\chi(0)_{ji}$. With real static self-energies the dressed mass matrix is therefore real symmetric and can be diagonalised by an orthogonal transformation,
\begin{equation}
    U_\chi^TM_\chi^2U_\chi
    =\operatorname{diag}\!\left(M_{\chi,1}^2,M_{\chi,2}^2,\ldots\right),
    \label{eq:scalar_mass_diagonalisation}
\end{equation}
with $U_\chi^TU_\chi=\mathbf 1$. The resummed tadpole then becomes
\begin{equation}
\begin{aligned}
    V_{\rm sc}^{\prime\,{\rm PD}}(h,T)
    ={}&\frac{T}{2}\mu_R^{2\epsilon}\sum_n
    \int\frac{\mathrm d^{d-1}\mathbf p}{(2\pi)^{d-1}}
    \Big( \sum_k
    \frac{\bigl[U_\chi^T(\partial_hm_\chi^2)U_\chi\bigr]_{kk}}
    {\omega_n^2+\mathbf p^2+M_{\chi,k}^2}\Big),
\end{aligned}
\label{eq:scalar_PD_matrix_tadpole}
\end{equation}
where $k$ runs over the scalar mass eigenmodes. The rotation thus also enters the contraction with the tree-level vertex. Neither $\Pi_\chi(0)$ nor $U_\chi$ is differentiated. The above equation can be written as
\begin{align}\begin{aligned}
     V_{\rm sc}^{\prime\,{\rm PD}}(h,T)=&\,\sum_{k,l}(U_{\chi})_{lk}^2\,\frac{T}{2}\,\mu_R^{2\epsilon}\sum_n
    \int\frac{\mathrm d^{d-1}\mathbf p}{(2\pi)^{d-1}}
   \left. \frac{\partial_h m_{\chi,l}^2}
    {\omega_n^2+\mathbf p^2+m_{\chi,l}^2}\right|_{m_{\chi,l}^2\rightarrow M_{\chi,k}^2}\\
    =&\,\sum_{k,l}(U_{\chi})_{lk}^2\, \left.{V^{(1)}}'_{\chi,l}\right|_{m_{\chi,l}^2\rightarrow M_{\chi,k}^2}.
\end{aligned}\end{align}
Here ${V^{(1)}}'_{\chi,l}$ denotes the background derivative of the standard one-loop potential contribution from a single real scalar field of mass $m_{\chi,l}$ (cf. Eq.~\eqref{eq:finite_temperature_functional_determinant}). The mass replacement $m_{\chi,l}^2(h,\xi) \to M_{\chi,k}^2$ is made after differentiation and acts on the mass factors and function arguments originating from propagators, while the mass derivatives are kept unchanged. The above identity suggests a renormalisation procedure in which the same relation between dressed and undressed tadpoles is preserved by the counterterms. As a consequence, the renormalised result can be expressed in terms of the Coleman--Weinberg potential renormalised in the $\overline{\rm MS}$ scheme and standard thermal contributions,  evaluated with the dressed masses $M^2_{\chi,k}$ and multiplied by appropriate mixing factors:
\begin{equation}
\begin{aligned}
 V_{\rm sc}^{\prime\,{\rm PD}}(h,T)
    =\sum_{k,l}{(U_\chi)}_{lk}^2
    \frac{\mathrm d}{\mathrm dh}\left[
    \frac{m_{\chi,l}^4(h,\xi)}{64\pi^2}
    \left(\log\frac{m_{\chi,l}^2(h,\xi)}{\mu_R^2}-\frac32\right)
    +\frac{T^4}{2\pi^2}
    J_{\rm B}\!\left(\frac{m_{\chi,l}^2(h,\xi)}{T^2}\right)
    \right]_{m_{\chi,l}^2\to M^2_{\chi,k}},
    \label{eq:scalar_PD_summary2}
\end{aligned}
\end{equation}
In the SM, the scalar self-energy is diagonal in the real basis $(G^1,G^2,h,G^0)$ introduced above. Electromagnetic charge conservation forbids charged--neutral mixing, while CP conservation forbids $h$--$G^0$ mixing. Hence
\begin{equation}
    \Pi_\chi(0)_{ij}=
    {\setlength{\arraycolsep}{0pt}
    \begin{pmatrix}
        \Pi_{G^\pm}(0)&0&0&0\\
        0&\Pi_{G^\pm}(0)&0&0\\
        0&0&\Pi_h(0)&0\\
        0&0&0&\Pi_{G^0}(0)
    \end{pmatrix}},
    \label{eq:Pizero_scalar}
\end{equation}
where $\Pi_{G^\pm}(0)$ is common to both charged modes. Since the matrix entering the propagator is already diagonal, no further change of basis is required, so that the mixing matrices are trivial. Then Eq.~\eqref{eq:scalar_PD_summary2} can be written as 
\begin{equation}
\begin{aligned}
    V_{\rm sc}^{\prime\,{\rm PD}}(h,T)
    =\sum_s n_s
    \frac{\mathrm d}{\mathrm dh}\left[
    \frac{m_s^4(h,\xi)}{64\pi^2}
    \left(\log\frac{m_s^2(h,\xi)}{\mu_R^2}-\frac32\right)
    +\frac{T^4}{2\pi^2}
    J_{\rm B}\!\left(\frac{m_s^2(h,\xi)}{T^2}\right)
    \right]_{m_s^2\to M_s^2}.
\end{aligned}
\label{eq:scalar_PD_Tadpole}
\end{equation}
Here the sum over $s$ runs over scalar species $s\in\{G^\pm,h,G^0\}$, whose multiplicities are $n_{G^\pm}=2$, $n_h=n_{G^0}=1$. The field-dependent scalar contribution then follows by integrating the tadpole,
\begin{equation}
    V_{\rm sc}^{\rm PD}(h,T)
    =\int_0^h\mathrm d\varphi\,
    V_{\rm sc}^{\prime\,{\rm PD}}(\varphi,T).
    \label{eq:scalar_PD_potential_reconstruction}
\end{equation}


\subsection{Gauge sector}
\label{subsec:selfenergydecomp}

We proceed as in the scalar sector and obtain the gauge-boson tadpole by differentiating the gauge contribution to Eq.~\eqref{eq:finite_temperature_functional_determinant} with respect to the background field $h$,
\begin{equation}
    V_{\rm g}^{\prime}(h,T)
    =\frac{T}{2}\mu_R^{2\epsilon}\sum_n
    \int\frac{\mathrm d^{d-1}\mathbf p}{(2\pi)^{d-1}}
    \operatorname{Tr}\!\left[
    \mathcal D_A^{(0)}(\mathbf p,\omega_n) \,\partial_hm_A^2(h)\right].
    \label{eq:Gauge_tadpole}
\end{equation}
The trace acts on internal gauge space and Euclidean space. The tree-level propagator in the $R_\xi$ gauge of Sec.~\ref{subsec:gauge_theory_setup} is
\begin{equation}
\begin{aligned}
    \mathcal D_A^{(0)}(\mathbf p,\omega_n)_{ab}^{\mu\nu}
    ={}&\left(\delta^{\mu\nu}-\frac{p^\mu p^\nu}{p^2}\right)
    \bigl(p^2+m_A^2(h)\bigr)^{-1}_{ab}+\xi \; \frac{p^\mu p^\nu}{p^2}
    \big(p^2+\xi m_A^2(h)\big)^{-1}_{ab},
\end{aligned}
\label{eq:gauge_tree_propagator}
\end{equation}
with $m_A^2(h)$ defined in Eq.~\eqref{eq:general_tree_mass_matrices_A}.

The heat bath singles out a preferred time direction, allowing the transverse and longitudinal modes to receive different thermal corrections. We therefore use the tensor basis of Appendix~\ref{appendixA}, following Ref.~\cite{Weldon:1996kb}. In the plasma rest frame, $u^\mu=(1,\mathbf 0)$, we introduce $ \widetilde p^\mu = ((p\cdot u) p^\mu-p^2 u^\mu)/|\mathbf p|$ and define the projection operators
\begin{equation}
\begin{aligned}
    P_T^{\mu\nu}
    &=\delta^{\mu\nu}
    -\frac{\widetilde p^\mu\widetilde p^\nu}{p^2}
    -\frac{p^\mu p^\nu}{p^2},
     \qquad
    P_L^{\mu\nu}
    =\frac{\widetilde p^\mu\widetilde p^\nu}{p^2},
    \qquad
    P_D^{\mu\nu}=\frac{p^\mu p^\nu}{p^2}.
\end{aligned}
\label{eq:gauge_thermal_projectors}
\end{equation}
Here $P_T$ projects onto the spatial directions perpendicular to $\mathbf p$, $P_L$ onto the remaining direction orthogonal to $p$, and $P_D$ onto the gauge direction along $p$. These operators are mutually orthogonal and idempotent,
\begin{equation}
    P_X^{\mu\rho}P_Y^{\rho\nu}
    =\delta_{XY}P_X^{\mu\nu},
    \qquad X,Y=T,L,D,
    \label{eq:gauge_projector_algebra}
\end{equation}
with traces $r_X\equiv(P_X)^{\mu\mu}$, explicitly given by
\begin{align}
     r_T= d-2, \quad r_L=&r_D=1.
     \label{eq:rX}
\end{align}
Since $P_T+P_L=\delta-P_D$, the first Lorentz structure in Eq.~\eqref{eq:gauge_tree_propagator} splits into the $T$ and $L$ sectors, while the second belongs to the $D$ sector:
\begin{equation}
\begin{aligned}
    \mathcal D_A^{(0)}(\mathbf p,\omega_n)_{ab}^{\mu\nu}
    ={}&\big(P_T^{\mu \nu} + P_L^{\mu \nu} \big)
    \bigl(p^2+m_A^2(h)\bigr)^{-1}_{ab}+\xi \; P_D^{\mu \nu}
    \big(p^2+\xi m_A^2(h)\big)^{-1}_{ab},
\end{aligned}
\label{eq:gauge_tree_propagator_2}
\end{equation}

The general self-energy can also contain a mixed tensor structure, which is discussed in detail in Appendix~\ref{appendixA}. Assuming a finite, isotropic infrared limit, the tensor decomposition simplifies, with
$\Pi_{A,D}(0)_{ab}=\Pi_{A,T}(0)_{ab}$. The thermal self-energy reduces to
\begin{equation}
    \Pi_{A}^{\mu\nu}(0)_{ab}
    =(P_T^{\mu\nu}+P_D^{\mu\nu})\Pi_{A,T}(0)_{ab}
    +P_L^{\mu\nu}\Pi_{A,L}(0)_{ab},
    \label{eq:gauge_static_insertion}
\end{equation}
where
\begin{equation}
    \Pi_{A,L}(0)_{ab}=\Pi_{A}^{00}(0)_{ab},
    \qquad
    \delta^{mn}\Pi_{A,T}(0)_{ab}=\Pi_{A}^{mn}(0)_{ab},
    \label{eq:gauge_static_selfenergies}
\end{equation}
with $m,n$ spatial indices.

For arbitrary one-loop self-energy insertions, the orthogonality of the projectors ensures that the geometric series of  insertions can  be resummed independently in each Lorentz sector. The tree-level propagator in Eq.~\eqref{eq:Gauge_tadpole} is therefore replaced by
\begin{equation}
\begin{aligned}
	 \mathcal D_A^{(0)}(\mathbf p,\omega_n)_{ab}^{\mu\nu} \longrightarrow
    \mathcal D_A(\mathbf p,\omega_n)_{ab}^{\mu\nu}
    ={}&P_T^{\mu\nu}
    \bigl(\omega_n^2+\mathbf p^2+M_{A,T}^2\bigr)^{-1}_{ab}+P_L^{\mu\nu}
    \bigl(\omega_n^2+\mathbf p^2+M_{A,L}^2\bigr)^{-1}_{ab}
    \\
    &+\xi P_D^{\mu\nu}
    \bigl(\omega_n^2+\mathbf p^2+\xi M_{A,D}^2\bigr)^{-1}_{ab},
\end{aligned}
\label{eq:resummed_propagator_off_diagonal}
\end{equation}
and
\begin{equation}
\begin{aligned}
    (M_{A,T}^2)_{ab}=(M_{A,D}^2)_{ab}
    &=m_A^2(h)_{ab}+\Pi_{A,T}(0)_{ab},
    \\
    (M_{A,L}^2)_{ab}
    &=m_A^2(h)_{ab}+\Pi_{A,L}(0)_{ab}.
\end{aligned}
\label{eq:gauge_dressed_mass_matrices}
\end{equation}
The factor $\xi$ is kept explicit in the $D$-sector propagator, such that $M_{A,D}^2=M_{A,T}^2$. By evaluating the self-energies in the resummed masses $M^2_{A,X}$, the above identities turn into gap equations that effectively resum nested insertions of one-loop self-energy corrections.

As in the scalar sector, in general the matrices $(M_{A,X}^2)_{ab}$ need not be diagonal in the chosen gauge basis. For the real  infrared self-energies considered here, these matrices are symmetric and can be diagonalised by orthogonal matrices $U_{A,X}$,
\begin{equation}
    U_{A,X}^TM_{A,X}^2U_{A,X}
    =\operatorname{diag}\!\left(M_{A,X,1}^2,M_{A,X,2}^2,\ldots\right),
    \label{eq:gauge_mass_diagonalisation_general}
\end{equation}
with $U_{A,X}^TU_{A,X}=\mathbf 1$. The transverse and longitudinal matrices generally require different rotations; since $M_{A,D}^2=M_{A,T}^2$, we take $U_{A,D}=U_{A,T}$. Then the trace in Eq.~\eqref{eq:Gauge_tadpole} can be easily evaluated, such that the tadpole becomes
\begin{equation}
    V_{\rm g}^{\prime\,{\rm PD}}(h,T)
    =\frac{T}{2}\mu_R^{2\epsilon}\sum_n
    \int\frac{\mathrm d^{d-1}\mathbf p}{(2\pi)^{d-1}}
    \left( \sum_{X,k}r_X \kappa_X(\xi)
    \frac{(U_{A,X})_{ak}(\partial_hm_A^2)_{ab}(U_{A,X})_{bk}}
    {\omega_n^2+\mathbf p^2+
    \kappa_X(\xi)M_{A,X,k}^2} \right),
    \label{eq:gauge_PD_matrix_tadpole}
\end{equation}
where $X=T,L,D$, $k$ labels the mass eigenmodes, the $r_X$ were defined in Eq.~\eqref{eq:rX}, and
$\kappa_X(\xi)$ denotes the sector-dependent prefactor that keeps track of the $\xi$-dependence:
\begin{equation}
    \kappa_X(\xi)=
    \begin{cases}
        1, & X=T,L,\\
        \xi, & X=D.
    \end{cases}
    \label{eq:gauge_sector_factor}
\end{equation}
The tree-level tadpole vertices remain undressed, and neither the self-energies nor the rotation matrices are differentiated. Starting from a tree-level basis in which $m^2_A$ is diagonal, with eigenvalues $m^2_{A,l}$, Eq.~\eqref{eq:gauge_PD_matrix_tadpole} can be written in terms of standard ${V^{(1)}}'_{A,l}$ contributions as in Eq.~\eqref{eq:finite_temperature_functional_determinant} multiplied by elements of mixing matrices and with a substitution of tree-level masses by the $M^2_{A,X,k}$. Choosing a renormalisation that preserves this relation between the dressed and undressed tadpoles, with the latter evaluated in the $\overline{\rm MS}$ scheme, leads to
\begin{align}\begin{aligned}
&V_\text{g}^{\prime\,{\rm PD}}(h,T)
 = \\
&\sum_{X,k,l} n_X ({U_{A,X}})_{lk}^2
\frac{\mathrm d}{\mathrm dh}
\bigg[
\frac{m_{A,X,l}^4(h)}{64\pi^2}
\Big(\log\frac{m_{A,X,l}^2(h)}{\mu_R^2}-c_X\Big)
+\frac{T^4}{2\pi^2}J_{\rm B}\!\Big(\frac{m_{A,X,l}^2(h)}{T^2}\Big)
\bigg]_{\displaystyle m_{A,l}^2\to M_{A,X,k}^2}.
\label{eq:gauge_PD_Tadpole_compact2}
\end{aligned}\end{align}
Above, we have defined  $m_{A,X,l}^2\equiv \kappa_X(\xi) m_{A,l}^2$
as well as $n_T=2,n_L=n_D=1$. The vacuum constants $c_X$ are given by $c_T=1/2$ and $c_L=c_D=3/2$, obtained by retaining the dimension-dependent transverse trace until the final $\overline{\rm MS}$ subtraction. For degenerate transverse and longitudinal masses, as happens at tree level, one recovers standard results in the literature~\cite{Quiros:1999jp}. The replacements in Eq.~\eqref{eq:gauge_PD_Tadpole_compact2} act on the underlying masses $m^2_{A,l}$  after differentiation, leaving the mass derivatives and the projection factors unchanged.

In the SM, we use the basis $(W^1,W^2,Z,A^\gamma)$, in which the tree-level gauge-boson mass matrix is diagonal, while the thermal self-energy tensor takes the form
\begin{equation}
    \Pi_{A}^{\mu\nu}(0)_{ab}=
    {\setlength{\arraycolsep}{2pt}
    \begin{pmatrix}
       \Pi_{W^\pm}^{\mu\nu}(0)& 0 &0&0\\
        0 & \Pi_{W^\pm}^{\mu\nu}(0) &0&0\\
        0&0&\Pi_Z^{\mu\nu}(0)&\Pi_{Z\gamma}^{\mu\nu}(0)\\
        0&0&\Pi_{\gamma Z}^{\mu\nu}(0)&\Pi_\gamma^{\mu\nu}(0)
    \end{pmatrix}},
    \label{eq:Pizero_gauge}
\end{equation}
where $\Pi_{W^\pm}^{\mu\nu}(0)$ is common to both charged modes and $\Pi_{Z\gamma}^{\mu\nu}(0)=\Pi_{\gamma Z}^{\mu\nu}(0)$. In the charged sector,  diagonalisation is trivial and Eq.~\eqref{eq:gauge_dressed_mass_matrices} reduces to a mass shift,
\begin{equation}
    M_{W,X}^2=m_W^2(h)+\Pi_{W^\pm,X}(0),
    \qquad X=T,L,
\end{equation}
with $M_{W,D}^2=M_{W,T}^2$. The neutral sector instead contains the mass matrices
\begin{equation}
    M_{\hat Z,X}^2=
    {\setlength{\arraycolsep}{2pt}
    \begin{pmatrix}
        m_Z^2(h)+\Pi_{Z,X}(0)&\Pi_{Z\gamma,X}(0)\\
        \Pi_{\gamma Z,X}(0)&\Pi_{\gamma,X}(0)
    \end{pmatrix}},
    \qquad X=T,L,
    \label{eq:gauge_SM_mass_blocks}
\end{equation}
with $M_{\hat Z,D}^2=M_{\hat Z,T}^2$. These $2\times2$ blocks are diagonalised as in Eq.~\eqref{eq:gauge_mass_diagonalisation_general}, with $M_{\hat Z,X,k}^2$ the corresponding mass-squared eigenvalues and $U_{\hat Z,X}$ the associated mixing matrices.

We label the neutral eigenmode with the larger tree-level $Z$ component as $Z$-like, and the orthogonal mode as photon-like. In the tree-level basis, the neutral tadpole vertex selects the $Z$ direction, since the photon mass vanishes (or equivalently, the neutral Higgs does not couple to the photon). The neutral contribution to Eq.~\eqref{eq:gauge_PD_matrix_tadpole} therefore carries the weights $(U_{\hat Z,X})_{Zk}^2$, so that both of the neutral thermal eigenmodes can contribute. For $V=W,Z$, we define
\begin{equation}
    m_{V,X}^2(h)=\kappa_X(\xi)m_V^2(h).
    \label{eq:gauge_SM_sector_tree_masses}
\end{equation}
The SM result of Eq.~\eqref{eq:gauge_PD_Tadpole_compact2} can then be expressed as the following sum of charged
and neutral contributions,
\begin{align}
\label{eq:gauge_PD_Tadpole_compact3}
&V_\text{g}^{\prime\,{\rm PD}}(h,T)
 = \\
&\sum_{X} n_X
\bigg\{
2\;
\frac{\mathrm d}{\mathrm dh}
\bigg[
\frac{m_{W,X}^4(h)}{64\pi^2}
\Big(\log\frac{m_{W,X}^2(h)}{\mu_R^2}-c_X\Big)
+\frac{T^4}{2\pi^2}J_{\rm B}\!\Big(\frac{m_{W,X}^2(h)}{T^2}\Big)
\bigg]_{\displaystyle m_W^2\to M_{W,X}^2}
\nonumber\\[0.2mm]
&+\sum_{k=1}^{2}(U_{\hat Z,X})_{Zk}^{2}\,
\frac{\mathrm d}{\mathrm dh}
\bigg[
\frac{m_{Z,X}^4(h)}{64\pi^2}
\Big(\log\frac{m_{Z,X}^2(h)}{\mu_R^2}-c_X\Big)
+\frac{T^4}{2\pi^2}J_{\rm B}\!\Big(\frac{m_{ Z,X}^2(h)}{T^2}\Big)
\bigg]_{\displaystyle m_Z^2\to M_{\hat Z,X,k}^2}
\bigg\}.\nonumber
\end{align}
The field-dependent gauge contribution then follows from
\begin{equation}
    V_{\rm g}^{\rm PD}(h,T)
    =\int_0^h\mathrm d\varphi\,
    V_{\rm g}^{\prime\,{\rm PD}}(\varphi,T).
    \label{eq:gauge_PD_potential_reconstruction}
\end{equation}


\subsection{Ghost sector}
\label{subsec:ghost_resummation}

We proceed as in the scalar and gauge sectors and obtain the ghost tadpole by differentiating the ghost contribution to Eq.~\eqref{eq:finite_temperature_functional_determinant} with respect to the background field $h$,
\begin{equation}
    V_{\rm gh}^{\prime}(h,T)
    =-T\mu_R^{2\epsilon}\sum_n
    \int\frac{\mathrm d^{d-1}\mathbf p}{(2\pi)^{d-1}}
    \operatorname{Tr}\!\left[
    \mathcal D_\eta^{(0)}(\mathbf p,\omega_n)\,
    \partial_hm_\eta^2(h,\xi)\right],
    \label{eq:ghost_tadpole}
\end{equation}
where the tree-level ghost propagator is
\begin{equation}
    \mathcal D_\eta^{(0)}(\mathbf p,\omega_n)_{ab}
    =\bigl(\omega_n^2+\mathbf p^2+m_\eta^2(h,\xi)\bigr)^{-1}_{ab},
    \label{eq:ghost_tree_propagator}
\end{equation}
and $m_\eta^2(h,\xi)=\xi m_A^2(h)$. The trace acts on ghost field space. The Matsubara frequencies are bosonic and the Grassmann nature of the fields gives the overall minus sign, while the independent ghost and antighost fields account for the absence of the factor $1/2$ present for real scalars.

Summing a geometric series of self-energy insertions replaces the propagator according to
\begin{equation}
    \mathcal D_\eta^{(0)}(\mathbf p,\omega_n)_{ab}
    \longrightarrow\mathcal D_\eta(\mathbf p,\omega_n)_{ab}
    =\bigl(\omega_n^2+\mathbf p^2+M_\eta^2\bigr)^{-1}_{ab},
    \label{eq:ghost_dressed_propagator}
\end{equation}
with
\begin{equation}
    (M_\eta^2)_{ab}
    =m_\eta^2(h,\xi)_{ab}+\Pi_{\eta}(0)_{ab},
    \label{eq:ghost_dressed_mass_matrix}
\end{equation}
where $\Pi_{\eta}(0)_{ab}$ is the thermal ghost self-energy matrix. Once more, the above can be promoted to a gap equation when evaluating the self-energies in the dressed masses $M^2_\varphi$ for all the species contributing to the one-loop correction. Unlike the scalar and gauge-boson mass matrices, $M_\eta^2$ need not be symmetric, since ghost and antighost fields are independent. If it is diagonalisable, we use an invertible matrix $U_\eta$,
\begin{equation}
    U_\eta^{-1}M_\eta^2U_\eta
    =\operatorname{diag}\!\left(M_{\eta,1}^2,M_{\eta,2}^2,\ldots\right).
    \label{eq:ghost_mass_diagonalisation_general}
\end{equation}
Evaluating the trace in Eq.~\eqref{eq:ghost_tadpole} with the dressed propagator then gives
\begin{equation}
    V_{\rm gh}^{\prime\,{\rm PD}}(h,T)
    =-T\mu_R^{2\epsilon}\sum_n
    \int\frac{\mathrm d^{d-1}\mathbf p}{(2\pi)^{d-1}}
    \left(\sum_k
    \frac{(U_\eta^{-1})_{ka}(\partial_hm_\eta^2)_{ab}(U_\eta)_{bk}}
    {\omega_n^2+\mathbf p^2+M_{\eta,k}^2}\right),
    \label{eq:ghost_PD_matrix_tadpole}
\end{equation}
where $k$ labels the ghost mass eigenmodes. The tree-level tadpole vertex is kept unchanged, and neither the self-energy nor the transformation matrices are differentiated. If no complete eigenbasis exists, the dressed inverse is used directly in Eq.~\eqref{eq:ghost_tadpole}. As in the previous cases, the above result can be related to the unresummed contributions to the potential in Eq.~\eqref{eq:finite_temperature_functional_determinant}. Choosing a renormalisation that preserves this relation leads to
\begin{align}
    \begin{aligned}
    &V_{\rm gh}^{\prime\,{\rm PD}}(h,T)
    =\\
    &-2\sum_{k,l}  (U_\eta^{-1})_{kl}(U_{\eta})_{lk}
    \frac{\mathrm d}{\mathrm dh}\bigg[
    \frac{m_{\eta,l}^4(h,\xi)}{64\pi^2}
    \Big(\log\frac{m_{\eta,l}^2(h,\xi)}{\mu_R^2}-\frac32\Big)
    +\frac{T^4}{2\pi^2}
    J_{\rm B}\!\Big(\frac{m_{\eta,l}^2(h,\xi)}{T^2}\Big)
    \bigg]_{\displaystyle m_{\eta,l}^2\to M_{\eta,k}^2}.
\end{aligned}
\label{eq:ghost_PD_Tadpole2}
\end{align}

In the static infrared limit, the SM ghost self-energies receive contributions only from scalar--ghost loops. In the tree-level basis $(\eta^1,\eta^2,\eta^Z,\eta^\gamma)$, the self-energy matrix takes the form
\begin{equation}
    \Pi_\eta(0)_{ab}=
     {\setlength{\arraycolsep}{1.5pt}
    \begin{pmatrix}
        \Pi_{\eta^\pm}(0)&0&0&0\\
        0&\Pi_{\eta^\pm}(0)&0&0\\
        0&0&\Pi_{\eta^Z}(0)&
        \Pi_{\bar\eta^Z\eta^\gamma}(0)\\
        0&0&0&0
    \end{pmatrix}},
    \label{eq:Pizero_ghost}
\end{equation}
where rows label antighosts and columns label ghosts, and $\Pi_{\eta^\pm}(0)$ is common to both charged modes. The vanishing photon-antighost row follows from $\mathcal{F}^\gamma=\partial^\mu A^\gamma_\mu$, whereas $\Pi_{\bar\eta^Z\eta^\gamma}(0)$ can be non-zero. The photon ghost therefore remains massless. In the charged sector, one has a trivial diagonalisation, with Eq.~\eqref{eq:ghost_dressed_mass_matrix} reducing to
\begin{equation}
    M_{\eta^\pm}^2
    =\xi m_W^2(h)+\Pi_{\eta^\pm}(0).
    \label{eq:ghost_charged_mass}
\end{equation}
In the neutral sector, the dressed mass matrix is
\begin{equation}
    M_{(\eta^Z\eta^\gamma)}^2=
    {\setlength{\arraycolsep}{1.5pt}
    \begin{pmatrix}
        M_{\eta^Z}^2 & \Pi_{\bar\eta^Z\eta^\gamma}(0)\\
        0 & 0
    \end{pmatrix}},
    \label{eq:ghost_neutral_mass_matrix}
\end{equation}
with $M_{\eta^Z}^2=\xi m_Z^2(h)+\Pi_{\eta^Z}(0)$. Since the tree-level photon-ghost mass is independent of $h$, the neutral tadpole involves only the $ZZ$ component of the dressed propagator. The mixing self-energy therefore does not enter the ghost tadpole, but is retained in internal neutral ghost propagators entering the coupled self-energies, as discussed in Appendix~\ref{app:neutral_mixing_replacements}.

Collecting the charged and neutral contributions, the renormalised vacuum and thermal tadpoles take the same form as in Eq.~\eqref{eq:scalar_PD_Tadpole},
\begin{equation}
\begin{aligned}
    V_{\rm gh}^{\prime\,{\rm PD}}(h,T)
    =-2\sum_{\eta}n_\eta
    \frac{\mathrm d}{\mathrm dh}\bigg[
    &\frac{m_\eta^4(h,\xi)}{64\pi^2}
    \Big(\log\frac{m_\eta^2(h,\xi)}{\mu_R^2}-\frac32\Big)
    +\frac{T^4}{2\pi^2}
    J_{\rm B}\!\Big(\frac{m_\eta^2(h,\xi)}{T^2}\Big)
    \bigg]_{\displaystyle m_\eta^2\to M_\eta^2},
\end{aligned}
\label{eq:ghost_PD_Tadpole3}
\end{equation}
where the sum runs over ghost species $\eta\in\{\eta^\pm,\eta^Z\}$, $m_{\eta^\pm}^2=\xi m_W^2$, $m_{\eta^Z}^2=\xi m_Z^2$ and $n_{\eta^\pm}=2$, $n_{\eta^Z}=1$. The field-dependent ghost contribution then follows from
\begin{equation}
    V_{\rm gh}^{\rm PD}(h,T)
    =\int_0^h\mathrm d\varphi\,
    V_{\rm gh}^{\prime\,{\rm PD}}(\varphi,T).
    \label{eq:ghost_PD_potential_reconstruction}
\end{equation}


\subsection{Gap equations}
\label{subsec:gap_equations}

From the previous sections, it is clear that the potential can be evaluated with the PD method once one knows the resummed mass eigenvalues $M^2_{\varphi,k}(h,T)$. These correspond to matrices that solve the following gap equations: 
\begin{align}
\label{eq:all_self_energies}
    \begin{aligned}
    (M_\chi^2)_{ij}
    =&~m_\chi^2(h,\xi)_{ij}
    +\Pi_{\chi} (0) \left(h,T;\{M_\varphi^2\}\right)_{ij},\\[0.3em]
       (M_{A,T}^2)_{ab} = &~(M_{A,D}^2)_{ab}
    =m_A^2(h)_{ab}
    +\Pi_A^{ii}(0) \left(h,T;\{M_\varphi^2\}\right)_{ab},
    \\[0.3em]
    (M_{A,L}^2)_{ab}
    =&~m_A^2(h)_{ab}
    +\Pi_A^{00} (0) \left(h,T;\{M_\varphi^2\}\right)_{ab},
    \\[0.3em]
    (M_\eta^2)_{ab}
    =&~m_\eta^2(h,\xi)_{ab}
    +\Pi_\eta(0)(h,T;\{M_\varphi^2\})_{ab}.
    \end{aligned}
\end{align}
Here no sum over the spatial index $i$ is implied. The self-energies depend on the dressed mass matrices of all species contributing at one loop, so that the gap equations form a coupled system. A subtlety is that self-energies are typically evaluated in terms of fields that diagonalise the tree-level mass matrices, using appropriate Feynman rules in such a basis. To evaluate them in terms of the dressed propagators, one can start with the tree-level Feynman rules and introduce mixing matrices implementing a change into a basis in which the dressed propagators are diagonal, so that one may use propagators with the mass eigenvalues $M^2_{\varphi,k}$. 

In the SM, the mixing effects only appear in the neutral gauge and ghost sectors. The correct generalisation of the self-energies in terms of the dressed mass eigenvalues can be carried out as follows, where we focus on the gauge sector. If a tree-level interaction couples to the tree-level mass eigenstates \(V_a=(Z,\gamma)_a\) through the combination \(q_a V_a\), then the relevant contraction in a loop is of the form
\begin{align}
    q_a\,\mathcal D_X(\mathbf p,\omega_n)_{ab}\,q_b .
\end{align}
Separating the Lorentz structure and the prefactor $\kappa_X(\xi)$, after the resummation the internal neutral propagator in sector $X$ can be expressed in terms of the diagonal resummed propagators as follows:
\begin{align}
    \mathcal D_X(\mathbf p,\omega_n)_{ab}
    =\sum_{k=1}^2
    \frac{(U_{\hat{Z},X})_{ak}(U_{\hat{Z},X})_{bk}}
    {\omega_n^2+\mathbf p^2+\kappa_X(\xi)M_{\hat{Z},X,k}^2},
    \label{eq:neutral_gap_propagator}
\end{align}
for $a,b\in\{Z,\gamma\}$. Note that this allows for off-diagonal correlations for $a\neq b$ that are not present at tree level. The corresponding contraction becomes
\begin{align}
    q_a\mathcal D_X(\mathbf p,\omega_n)_{ab} \, q_b
    =\sum_{k=1}^2
    \frac{q_{X,k}^2}
    {\omega_n^2+\mathbf p^2+\kappa_X(\xi)M_{\hat{Z},X,k}^2},
    \label{eq:neutral_gap_vertex_projection}
\end{align}
with $q_{X,k}\equiv q_a(U_{\hat Z,X})_{ak}$. This treatment ensures that the diagonalisation also projects the tree-level vertices onto the dressed propagating modes. Off-diagonal propagators $\mathcal D_X(\mathbf p,\omega_n)_{Z\gamma}$, which vanish at tree level, contribute when the dressed neutral mass matrix is non-diagonal.

The explicit SM self-energies entering Eqs.~\eqref{eq:all_self_energies} are given in Sec.~\ref{sec:4}, with vertices expressed in the tree-level basis. Whenever an internal neutral gauge boson appears, the propagator projections described above are applied. The explicit replacements for internal neutral gauge-boson and ghost propagators are collected in Appendix~\ref{app:neutral_mixing_replacements}. For $U_{\hat{Z},X}=\mathbf 1_2$, the gauge-boson projections reduce to the separate $Z$- and photon-propagator contributions. Similarly, when the off-diagonal neutral ghost self-energy vanishes, the neutral ghost propagator reduces to its diagonal form.


\section{SM self-energies beyond the high-temperature approximation}
\label{sec:4}

\begin{figure}[!t]
    \centering
        {\includegraphics[width=0.85\linewidth]{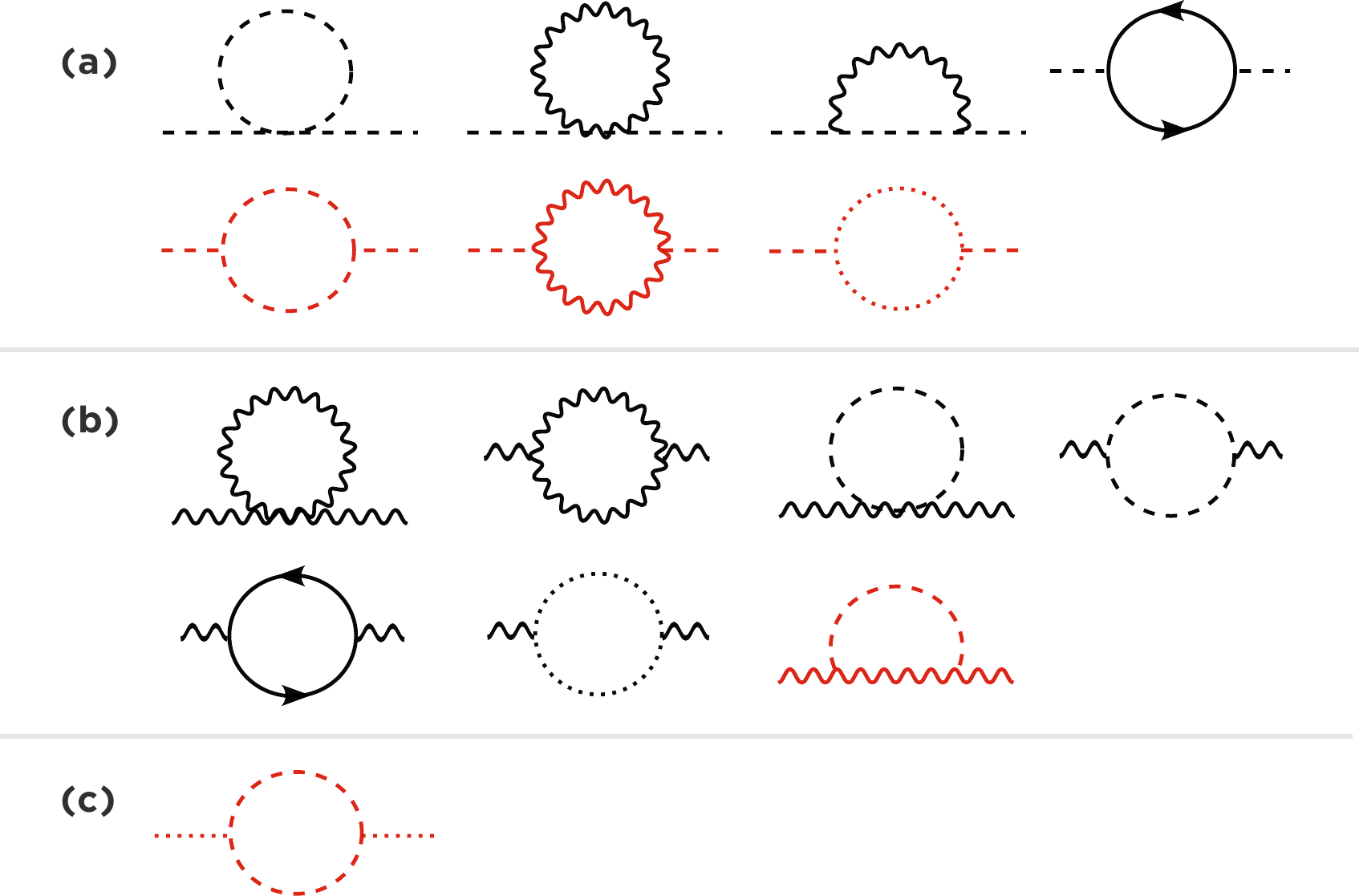}
        }
    \caption{Diagrams that contribute to the (a) scalar, (b) gauge-boson and (c) ghost self-energies. Diagrams highlighted in red contain two dimensionful cubic vertices and contribute beyond the leading \(\mathcal{O}(T^2)\) thermal masses.}
    \label{fig:SEall}
\end{figure}

We now give the one-loop thermal SM self-energies entering the coupled gap equations of Sec.~\ref{subsec:gap_equations}, retaining the full mass dependence of the thermal loop functions. In scalar theories, the one-loop mass corrections can be obtained straightforwardly from the curvature of the effective potential, as used in earlier PD studies~\cite{Boyd:1993tz,Curtin:2016urg,Bahl:2024bcg},
\begin{align}
    \Delta m_{ij}^2
    &=\frac{\partial^2}{\partial\bar\Phi_i\,\partial\bar\Phi_j}
    \left[V_{\rm CW}(\bar\Phi)+V_T(\bar\Phi,T)\right].
    \label{eq:scalar_loop_curvature}
\end{align}
This shortcut does not determine the gauge-boson self-energies, since the scalar-background potential does not resolve their Lorentz structure. We therefore evaluate all sectors diagrammatically, including the separate temporal and spatial gauge components in Eq.~\eqref{eq:gauge_static_selfenergies}. Fig.~\ref{fig:SEall} gives an overview of the contributing self-energy diagrams.

The tadpole integral $I_1$ introduced in Eq.~\eqref{eq:I1_definition} provides the starting point for all thermal loop functions used below. Its bosonic and fermionic thermal parts are
\begin{align}
    I_1^{\rm B /\rm F}(m^2)
    =\frac{T^2}{\pi^2}
    J_{\rm B /\rm F}'\!\left(\frac{m^2}{T^2}\right),
    \label{eq:I1-definition}
\end{align}
with $J_{\rm B /\rm F}$ defined in Eq.~\eqref{eq:JB} and the prime denoting differentiation with respect to the dimensionless argument. All remaining thermal functions can be derived from $I_1^{\rm B /\rm F}$ through mass differentiation and integration, combined with algebraic reductions. Their definitions are collected in Appendix~\ref{sec:appendix_B}. Throughout this section, these functions denote thermal contributions, their mass arguments are squared masses, and their explicit dependence on $T$ is suppressed.

The expressions below are written in the tree-level field bases introduced in Sec.~\ref{subsec:gauge_theory_setup}. When used in the gap equations, only masses originating from bosonic propagators are dressed, while the vertices remain at tree level. Internal neutral gauge and ghost propagators are treated with the mixing replacements collected in Appendix~\ref{app:neutral_mixing_replacements}.

For the explicit expressions below and the numerical analysis, we specialise to Feynman--'t Hooft gauge, $\xi=1$~\cite{Fujikawa:1972fe}. The tree-level vector propagator is then proportional to $\delta_{\mu\nu}$, while the dressed propagator retains the thermal tensor decomposition. The effective potential away from its extrema depends on the gauge parameter. Nielsen identities relate this dependence to derivatives of the potential and imply that its value at an exact stationary point is gauge
independent~\cite{Nielsen:1975fs,Fukuda:1975di}. In a strict perturbative expansion, this cancellation requires a consistent treatment of both the potential and its extrema~\cite{Patel:2011th}. A finite-order calculation supplemented by partial resummation does not automatically preserve these cancellations and can retain residual gauge dependence. A systematic assessment of the gauge-parameter dependence of the present calculation is left to future work.


\subsection{Scalars}

The scalar self-energies receive contributions from the diagrams in Fig.~\ref{fig:SEall}(a). The upper-row topologies contribute already to the leading thermal masses of order $T^2$. The bubbles in the lower row instead contain two dimensionful cubic vertices and are subleading in the strict high-temperature expansion. We retain both classes to include the full mass dependence of the one-loop thermal self-energies entering the gap equations.

The scalar self-energy matrix has the diagonal form given in Eq.~\eqref{eq:Pizero_scalar} in the real basis $(G^1,G^2,h,G^0)$. The two charged modes have the same self-energy, leaving the three independent entries $\Pi_h(0)$, $\Pi_{G^\pm}(0)$ and $\Pi_{G^0}(0)$. Reducing the diagrams to the thermal functions $I_i^{\rm B /\rm F}$ gives
\begingroup
\allowdisplaybreaks[2]
\begin{align}
\Pi_h(0)
={}&
g^2\Big[
\frac12 (2 I_1^{\rm B}(m^2_{W,T})
+ I_1^{\rm B}(m^2_{W,L})
+ I_1^{\rm B}(m^2_{W,D}) )
-\frac12 I_5^{\rm B}(m^2_{W,D},m^2_{G^\pm})
\Big]
\nonumber\\
&+
\frac{g^2}{c_w^2}
\Big[
\frac14(2 I_1^{\rm B}(m^2_{Z,T})
+ I_1^{\rm B}(m^2_{Z,L})
+ I_1^{\rm B}(m^2_{Z,D}))
-\frac14 I_5^{\rm B}(m^2_{Z,D},m^2_{G^0})
\Big]
\nonumber\\
&+
\lambda\Big[
\frac32 I_1^{\rm B}(m^2_h)
+I_1^{\rm B}(m^2_{G^\pm})
+\frac12 I_1^{\rm B}(m^2_{G^0})
\Big] - 6\sum_{q=t,b}
y_q^2
\Big[
I_1^{\rm F}(m_q^2)-2m_q^2 I_3^{\rm F}(m_q^2)
\Big]
\nonumber\\
&-
g^4h^2 \Big[
\frac14(
2 I_3^{\rm B}(m^2_{W,T})
+ I_3^{\rm B}(m^2_{W,L})
+ I_3^{\rm B}(m^2_{W,D}) ) - \frac18 I_3^{\rm B}(m^2_{\eta^\pm})
\nonumber\\
&\hspace{1.4cm}
- \frac{1}{16c_w^4}I_3^{\rm B}(m^2_{\eta^Z})
+ \frac{1}{8c_w^4}
(
2 I_3^{\rm B}(m^2_{Z,T})
+ I_3^{\rm B}(m^2_{Z,L})
+ I_3^{\rm B}(m^2_{Z,D})
)
\Big]
\nonumber\\
&-
\lambda^2h^2
\Big[
\frac92 I_3^{\rm B}(m^2_h)
+I_3^{\rm B}(m^2_{G^\pm})
+\frac12 I_3^{\rm B}(m^2_{G^0})
\Big] ,
\label{eq:Pi-h-full}
\\[1.0em]
\Pi_{G^\pm}(0)
={}&
g^2 \Big[
\frac12(
2 I_1^{\rm B}(m^2_{W,T})
+ I_1^{\rm B}(m^2_{W,L})
+ I_1^{\rm B}(m^2_{W,D})
)
-\frac14 I_5^{\rm B}(m^2_{W,D},m^2_h)
\nonumber\\
&\hspace{0.5cm}
-\frac14 I_5^{\rm B}(m^2_{W,D},m^2_{G^0})
\Big]
\nonumber\\
&+
\frac{g^2(c_w^2-s_w^2)^2}{c_w^2}
\Big[
\frac14(
2 I_1^{\rm B}(m^2_{Z,T})
+ I_1^{\rm B}(m^2_{Z,L})
+ I_1^{\rm B}(m^2_{Z,D})
)
-\frac14 I_5^{\rm B}(m^2_{G^\pm},m^2_{Z,D})
\Big]
\nonumber\\
&+
e^2 \Big[
2 I_1^{\rm B}(m^2_{\gamma,T})
+ I_1^{\rm B}(m^2_{\gamma,L})
+ I_1^{\rm B}(m^2_{\gamma,D})
-I_5^{\rm B}(m^2_{G^\pm},m^2_{\gamma,D})
\Big]
\nonumber\\
&+
\lambda\Big[
\frac12 I_1^{\rm B}(m^2_h)
+2I_1^{\rm B}(m^2_{G^\pm})
+\frac12 I_1^{\rm B}(m^2_{G^0})
\Big]
\nonumber\\
&-
6\left(y_t^2+y_b^2\right)I_5^{\rm F}(m^2_t,m^2_b)
-12y_ty_b\,m_tm_b\,I_7^{\rm F}(m^2_t,m^2_b)
-\lambda^2h^2 I_7^{\rm B}(m^2_h,m^2_{G^\pm})
\nonumber\\
&-
g^4h^2 \Big[
\frac{s_w^2}{4}
(
2I_7^{\rm B}(m^2_{W,T},m^2_{\gamma,T})
+ I_7^{\rm B}(m^2_{W,L},m^2_{\gamma,L})
+ I_7^{\rm B}(m^2_{W,D},m^2_{\gamma,D})
)
\nonumber\\
&\hspace{1.3cm}
+ \frac{s_w^4}{4c_w^2}
(
2I_7^{\rm B}(m^2_{W,T},m^2_{Z,T})
+ I_7^{\rm B}(m^2_{W,L},m^2_{Z,L})
+ I_7^{\rm B}(m^2_{W,D},m^2_{Z,D})
)
\nonumber\\
&\hspace{1.3cm}
+\frac18 \frac{c_w^2-s_w^2}{c_w^2}
I_7^{\rm B}(m^2_{\eta^\pm},m^2_{\eta^Z})
\Big] ,
\label{eq:Pi-Gpm-full}
\\[1.0em]
\Pi_{G^0}(0)
={}&
g^2\Big[
\frac12(
2 I_1^{\rm B}(m^2_{W,T})
+ I_1^{\rm B}(m^2_{W,L})
+ I_1^{\rm B}(m^2_{W,D})
)
-\frac12 I_5^{\rm B}(m^2_{W,D},m^2_{G^\pm})
\Big]
\nonumber\\
&+
\frac{g^2}{c_w^2}
\Big[
\frac14(
2 I_1^{\rm B}(m^2_{Z,T})
+ I_1^{\rm B}(m^2_{Z,L})
+ I_1^{\rm B}(m^2_{Z,D})
)
-\frac14 I_5^{\rm B}(m^2_{Z,D},m^2_h)
\Big]
\nonumber\\
&+
\lambda\Big[
\frac12 I_1^{\rm B}(m^2_h)
+I_1^{\rm B}(m^2_{G^\pm})
+\frac32 I_1^{\rm B}(m^2_{G^0})
\Big]
\nonumber\\
&- 6\sum_{q=t,b}
y_q^2 I_1^{\rm F}(m^2_q)
\nonumber\\
&-
\frac18 g^4h^2 I_3^{\rm B}(m^2_{\eta^\pm})
-\lambda^2h^2 I_7^{\rm B}(m^2_h,m^2_{G^0}) .
\label{eq:Pi-G0-full}
\end{align}
\endgroup
The Yukawa contributions follow the conventions of Sec.~\ref{subsec:gauge_theory_setup}: the neutral entries are summed over $q=t,b$, while the charged Goldstone entry contains the pair $(t,b)$.

At leading order in the high-temperature expansion, the three scalar self-energies coincide:
\begin{align}
\Pi_h(0)
=
\Pi_{G^\pm}(0)
=
\Pi_{G^0}(0)
\overset{T\gg m}{=}
\left(
\frac{3}{16}g^2
+\frac{1}{16}g'^2
+\frac{1}{4}\lambda
+\frac{1}{4}(y_t^2+y_b^2)
\right)T^2 .
\label{eq:scalar-self-energies-highT}
\end{align}
This reproduces the standard leading thermal mass in our potential normalisation~\cite{Carrington:1991hz}.


\subsection{Gauge bosons}
\label{sec:gbselfenergies}

The gauge-boson self-energies receive contributions from the diagrams in Fig.~\ref{fig:SEall}(b). We list the temporal components $\Pi_A^{00}(0)_{ab}$ and one spatial diagonal component $\Pi_A^{ii}(0)_{ab}$, with no sum over $i$, entering the longitudinal and transverse gap equations, respectively, in Eqs.~\eqref{eq:all_self_energies}. The $D$ sector follows from $M_{A,D}^2=M_{A,T}^2$ within the prescription introduced in Sec.~\ref{subsec:selfenergydecomp}. In the real tree-level basis $(W^1,W^2,Z,A^\gamma)$, the independent entries are $\Pi_{W^\pm}^{\mu\nu}(0)$, $\Pi_Z^{\mu\nu}(0)$, $\Pi_\gamma^{\mu\nu}(0)$ and $\Pi_{Z\gamma}^{\mu\nu}(0)=\Pi_{\gamma Z}^{\mu\nu}(0)$. The off-diagonal neutral entry is retained when constructing the mass matrices in Eq.~\eqref{eq:gauge_SM_mass_blocks}.

For the neutral-current fermion loops, we define
\begin{align}
g_V^f
=
\frac12 T_3^f-Q_f s_w^2,
\qquad
g_A^f
=
\frac12 T_3^f,
\end{align}
and
\begin{align}
C_f=(g_V^f)^2+(g_A^f)^2,
\qquad
D_f=(g_V^f)^2-(g_A^f)^2 .
\end{align}
Here $T_3^f$ is the weak isospin of the left-handed fermion component and $Q_f$ its electric charge. The sums over $f$ include all SM fermions, with $N_c^f=3$ for quarks and $N_c^f=1$ for leptons. Following Sec.~\ref{subsec:gauge_theory_setup}, only $t$ and $b$ retain their field-dependent masses; the remaining fermions are treated as massless. The sum $\sum_{(u,d)}$ runs over the quark and lepton weak doublets of all three generations, with $N_c$ denoting the corresponding colour multiplicity.

Reducing the diagrams to the thermal functions introduced above gives the temporal components

\begingroup
\allowdisplaybreaks[2]
\begin{align}
\Pi_{W^\pm}^{00}(0)
={}&
g^2c_w^2
\Big[
-8I_6^{\rm B}(m^2_{Z,T},m^2_{W,T})
-4I_6^{\rm B}(m^2_{Z,L},m^2_{W,L})
\nonumber\\
&  \ \qquad
-3I_8^{\rm B}(m^2_{Z,L},m^2_{W,D})
-3I_8^{\rm B}(m^2_{Z,D},m^2_{W,L})
\nonumber\\
&  \ \qquad
+2I_1^{\rm B}(m^2_{Z,T})
+I_6^{\rm B}(m^2_{Z,L},0)
+3I_8^{\rm B}(m^2_{Z,D},0)
+2I_6^{\rm B}(m^2_{\eta^Z},m^2_{\eta^\pm})
\Big]
\nonumber\\
&+
e^2
\Big[
-8I_6^{\rm B}(m^2_{\gamma,T},m^2_{W,T})
-4I_6^{\rm B}(m^2_{\gamma,L},m^2_{W,L})
\nonumber\\
&  \ \qquad
-3I_8^{\rm B}(m^2_{\gamma,L},m^2_{W,D})
-3I_8^{\rm B}(m^2_{\gamma,D},m^2_{W,L})
\nonumber\\
&  \ \qquad
+2I_1^{\rm B}(m^2_{\gamma,T})
+I_6^{\rm B}(m^2_{\gamma,L},0)
+3I_8^{\rm B}(m^2_{\gamma,D},0)
+2I_6^{\rm B}(m^2_{\eta^\gamma},m^2_{\eta^\pm})
\Big]
\nonumber\\
&+
g^2
\Big[
2I_1^{\rm B}(m^2_{W,T})
+I_6^{\rm B}(m^2_{W,L},0)
+3I_8^{\rm B}(m^2_{W,D},0)
+\frac14I_1^{\rm B}(m^2_h)
\nonumber\\
&  \ \qquad
+\frac14I_1^{\rm B}(m^2_{G^0})
+\frac12I_1^{\rm B}(m^2_{G^\pm})
-I_6^{\rm B}(m^2_h,m^2_{G^\pm})
-I_6^{\rm B}(m^2_{G^0},m^2_{G^\pm})
\Big]
\nonumber\\
&+
g^2\sum_{(u,d)}N_c
\Big[
2I_6^{\rm F}(m^2_u,m^2_d)-I_5^{\rm F}(m^2_u,m^2_d)
\Big]
\nonumber\\
&-
\frac{g^4h^2}{4}
\Big[
3\mathcal{I}_8^{\rm B}(m^2_{W,L},m^2_h)
+\mathcal{I}_6^{\rm B}(m^2_{W,D},m^2_h)
+3 s_w^2 \mathcal{I}_8^{\rm B}(m^2_{\gamma,L},m^2_{G^\pm})
\nonumber\\
& \qquad  \quad \  \ 
+s_w^2\mathcal{I}_6^{\rm B}(m^2_{\gamma,D},m^2_{G^\pm})
+3 \frac{s_w^4}{c_w^2}\mathcal{I}_8^{\rm B}(m^2_{Z,L},m^2_{G^\pm})
+\frac{s_w^4}{c_w^2} \mathcal{I}_6^{\rm B}(m^2_{Z,D},m^2_{G^\pm}) \Big]
\label{eq:Pi-W-00-full}
\\[1.0em]
\Pi_Z^{00}(0)
={}&
g^2c_w^2
\Big[
-8I_2^{\rm B}(m^2_{W,T})
-4I_2^{\rm B}(m^2_{W,L})
-6I_8^{\rm B}(m^2_{W,L},m^2_{W,D})
+4I_1^{\rm B}(m^2_{W,T})
\nonumber\\
&  \ \qquad
+2I_6^{\rm B}(m^2_{W,L},0)
+6I_8^{\rm B}(m^2_{W,D},0)
+2I_2^{\rm B}(m^2_{\eta^\pm})
\Big]
\nonumber\\
&+
\frac{g^2}{c_w^2}
\Big[
\frac14I_1^{\rm B}(m^2_h)
+\frac14I_1^{\rm B}(m^2_{G^0})
-I_6^{\rm B}(m^2_h,m^2_{G^0})
\Big]
\nonumber\\
&+
g^2\frac{(c_w^2-s_w^2)^2}{c_w^2}
\Big[
\frac12I_1^{\rm B}(m^2_{G^\pm})
-I_2^{\rm B}(m^2_{G^\pm})
\Big]
\nonumber\\
&+
4\frac{g^2}{c_w^2}
\sum_f N_c^f
\Big[
2C_f I_2^{\rm F}(m^2_f)
-C_f I_1^{\rm F}(m^2_f)
+(C_f-D_f)m_f^2 I_3^{\rm F}(m^2_f)
\Big]
\nonumber\\
&-
\frac{g^4h^2}{4}
\Big[
\frac{3}{c_w^4}\mathcal{I}_8^{\rm B}(m^2_{Z,L},m^2_h)
+\frac{1}{c_w^4} \mathcal{I}_6^{\rm B}(m^2_{Z,D},m^2_h)
+6 \frac{s_w^4}{c_w^2}\mathcal{I}_8^{\rm B}(m^2_{W,L},m^2_{G^\pm})
\nonumber\\
& \qquad \quad \ \ +2 \frac{s_w^4}{c_w^2}\mathcal{I}_6^{\rm B}(m^2_{W,D},m^2_{G^\pm})
\Big],
\label{eq:Pi-Z-00-full}
\\[1.0em]
\Pi_\gamma^{00}(0)
={}&
e^2
\Big[
-8I_2^{\rm B}(m^2_{W,T})
-4I_2^{\rm B}(m^2_{W,L})
-6I_8^{\rm B}(m^2_{W,L},m^2_{W,D})
+4I_1^{\rm B}(m^2_{W,T})
\nonumber\\
&  \ \qquad
+2I_6^{\rm B}(m^2_{W,L},0)
+6I_8^{\rm B}(m^2_{W,D},0)
+2I_2^{\rm B}(m^2_{\eta^\pm})
\nonumber\\
&  \ \qquad
-4I_2^{\rm B}(m^2_{G^\pm})
+2I_1^{\rm B}(m^2_{G^\pm})
\Big]
\nonumber\\
&+
4 e^2 \sum_f N_c^f Q_f^2
\Big[
2I_2^{\rm F}(m^2_f)-I_1^{\rm F}(m^2_f)
\Big]
\nonumber\\
&-
\frac{g^4s_w^2h^2}{2}
\Big[
3\mathcal{I}_8^{\rm B}(m^2_{W,L},m^2_{G^\pm})
+\mathcal{I}_6^{\rm B}(m^2_{W,D},m^2_{G^\pm})
\Big],
\label{eq:Pi-gamma-00-full}
\\[1.0em]
\Pi_{Z\gamma}^{00}(0)
={}&
egc_w
\Big[
-8I_2^{\rm B}(m^2_{W,T})
-4I_2^{\rm B}(m^2_{W,L})
-6I_8^{\rm B}(m^2_{W,L},m^2_{W,D})
+4I_1^{\rm B}(m^2_{W,T})
\nonumber\\
&  \ \qquad
+2I_6^{\rm B}(m^2_{W,L},0)
+6I_8^{\rm B}(m^2_{W,D},0)
+2I_2^{\rm B}(m^2_{\eta^\pm})
\Big]
\nonumber\\
&+
eg\frac{c_w^2-s_w^2}{c_w}
\Big[
I_1^{\rm B}(m^2_{G^\pm})
-2I_2^{\rm B}(m^2_{G^\pm})
\Big]
\nonumber\\
&+
4 \frac{eg}{c_w} \sum_f N_c^f Q_f g_V^f
\Big[
2I_2^{\rm F}(m^2_f)-I_1^{\rm F}(m^2_f)
\Big]
\nonumber\\
&+
\frac{g^4s_w^3h^2}{2c_w}
\Big[
3\mathcal{I}_8^{\rm B}(m^2_{W,L},m^2_{G^\pm})
+\mathcal{I}_6^{\rm B}(m^2_{W,D},m^2_{G^\pm})
\Big].
\label{eq:Pi-Zgamma-00-full}
\end{align}
\endgroup

The corresponding spatial components are
\begingroup
\allowdisplaybreaks[2]
\begin{align}
\Pi_{W^\pm}^{ii}(0)
={}&
g^2c_w^2
\Big[
-8I_8^{\rm B}(m^2_{Z,T},m^2_{W,T})
-4I_8^{\rm B}(m^2_{Z,L},m^2_{W,L})
-\frac23 I_5^{\rm B}(m^2_{Z,T},m^2_{W,D})
\nonumber\\
&  \ \qquad
-\frac23 I_5^{\rm B}(m^2_{Z,D},m^2_{W,T})
-\frac13 I_6^{\rm B}(m^2_{Z,L},m^2_{W,D})
-\frac13 I_6^{\rm B}(m^2_{Z,D},m^2_{W,L})
\nonumber\\
&  \ \qquad
+\frac43 I_1^{\rm B}(m^2_{Z,T})
+I_1^{\rm B}(m^2_{Z,L})
-\frac13 I_6^{\rm B}(m^2_{Z,L},0)
+I_1^{\rm B}(m^2_{Z,D})
-I_8^{\rm B}(m^2_{Z,D},0)
\nonumber\\
&  \ \qquad
+2I_8^{\rm B}(m^2_{\eta^Z},m^2_{\eta^\pm})
\Big]
\nonumber\\
&+
e^2
\Big[
-8I_8^{\rm B}(m^2_{\gamma,T},m^2_{W,T})
-4I_8^{\rm B}(m^2_{\gamma,L},m^2_{W,L})
-\frac23 I_5^{\rm B}(m^2_{\gamma,T},m^2_{W,D})
\nonumber\\
&  \ \qquad
-\frac23 I_5^{\rm B}(m^2_{\gamma,D},m^2_{W,T})
-\frac13 I_6^{\rm B}(m^2_{\gamma,L},m^2_{W,D})
-\frac13 I_6^{\rm B}(m^2_{\gamma,D},m^2_{W,L})
\nonumber\\
&  \ \qquad
+\frac43 I_1^{\rm B}(m^2_{\gamma,T})
+I_1^{\rm B}(m^2_{\gamma,L})
-\frac13 I_6^{\rm B}(m^2_{\gamma,L},0)
+I_1^{\rm B}(m^2_{\gamma,D})
-I_8^{\rm B}(m^2_{\gamma,D},0)
\nonumber\\
&  \ \qquad
+2I_8^{\rm B}(m^2_{\eta^\gamma},m^2_{\eta^\pm})
\Big]
\nonumber\\
&+
g^2
\Big[
\frac43 I_1^{\rm B}(m^2_{W,T})
+I_1^{\rm B}(m^2_{W,L})
-\frac13 I_6^{\rm B}(m^2_{W,L},0)
+I_1^{\rm B}(m^2_{W,D})
-I_8^{\rm B}(m^2_{W,D},0)
\nonumber\\
&  \ \qquad
+\frac14I_1^{\rm B}(m^2_h)
+\frac14I_1^{\rm B}(m^2_{G^0})
+\frac12I_1^{\rm B}(m^2_{G^\pm})
-I_8^{\rm B}(m^2_h,m^2_{G^\pm})
-I_8^{\rm B}(m^2_{G^0},m^2_{G^\pm})
\Big]
\nonumber\\
&+
g^2\sum_{(u,d)}N_c
\Big[
2I_8^{\rm F}(m^2_u,m^2_d)-I_5^{\rm F}(m^2_u,m^2_d)
\Big]
\nonumber\\
&-
\frac{g^4h^2}{4}
\Big[
\frac23 I_7^{\rm B}(m^2_{W,T},m^2_h)
+\frac13 \mathcal{I}_6^{\rm B}(m^2_{W,L},m^2_h)
+\mathcal{I}_8^{\rm B}(m^2_{W,D},m^2_h)
\nonumber\\
& \qquad \quad \ \
+\frac23s_w^2 I_7^{\rm B}(m^2_{\gamma,T},m^2_{G^\pm})
+\frac13 s_w^2 \mathcal{I}_6^{\rm B}(m^2_{\gamma,L},m^2_{G^\pm})
+s_w^2 \mathcal{I}_8^{\rm B}(m^2_{\gamma,D},m^2_{G^\pm})
\nonumber\\
&  \qquad \quad \ \
+ \frac23 \frac{s_w^4}{c_w^2} I_7^{\rm B}(m^2_{Z,T},m^2_{G^\pm})
+\frac13 \frac{s_w^4}{c_w^2} \mathcal{I}_6^{\rm B}(m^2_{Z,L},m^2_{G^\pm})
+\frac{s_w^4}{c_w^2} \mathcal{I}_8^{\rm B}(m^2_{Z,D},m^2_{G^\pm})
\Big],
\label{eq:Pi-W-ii-full}
\\[1.0em]
\Pi_Z^{ii}(0)
={}&
g^2c_w^2
\Big[
-8I_4^{\rm B}(m^2_{W,T})
-4I_4^{\rm B}(m^2_{W,L})
-\frac43 I_5^{\rm B}(m^2_{W,T},m^2_{W,D})
-\frac23 I_6^{\rm B}(m^2_{W,L},m^2_{W,D})
\nonumber\\
&  \ \qquad
+\frac83 I_1^{\rm B}(m^2_{W,T})
+2 I_1^{\rm B}(m^2_{W,L})
-\frac23 I_6^{\rm B}(m^2_{W,L},0)
+2I_1^{\rm B}(m^2_{W,D})
\nonumber\\
&  \ \qquad
-2I_8^{\rm B}(m^2_{W,D},0)
+2I_4^{\rm B}(m^2_{\eta^\pm})
\Big]
\nonumber\\
&+
\frac{g^2}{c_w^2}
\Big[
\frac14I_1^{\rm B}(m^2_h)
+\frac14I_1^{\rm B}(m^2_{G^0})
-I_8^{\rm B}(m^2_h,m^2_{G^0})
\Big]
\nonumber\\
&+
g^2\frac{(c_w^2-s_w^2)^2}{c_w^2}
\Big[
\frac12I_1^{\rm B}(m^2_{G^\pm})
-I_4^{\rm B}(m^2_{G^\pm})
\Big]
\nonumber\\
&+
4\frac{g^2}{c_w^2}
\sum_f N_c^f
\Big[
2C_f I_4^{\rm F}(m^2_f)
-C_f I_1^{\rm F}(m^2_f)
+(C_f-D_f)m_f^2 I_3^{\rm F}(m^2_f)
\Big]
\nonumber\\
&-
\frac{g^4h^2}{4}
\Big[
\frac23 \frac{1}{c_w^4}I_7^{\rm B}(m^2_{Z,T},m^2_h)
+\frac13 \frac{1}{c_w^4}\mathcal{I}_6^{\rm B}(m^2_{Z,L},m^2_h)
+ \frac{1}{c_w^4} \mathcal{I}_8^{\rm B}(m^2_{Z,D},m^2_h)
\nonumber\\
&  \qquad \quad \ \
+\frac43 \frac{s_w^4}{c_w^2} I_7^{\rm B}(m^2_{W,T},m^2_{G^\pm})
+\frac23 \frac{s_w^4}{c_w^2} \mathcal{I}_6^{\rm B}(m^2_{W,L},m^2_{G^\pm})
+2\frac{s_w^4}{c_w^2}\mathcal{I}_8^{\rm B}(m^2_{W,D},m^2_{G^\pm})
\Big],
\label{eq:Pi-Z-ii-full}
\\[1.0em]
\Pi_\gamma^{ii}(0)
={}&
e^2
\Big[
-8I_4^{\rm B}(m^2_{W,T})
-4I_4^{\rm B}(m^2_{W,L})
-\frac43 I_5^{\rm B}(m^2_{W,T},m^2_{W,D})
-\frac23 I_6^{\rm B}(m^2_{W,L},m^2_{W,D})
\nonumber\\
&  \ \qquad
+\frac83 I_1^{\rm B}(m^2_{W,T})
+2I_1^{\rm B}(m^2_{W,L})
-\frac23 I_6^{\rm B}(m^2_{W,L},0)
+2I_1^{\rm B}(m^2_{W,D})
-2I_8^{\rm B}(m^2_{W,D},0)
\nonumber\\
&  \ \qquad
+2I_4^{\rm B}(m^2_{\eta^\pm})
-4I_4^{\rm B}(m^2_{G^\pm})
+2I_1^{\rm B}(m^2_{G^\pm})
\Big]
\nonumber\\
&+
4 e^2 \sum_f N_c^f Q_f^2
\Big[
2I_4^{\rm F}(m^2_f)-I_1^{\rm F}(m^2_f)
\Big]
\nonumber\\
&-
\frac{g^4s_w^2h^2}{2}
\Big[
\frac23 I_7^{\rm B}(m^2_{W,T},m^2_{G^\pm})
+\frac13 \mathcal{I}_6^{\rm B}(m^2_{W,L},m^2_{G^\pm})
+\mathcal{I}_8^{\rm B}(m^2_{W,D},m^2_{G^\pm})
\Big],
\label{eq:Pi-gamma-ii-full}
\\[1.0em]
\Pi_{Z\gamma}^{ii}(0)
={}&
egc_w
\Big[
-8I_4^{\rm B}(m^2_{W,T})
-4I_4^{\rm B}(m^2_{W,L})
-\frac43 I_5^{\rm B}(m^2_{W,T},m^2_{W,D})
-\frac23 I_6^{\rm B}(m^2_{W,L},m^2_{W,D})
\nonumber\\
&  \ \qquad
+\frac83 I_1^{\rm B}(m^2_{W,T})
+2I_1^{\rm B}(m^2_{W,L})
-\frac23 I_6^{\rm B}(m^2_{W,L},0)
+2I_1^{\rm B}(m^2_{W,D})
-2I_8^{\rm B}(m^2_{W,D},0)
\nonumber\\
&  \ \qquad
+2I_4^{\rm B}(m^2_{\eta^\pm})
\Big]
\nonumber\\
&+
\frac{eg}{c_w}(c_w^2-s_w^2)
\Big[
I_1^{\rm B}(m^2_{G^\pm})
-2I_4^{\rm B}(m^2_{G^\pm})
\Big]
\nonumber\\
&+
4 \frac{eg}{c_w} \sum_f N_c^f Q_f g_V^f
\Big[
2I_4^{\rm F}(m^2_f)-I_1^{\rm F}(m^2_f)
\Big]
\nonumber\\
&+
\frac{g^4s_w^3h^2}{2c_w}
\Big[
\frac23 I_7^{\rm B}(m^2_{W,T},m^2_{G^\pm})
+\frac13 \mathcal{I}_6^{\rm B}(m^2_{W,L},m^2_{G^\pm})
+\mathcal{I}_8^{\rm B}(m^2_{W,D},m^2_{G^\pm})
\Big].
\label{eq:Pi-Zgamma-ii-full}
\end{align}
\endgroup

As a check, we consider the leading high-temperature limit. Rotating the temporal self-energy matrix from the real tree-level basis $(W^1,W^2,Z,A^\gamma)$ back to the gauge basis $(W^1,W^2,W^3,B)$ gives
\begin{align}
\Pi_A^{00}(0)_{ab}
\overset{T\gg m}{=}
\frac{11}{6}T^2\,
\mathrm{diag}\left(g^2,g^2,g^2,g'^2\right)_{ab}.
\label{eq:gauge-self-energies-highT-gaugebasis}
\end{align}
This reproduces the standard leading Debye masses~\cite{Espinosa:1992kf,Espinosa:1993bs,Carrington:1991hz}. The spatial contributions cancel in the combined result at the same order,
\begin{align}
\Pi_A^{ii}(0)_{ab}
\overset{T\gg m}{=}
0,
\qquad \text{no sum over } i .
\label{eq:gauge-self-energies-highT-spatial}
\end{align}
At this order, only the longitudinal sector receives a thermal mass correction. Beyond this limit, we retain the mass-dependent spatial contributions in the transverse and $D$-sector matrices according to
Eq.~\eqref{eq:gauge_dressed_mass_matrices}.

The neutral longitudinal block can also be diagonalised analytically at leading high temperature:
\begin{align}
M_{\hat{Z},L,1/2}^2
=
\frac12
\left[
m_Z^2
+\frac{11}{6}(g^2+g^{\prime 2})T^2
\pm
\sqrt{
m_Z^4
+\frac{11}{3}
\frac{(g^{\prime 2}-g^2)^2}{g^2+g^{\prime 2}}
m_Z^2T^2
+\frac{121}{36}(g^{\prime 2}-g^2)^2T^4
}
\right].
\label{eq:neutral-longitudinal-highT-eigenvalues}
\end{align}
This agrees with the high-temperature result of Ref.~\cite{Espinosa:1993bs}. For $T=0$, the eigenvalues reduce to the tree-level squared masses of the $Z$ boson and photon, $m_Z^2$ and $0$.
For the full mass-dependent calculation, the neutral mass matrices and vertex projections are determined self-consistently using the numerical procedure described in Sec.~\ref{sec:numerical_setup} and the internal-propagator replacements in Appendix~\ref{app:neutral_mixing_replacements}.


\subsection{Ghosts}

The ghost sector is structurally simpler than the gauge-boson sector. At vanishing external momentum, the gauge--ghost loop vanishes because the derivative coupling supplies the external antighost momentum. The remaining contributions are scalar--ghost bubbles, shown in Fig.~\ref{fig:SEall}(c). They contain two dimensionful cubic vertices and are subleading in the strict high-temperature expansion.

The neutral self-energy has the upper-triangular structure of Eq.~\eqref{eq:Pizero_ghost}. In particular, $\Pi_{\eta^\gamma}(0)=\Pi_{\bar\eta^\gamma\eta^Z}(0)=0$, so the photon ghost remains massless. The non-vanishing entries are
\begin{align}
\Pi_{\eta^\pm}(0)
={}&
\frac{g^4h^2}{16}
\Big[
- I_7^{\rm B}(m^2_h,m^2_{\eta^\pm})
+ I_7^{\rm B}(m^2_{G^0},m^2_{\eta^\pm})
+\frac{c_w^2-s_w^2}{c_w^2}
I_7^{\rm B}(m^2_{G^\pm},m^2_{\eta^Z})
\Big],
\label{eq:Pi-eta-pm-full}
\\[1.0em]
\Pi_{\eta^Z}(0)
={}&
\frac{g^4h^2}{16}
\Big[
-\frac{1}{c_w^4}I_7^{\rm B}(m^2_h,m^2_{\eta^Z})
+2\frac{c_w^2-s_w^2}{c_w^2}
I_7^{\rm B}(m^2_{G^\pm},m^2_{\eta^\pm})
\Big],
\label{eq:Pi-eta-Z-full}
\\[1.0em]
\Pi_{\bar\eta^Z\eta^\gamma}(0)
={}&
\frac{g^4h^2}{4} \frac{s_w}{c_w}
I_7^{\rm B}(m^2_{G^\pm},m^2_{\eta^\pm}).
\label{eq:Pi-eta-Zgamma-full}
\end{align}
Only $\Pi_{\eta^\pm}(0)$ and $\Pi_{\eta^Z}(0)$ enter
the diagonal entries of the ghost mass matrices in
Eqs.~\eqref{eq:ghost_charged_mass} and \eqref{eq:ghost_neutral_mass_matrix}. The off-diagonal entry $\Pi_{\bar\eta^Z\eta^\gamma}(0)$ drops out of the ghost tadpole trace and the scalar ghost-loop contributions. It is nevertheless retained in the internal neutral ghost propagator entering the charged $W$-boson self-energies, using the replacement rules in Appendix~\ref{app:neutral_mixing_replacements}.


\section{Results}
\label{sec:5}

We now apply the PD framework to the electroweak SM and examine the convergence of the gap equations, the resummed effective potential and the resulting thermodynamic observables.

\subsection{Numerical setup}
\label{sec:numerical_setup}

We implement the PD resummation numerically in \textsc{Mathematica}. For each background field $h$ and temperature $T$, the coupled gap equations of Eqs.~\eqref{eq:all_self_energies} are solved using the simultaneous iteration
\begin{align}
    M_{i,(n+1)}^2(h,T)
    =m_i^2(h)
    +\Pi_i(0)\!\Big(h,T;\{M_{j,(n)}^2(h,T)\}\Big),
    \label{eq:gap_iteration}
\end{align}
starting from $M_{i,(0)}^2(h,T)=m_i^2(h)$. This local algebraic formulation follows Ref.~\cite{Curtin:2022ovx}. In the earlier OPD construction of Ref.~\cite{Curtin:2016urg}, differentiating the already dressed tadpole introduces field derivatives of the self-energies into the gap equations; the later study reported numerical artefacts from this differential formulation in its scalar benchmark. All self-energies are evaluated with the spectrum from the previous iteration, and the scalar, gauge-boson and ghost masses are updated simultaneously. The same update is applied to the matrix entries in sectors with mixing. The neutral gauge mass matrices are diagonalised at every iteration, retaining the field- and temperature-dependent $Z$--photon mixing. Only the thermal self-energies from Sec.~\ref{sec:4} enter our gap equations, while fermion loops are included with tree-level masses.

For the numerical evaluation, the gap equations and the dressed tadpole are evaluated on uniform two-dimensional $(h,T)$ grids. For the broad-field potential comparisons, we use $h/m_W\in \{0,0.01,\ldots,5\}$ and $T/m_W\in \{0.001,0.002,\ldots,2.4\}$. The small $T$ steps are necessary to determine the free energy in the later section. To resolve the small-field barrier structure and its dependence on the iteration number in Sec.~\ref{sec:potential_results}, we use a separate grid with $h/m_W\in \{0,10^{-3},\ldots,0.4\}$ and $T/m_W\in \{1.8,1.8001,\ldots,1.9\}$. The PD potentials are reconstructed by numerically integrating the renormalised Coleman--Weinberg and thermal tadpoles along the $h$ direction, with an additional integration constant as in Eq.~\eqref{eq:PD_Normalisation}. The thermal functions in these tadpoles are evaluated without a high-temperature expansion, as in Eqs.~\eqref{eq:scalar_PD_Tadpole}, \eqref{eq:gauge_PD_Tadpole_compact3} and \eqref{eq:ghost_PD_Tadpole3}. The resulting potential is interpolated in $h$ and $T$, allowing $V^{\rm PD}_{\rm eff}(h,T)$ to be evaluated at arbitrary points within the grid.

For negative scalar squared masses, the perturbative effective potential can develop an imaginary part, reflecting unstable fluctuations about the chosen background. It was shown that this imaginary part is related to the decay rate per unit volume of the associated quantum state~\cite{Weinberg:1987vp}. Following the finite-temperature treatment of Ref.~\cite{Espinosa:1992gq}, we use the real part of the potential throughout the numerical analysis. In the code, the bosonic thermal functions and their derivatives are evaluated as the real parts of their analytic continuations for $M^2/T^2<0$. The gap equations are iterated with real self-energies, retaining the sign of each squared mass, and the PD potential is reconstructed from the real dressed tadpole and a real FD integration constant. Regarding the latter, we use Eq.~\eqref{eq:PD_Normalisation} yet carrying the matching at a small reference field $h_{\rm ref}/m_W=10^{-6}$ instead of at $h=0$ in order to avoid numerical singularities at the origin of field space,
\begin{equation}
    V^{\rm PD}_{\rm eff}(h,T)
    =
    V^{\rm FD}_{\rm eff}(h_{\rm ref},T)
    +
    \int_{h_{\rm ref}}^{h} {\rm d}h'\,
    \frac{\partial V^{\rm PD}_{\rm eff}(h',T)}{\partial h'} \, .
    \label{eq:PD_Normalisation2}
\end{equation}

The SM parameters entering the calculation are obtained from physical electroweak inputs and evolved to the renormalisation scale of the effective potential. We take $m_W=80.363\,\mathrm{GeV}$, $m_Z=91.188\,\mathrm{GeV}$, $m_h=125.130\,\mathrm{GeV}$, $m_t=172.600\,\mathrm{GeV}$, $m_b=4.186\,\mathrm{GeV}$, $m_\tau=1.777\,\mathrm{GeV}$ and $G_F=1.166\times10^{-5}\,\mathrm{GeV}^{-2}$, $\alpha_{\rm em}^{-1}(m_Z)=127.930$, $\alpha_s(m_Z)=0.118$, $\sin^2\theta_W=0.231$ as boundary data \cite{ParticleDataGroup:2026mpi}. The electroweak vacuum expectation value is extracted from $G_F$ including the one-loop radiative correction, while the gauge couplings are fixed at $\mu_R=m_Z$ and evolved with the two-loop SM RGEs, supplemented by the three-loop contribution to the running of $g_3$. At the top-mass scale, the top and bottom Yukawa couplings are matched to the corresponding fermion masses including QCD corrections through $\mathcal{O}(\alpha_s^2)$, the $\mathcal{O}(\alpha_s^3)$ contribution for the top quark, and the one-loop QED and weak threshold corrections~\cite{Hempfling:1994ar,Chetyrkin:1999qi,Melnikov:2000qh}.

Since the Higgs-sector parameters enter the RG evolution themselves, their determination is performed iteratively. In each iteration, $\lambda$ and $\mu^2$ are adjusted such that the one-loop zero-temperature effective potential satisfies
\begin{equation}
    \left.
    \frac{\partial V_{\rm eff}^{T=0}}{\partial h}
    \right|_{h=v}=0,
    \qquad
    \left.
    \frac{\partial^2 V_{\rm eff}^{T=0}}{\partial h^2}
    \right|_{h=v}=m_h^2 .
    \label{eq:SM_higgs_matching}
\end{equation}
The second condition therefore identifies the Higgs input with the zero-momentum curvature mass; momentum-dependent pole-mass corrections are not included in the present matching. The RG evolution and matching are repeated until the Higgs-sector parameters have converged. The resulting RG trajectory is then used to evaluate all running parameters at the benchmark scale $\mu_R=m_W$. For the calculations, we use $m_W$ also to normalise fields, masses and potentials. This reference value should not be confused with the field-dependent tree-level mass $m_W(h)$.


\subsection{Convergence of the gap equations and thermal masses}
\label{sec:gap_results}

\begin{figure}[t]
    \centering
    \includegraphics[]
    {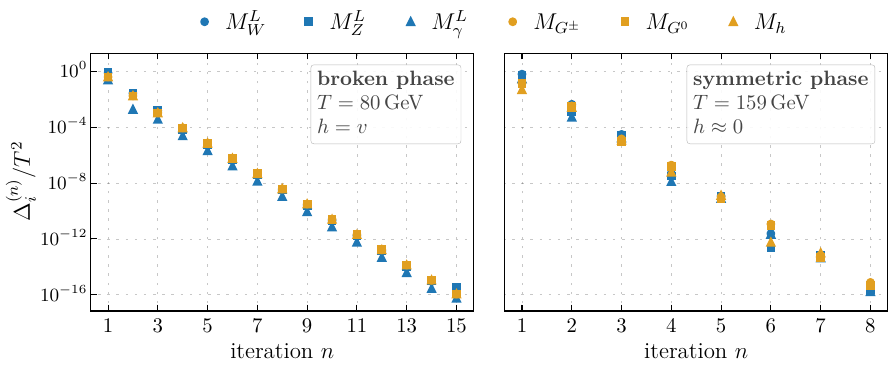}
    \caption{Convergence of the coupled gap equations in the $\mathrm{PD}_{\rm(sc+L)}$ prescription at $T=80\,\mathrm{GeV}$ and $h=v\simeq240\,\mathrm{GeV}$ (left), and $T=159\,\mathrm{GeV}$ and $h\simeq0$ (right). Shown are the absolute changes between successive squared-mass iterates, $\Delta_i^{(n)}=|M_{i,(n)}^2-M_{i,(n-1)}^2|$, normalised to $T^2$, for the scalar (orange) and longitudinal gauge-boson (blue) modes. At both benchmarks, the changes decrease to the level of machine precision.}
    \label{fig:gap_convergence}
\end{figure}

We first examine how the dressed masses approach a stable result under the simultaneous updates of Eq.~\eqref{eq:gap_iteration}. For each mode $i$, we monitor the change between successive squared-mass iterates,
\begin{equation}
    \Delta_i^{(n)}
    \equiv
    \big|M_{i,(n)}^2-M_{i,(n-1)}^2\big| .
    \label{eq:gap_iteration_change}
\end{equation}
Fig.~\ref{fig:gap_convergence} shows the dimensionless quantity $\Delta_i^{(n)}/T^2$ for the scalar and longitudinal gauge-boson modes at two representative benchmark points. The broken-phase benchmark (left) corresponds to $T=80~\mathrm{GeV}$ and $h=v\simeq240\,\rm{GeV}$, while the symmetric-phase benchmark (right) is evaluated at $T=159\,\mathrm{GeV}$ and $h\approx0$. At both benchmark points, the changes decrease by several orders of magnitude within only a few iterations, eventually reaching $\Delta_i^{(n)}/T^2\sim10^{-16}$, consistent with the round-off level of the machine-precision arithmetic used here. This level is reached after about eight iterations at the symmetric-phase benchmark, roughly half the number required at the broken-phase benchmark. The required iteration count nevertheless depends not only on the convergence of the masses, but also on the quantity being resolved. Small remaining changes in the dressed spectrum can still affect shallow structures in the effective potential, whose slope involves cancellations between tree-level and loop contributions. We therefore examine the convergence of the reconstructed potential separately in Sec.~\ref{sec:potential_results}.

We next consider the field dependence of the dressed thermal masses. To distinguish the successive extensions of the gap resummation used below, we adopt the following notation: We denote by $\mathrm{TFD}$ the conventional truncated full-dressing prescription, in which all thermal self-energies are kept at leading $\mathcal{O}(T^2)$. The notation $\mathrm{PD}_{\rm(sc)}$ refers to partial dressing with self-consistently determined scalar masses, while the gauge-boson self-energies remain at leading $\mathcal{O}(T^2)$. We denote by $\mathrm{PD}_{\rm(sc+L)}$ the extension in which the longitudinal gauge-boson sector is also treated self-consistently, while the transverse gauge-boson masses remain at tree level. Finally, $\mathrm{PD}_{\rm(sc+g)}$ additionally includes the transverse gauge-boson gap equations; ghost dressing is included self-consistently in both gauge-sector extensions. All three PD prescriptions retain the full mass dependence of the thermal loop functions in the self-energies that are iterated. Our main comparison concerns $\mathrm{PD}_{\rm(sc)}$ and $\mathrm{PD}_{\rm(sc+L)}$, with $\mathrm{PD}_{\rm(sc+g)}$ included as an exploratory extension in view of the nonperturbative infrared sensitivity of the transverse sector at small background fields.

\begin{figure}[t!]
    \centering
    \includegraphics[width=\textwidth]
    {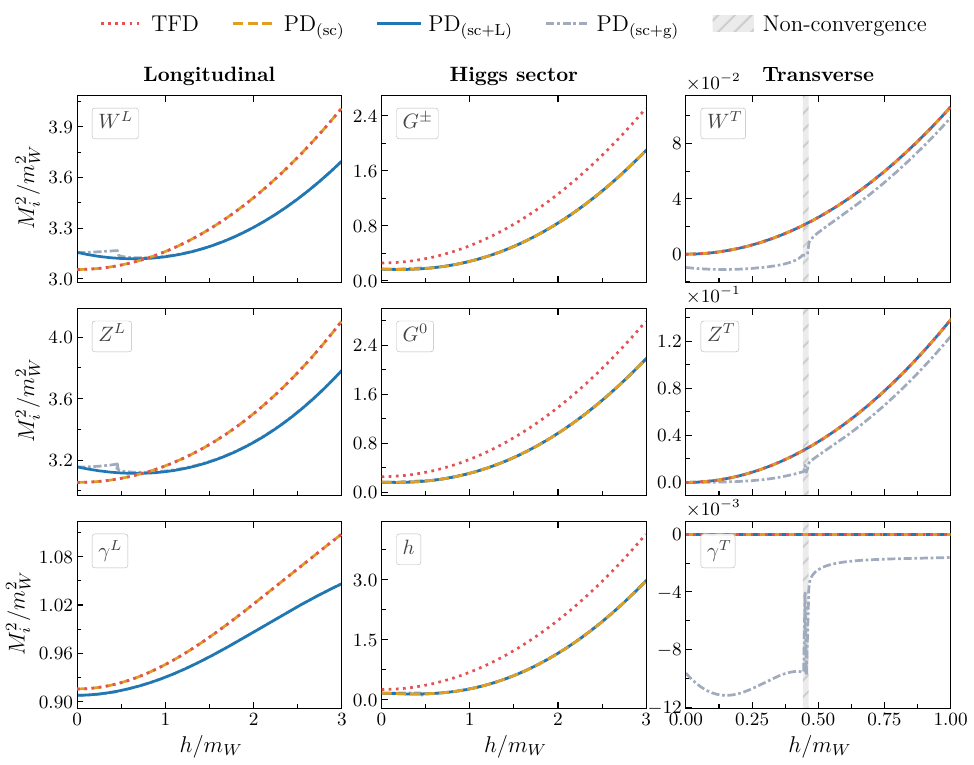}
    \caption{Field-dependent dressed squared masses at $T=159~\mathrm{GeV}$ for TFD, $\mathrm{PD}_{\rm(sc)}$, $\mathrm{PD}_{\rm(sc+L)}$ and $\mathrm{PD}_{\rm(sc+g)}$. From left to right, the columns show the longitudinal gauge, scalar and transverse gauge modes. All squared masses are normalised to $m_W^2$. The TFD and $\mathrm{PD}_{\rm(sc)}$ gauge curves overlap because both retain only the leading $\mathcal{O}(T^2)$ self-energies. Their transverse masses also coincide with $\mathrm{PD}_{\rm(sc+L)}$, which keeps these modes at tree level. The hatched band marks nonconvergence of the coupled $\mathrm{PD}_{\rm(sc+g)}$ gap iteration.}
    \label{fig:thermal_masses}
\end{figure}

Fig.~\ref{fig:thermal_masses} compares $\mathrm{PD}_{\rm(sc)}$ and $\mathrm{PD}_{\rm(sc+L)}$ with the leading $\mathcal{O}(T^2)$ thermal masses underlying the TFD prescription. In $\mathrm{PD}_{\rm(sc)}$, the longitudinal gauge-boson self-energies remain independent of dressed mass arguments, although the resulting thermally corrected gauge-boson masses still enter the scalar gap equations. In $\mathrm{PD}_{\rm(sc+L)}$, the longitudinal gauge-boson self-energies themselves are evaluated with the dressed spectrum, such that the scalar and longitudinal gauge sectors are treated self-consistently without relying on a high-temperature expansion of the thermal loop functions.

The comparison demonstrates qualitatively different results for the scalar- and gauge-sector corrections. In the scalar sector, $\mathrm{PD}_{\rm(sc)}$ and $\mathrm{PD}_{\rm(sc+L)}$ give almost indistinguishable results over the field range shown, whereas both deviate substantially from the leading $\mathcal{O}(T^2)$ approximation. For this benchmark, the dominant correction to the scalar spectrum is therefore already captured by the mass-dependent scalar self-energies included in $\mathrm{PD}_{\rm(sc)}$. Extending the self-consistent treatment to the longitudinal gauge sector induces only a small additional feedback on the Higgs and Goldstone masses.

The opposite pattern is found in the longitudinal gauge sector. Since the gauge-boson self-energies remain at leading high temperature in $\mathrm{PD}_{\rm(sc)}$, the corresponding $W^L$, $Z^L$ and $\gamma^L$ masses coincide with the TFD result. The $\mathrm{PD}_{\rm(sc+L)}$ solution instead generates an additional field dependence. For $W^L$ and $Z^L$, the masses are slightly enhanced close to the origin, cross the leading-high-temperature result around $h/m_W\sim0.8$, and fall increasingly below it at larger background fields. The effect is more moderate for the longitudinal photon, whose mass remains below the leading result throughout. At larger fields, this suppression reflects the gradual reduction of thermal contributions as the masses entering the loop functions increase relative to the temperature. In particular, the crossing between the $W^L$ and $Z^L$ curves and the corresponding results in the high-temperature approximation shows that the corrections beyond the high-temperature approximation cannot be represented by a field-independent shift of the leading Debye masses.

At larger background fields, where the $\mathrm{PD}_{\rm(sc+g)}$ iteration converges and the effective couplings in the longitudinal and transverse sectors remain perturbative, the scalar and longitudinal gauge-boson masses remain close to their $\mathrm{PD}_{\rm(sc+L)}$ values. The additional transverse dressing therefore induces only a small feedback on these sectors. The masses of the transverse gauge boson excitations do not deviate strongly from their values in the high-temperature approximation. While for large $h$ one can obtain negative values of the transverse photon mass, this is likely due to numerical errors in the diagonalisation which propagate throughout the iterative procedure, as the absolute value of the $\gamma^T$ mass remains much below those of the $Z^T$ and $W^T$ excitations.  Towards the origin, however, the absence of a leading $\mathcal{O}(g^2T^2)$ transverse screening mass squared leaves the non-Abelian sector infrared sensitive, with a loop-expansion parameter scaling as $\mathrm{(loop/tree)} \sim g^2T/M_T\sim 1$, as discussed in Sec.~\ref{sec:gauge_power_counting}.
Numerically, the iterative procedure fails to converge for $0.442\lesssim h/m_W\lesssim0.463$ at $T=159~\mathrm{GeV}$, even after $10^4$ updates, and for smaller values of $h$ the $W^T$ excitations become sizably tachyonic. Such tachyonic masses, which are unphysical, represent a breakdown of the resummation, as expected from the emergence of a strong effective coupling in the transverse sector. Continuation tests reveal turning points of the relevant solution branches, driven by strong transverse gauge-sector feedback; other formal roots exist but are repulsive under the iteration. These findings expose limitations of the present resummation: neither numerical convergence nor its failure establishes a controlled description of nonperturbative magnetic screening.

At this benchmark, extending $\mathrm{PD}_{\rm(sc)}$ to $\mathrm{PD}_{\rm(sc+L)}$ therefore modifies the longitudinal screening masses while inducing only a small feedback on the scalar spectrum. We now examine how these spectral differences affect the reconstructed effective potential, particularly the structures in the transition region, whose sensitivity cannot be inferred from the mass comparison alone.

\begin{figure}[t]
    \centering
    \includegraphics[]
    {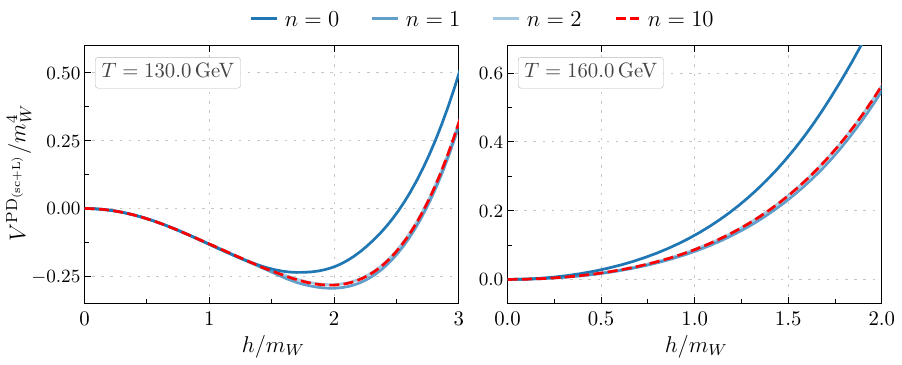}
    \caption{Convergence of the $\mathrm{PD}_{\rm(sc+L)}$ effective potential at $T=130~\mathrm{GeV}$ (left panel) and $T=160~\mathrm{GeV}$ (right panel), representative of the broken and symmetric phases, respectively. The curves show the iterations $n=0,1,2,10$ of the coupled gap equations, with each potential shifted such that $V_{(n)}(0,T)=0$ and normalised to $m_W^4$. The red dashed curve corresponds to $n=10$. In both panels, the overall potential profile is already stable by $n\simeq2$ on the scale shown.}
    \label{fig:potential_convergence}
\end{figure}

\subsection{Resummed effective potential}
\label{sec:potential_results}

We now turn to the effective potential reconstructed from the dressed tadpole. Since a field-independent contribution does not affect the potential shape at fixed temperature, all potentials shown in this subsection are shifted such that $V(0,T)=0$. We continue to denote the resulting quantity simply by $V(h,T)$. The absolute normalisation becomes relevant only for the thermodynamic observables discussed in Sec.~\ref{sec:dof_results}.

We first verify that the convergence of the gap equations established in Sec.~\ref{sec:gap_results} is inherited by the reconstructed potential. Fig.~\ref{fig:potential_convergence} shows the $\mathrm{PD}_{\rm(sc+L)}$ potential for successive iterations of the coupled gap equations at $T=130~\mathrm{GeV}$ and $T=160~\mathrm{GeV}$, representative of the broken and symmetric phases, respectively. The first iterations visibly modify the position and depth of the broken-phase minimum, whereas the potential profile stabilises rapidly: in both panels, the $n=2$ and $n=10$ curves are nearly indistinguishable on the scale shown. Together with the convergence of the individual thermal masses in
Fig.~\ref{fig:gap_convergence}, this supports the numerical stability of the reconstructed potential on the scale relevant for its overall shape. Convergence is obtained for most of the sampled benchmark points, with the remaining difficulties localised in field intervals of the broken-phase benchmarks where a dressed scalar squared mass approaches zero. In these intervals, the infrared sensitivity of the bosonic loop functions can slow or prevent convergence of the gap iteration.

\begin{figure}[t!]
    \centering
    \includegraphics[]
    {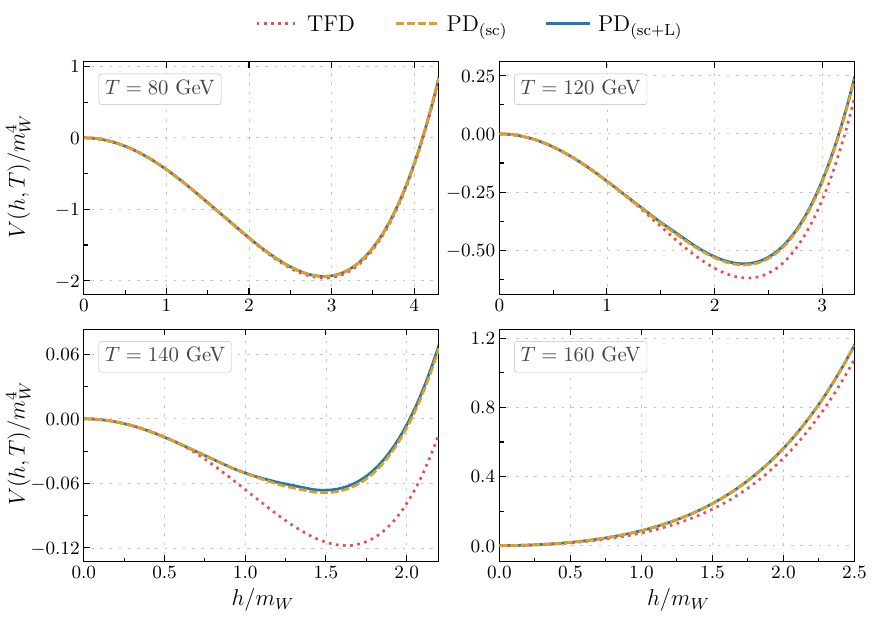}
    \caption{Comparison of the TFD, $\mathrm{PD}_{\rm(sc)}$, and $\mathrm{PD}_{\rm(sc+L)}$ effective potentials at different benchmark temperatures. The first three panels show the potentials in the broken phase at $T=80\,\mathrm{GeV}$, $120\,\mathrm{GeV}$, and $140\,\mathrm{GeV}$, while the last panel illustrates their behaviour in the symmetric phase at $T=160\,\mathrm{GeV}$. While the two PD prescriptions remain close throughout, with modest differences at intermediate temperatures, TFD predicts a deeper broken-phase minimum, with the discrepancy most apparent near the transition temperature.}
    \label{fig:potential_comparison}
\end{figure}

Having established the numerical stability of the reconstructed potential, we compare the resummation prescriptions in Fig.~\ref{fig:potential_comparison}.
The $\mathrm{PD}_{\rm(sc)}$ and $\mathrm{PD}_{\rm(sc+L)}$ potentials nearly coincide at $T=80\,\mathrm{GeV}$ and $160\,\mathrm{GeV}$. At intermediate temperatures, the additional longitudinal gauge-sector dressing makes the broken-phase minimum slightly shallower, with little change in its position. Their close agreement therefore indicates that this additional feedback has only a modest effect on the overall potential shape at these benchmarks. TFD instead predicts a deeper broken-phase minimum, with the difference most apparent at $T=140\,\mathrm{GeV}$. This illustrates the limitations of retaining only the leading $\mathcal{O}(T^2)$ thermal self-energies when the masses entering the plasma become comparable to or larger than the temperature. The full thermal functions employed in the PD self-energies account for the corresponding suppression of heavy thermal modes. Nevertheless, at $T=80\,\mathrm{GeV}$ all three potentials remain close on the plotted scale.

\begin{figure}[t]
    \centering
    \includegraphics[]
    {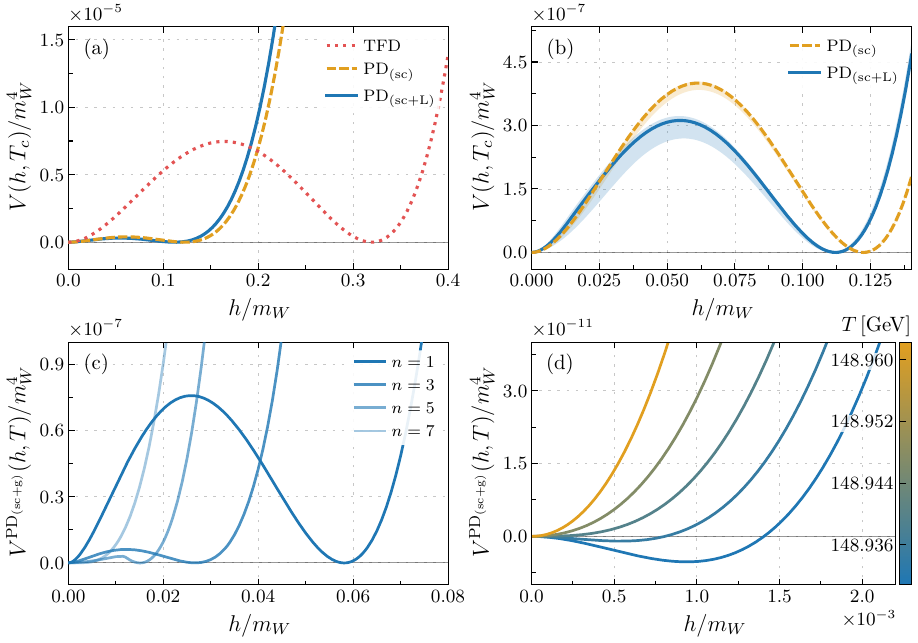}
    \caption{Small-field structure of the resummed effective potential. (a) TFD, $\mathrm{PD}_{\rm(sc)}$ and $\mathrm{PD}_{\rm(sc+L)}$ at their respective critical temperatures, $T_c = 151.475$, $148.195$ and $148.210\,\mathrm{GeV}$. (b) Magnification of the two PD critical potentials. The central curves use an endpoint- and slope-matched quadratic tadpole reconstruction; shaded envelopes span the tested matching points, fitting windows and polynomial orders, with each reconstruction evaluated at its own equal-depth temperature. (c) Finite-iteration $\mathrm{PD}_{\rm(sc+g)}$ potentials for $n=1,3,5,7$, at respective temperatures $T=150.972,149.344,149.092$ and $149.002\,\mathrm{GeV}$. (d) Residual-converged $\mathrm{PD}_{\rm(sc+g)}$ potentials near the crossing of zero of the effective mass at the origin for $T\simeq148.939\,\mathrm{GeV}$. The non-zero minimum approaches the origin without a resolved intervening barrier on this branch.}
    \label{fig:potential_profile}
\end{figure}

These modest changes in the overall potential can nevertheless have a qualitative impact on the phase structure. Near the transition, a weak first-order transition is characterised by a shallow barrier between the symmetric and broken minima, whose height is far below the scales resolved in Fig.~\ref{fig:potential_comparison}. Even small corrections may therefore determine whether this barrier survives. To examine this sensitivity, we compare the small-field potentials obtained with $\mathrm{PD}_{\rm(sc)}$, $\mathrm{PD}_{\rm(sc+L)}$ and $\mathrm{PD}_{\rm(sc+g)}$ as gauge-sector self-energies are successively included.

At the respective critical temperatures, scalar dressing in $\mathrm{PD}_{\rm(sc)}$ and $\mathrm{PD}_{\rm(sc+L)}$ substantially lowers the barrier and shifts the broken minimum towards the origin relative to TFD, as shown in panels (a) and (b) of Fig.~\ref{fig:potential_profile}. The barrier height decreases from approximately $7.5\times10^{-6}m_W^4$ in TFD to $4.0\times10^{-7}m_W^4$ in $\mathrm{PD}_{\rm(sc)}$ and $3.1\times10^{-7}m_W^4$ in $\mathrm{PD}_{\rm(sc+L)}$. The corresponding broken minima move from $h/m_W\simeq0.32$ to approximately $h/m_W\simeq0.12$ and $h/m_W\simeq0.11$, respectively. Including longitudinal dressing therefore lowers the barrier height by a further $22\%$ relative to $\mathrm{PD}_{\rm(sc)}$, while retaining the barrier between the symmetric and broken minima. Thus, the close agreement of the prescriptions on the broad-field scale does not imply agreement on the much smaller structure separating the symmetric and broken minima.

At some of the smallest sampled field values, the gap iteration does not converge. The converged outer profiles extend down to $h/m_W=0.025$ for $\mathrm{PD}_{\rm(sc)}$ and $h/m_W=0.018$ for $\mathrm{PD}_{\rm(sc+L)}$, leaving a small interval near the origin, $h\lesssim2\,\mathrm{GeV}$, to be reconstructed. Although the tadpole is small in this interval, uncertainty in its integral shifts the potential at all larger fields relative to the origin, thereby affecting the barrier height. For the thick curves in panel~(b), we describe the tadpole in this interval by a quadratic function matched in value and slope to the converged outer samples. The shaded bands span variations of the matching point, fitting window and quadratic or cubic polynomial ansatz. Each reconstruction is evaluated at its own critical temperature, where the two minima have equal depth. The bands indicate sensitivity to this reconstruction procedure; they do not represent statistical confidence intervals or the full theoretical uncertainty. The first-order structure persists across the tested reconstructions for both PD prescriptions.

Including transverse gauge-sector dressing progressively suppresses the barrier as the coupled gap equations are iterated. For the odd iterations $n=1,3,5,7$ shown in Fig.~\ref{fig:potential_profile}(c), with the temperature adjusted for each iterate, the residual barrier moves towards the origin and its height decreases by more than three orders of magnitude between $n=1$ and $n=7$. To determine whether a barrier survives convergence, we examine solutions that satisfy the coupled gap equations.

The resulting profiles in Fig.~\ref{fig:potential_profile}(d) show a non-zero minimum below $T\simeq148.939\,\mathrm{GeV}$, which approaches the origin continuously as the temperature increases. Above this value, the origin is the only minimum within the resolved field interval. No intervening barrier or pair of separated minima is resolved across these profiles. The characteristic temperature is therefore identified by the change in sign of the curvature at the origin --- or equivalently, the effective Higgs mass at $h=0$ --- rather than by the degeneracy of two distinct minima. Within our numerical resolution, this branch thus evolves continuously without a resolved barrier, with a critical temperature of $T_c=148.9\,\mathrm{GeV}$.

We further assess the sensitivity of this result to the zero-momentum approximation in Appendix~\ref{app:momentum_dependence}, where we incorporate the leading momentum-dependent contributions to selected scalar self-energies into the dressed tadpole, keeping the gap-equation solutions fixed. These corrections modify the potential near the transition and slightly shift the transition temperature, but do not restore a barrier within our numerical resolution.

Taken together, these comparisons show that longitudinal dressing lowers the reconstructed barrier, while additional transverse-sector feedback leaves no resolved barrier on the converged branch. At these small fields, however, the effective coupling of the non-Abelian transverse modes becomes large, and superdaisy resummation alone does not provide a controlled description of this sector, as discussed in Sec.~\ref{sec:gauge_power_counting}. Numerical convergence therefore does not establish the physical reliability of these solutions. The observed barrier suppression highlights the sensitivity of the shallow potential to the treatment of transverse gauge modes, while its trend towards barrier-free evolution is qualitatively consistent with the SM crossover with $159.5\pm1.5\,\mathrm{GeV}$ (within 7\% of the $\rm PD_{(\rm s+g)}$ estimate) established nonperturbatively~\cite{DOnofrio:2015gop}. A determination of the physical transition nevertheless requires control over the nonperturbative transverse dynamics, as well as the truncation, gauge and renormalisation-scale dependence.


\subsection{Entropy degrees of freedom}
\label{sec:dof_results}
The resummed effective potential also determines the thermodynamic properties of the electroweak plasma. We focus on the entropy density and the associated effective number of degrees of freedom. Their evaluation requires the absolute free energy, including its temperature-dependent normalisation. Denoting the equilibrium free-energy density by $\mathcal{F}(T)=V(h_{\rm min}(T),T)$, the entropy density is given by $s(T)=-{\rm d}\mathcal{F}/{\rm d}T$. We define the corresponding effective number of entropy degrees of freedom as
\begin{equation}
    g_{\star s}(T)
    =
    \frac{45}{2\pi^2 T^3}\,s(T).
    \label{eq:entropy_degrees_of_freedom}
\end{equation}
At a stationary minimum, $\partial V/\partial h=0$, so the implicit temperature dependence of $h_{\min}(T)$ does not contribute to the derivative on a smooth equilibrium branch. The derivative must retain the temperature dependence of the dressed masses and, for PD, of the field-independent normalisation fixed in Eq.~\eqref{eq:PD_Normalisation2}.

\begin{figure}[t]
    \centering
    \includegraphics[]
    {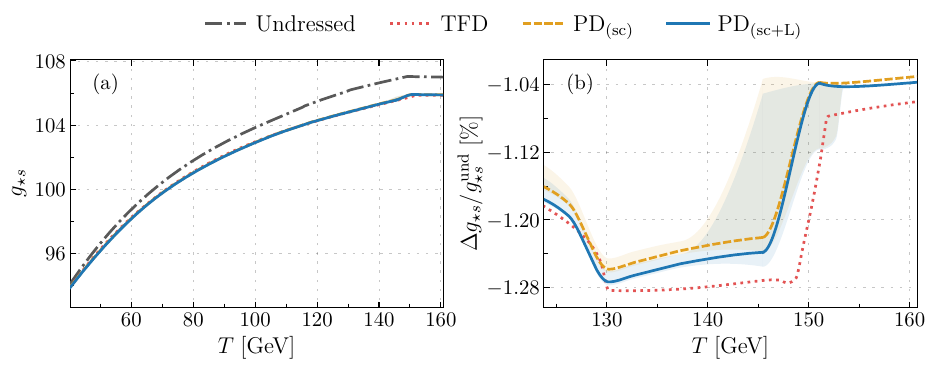}
    \caption{
        Effective number of entropy degrees of freedom for the undressed, TFD, $\mathrm{PD}_{\rm(sc)}$ and $\mathrm{PD}_{\rm(sc+L)}$ prescriptions. (a) Absolute values of $g_{\star s}(T)$. (b) Relative differences with respect to the undressed result, expressed in percent, in the transition region. The PD results retain the temperature-dependent FD normalisation. Shaded envelopes show sensitivity to the tested tadpole reconstructions at unconverged benchmark points.
    }
    \label{fig:dof_comparison}
\end{figure}

Fig.~\ref{fig:dof_comparison}(a) compares the undressed, TFD, $\mathrm{PD}_{\rm(sc)}$ and $\mathrm{PD}_{\rm(sc+L)}$ results. All curves increase with temperature as more electroweak particles become thermally excited. Conversely, cooling the plasma suppresses the top, Higgs, $W^\pm$
and $Z$ contributions, reducing the ideal-gas counting from $106.75$ towards $86.25$. Thermal resummation lowers the entropy by approximately one percent near the transition, while the three resummed curves remain close on this scale.

The main deviations arise around the transition temperature. These differences are resolved more clearly in Fig.~\ref{fig:dof_comparison}(b). It shows the relative differences from the undressed result,
\begin{align}
    \frac{\Delta g_{\star s}}{g_{\star s}^{\rm und}}
    \equiv
    \frac{g_{\star s}-g_{\star s}^{\rm und}}
         {g_{\star s}^{\rm und}}.
\end{align}
For $130\, \rm{GeV} \lesssim T \lesssim 150 \, \rm{GeV}$ the central curves lie approximately $1.2$--$1.3\%$ below the undressed result. Towards the upper end of the displayed interval, $150\, \rm{GeV} \lesssim T \lesssim 160 \, \rm{GeV} $, this suppression decreases slightly. The variation in TFD is
displaced towards higher temperatures relative to PD, qualitatively consistent with its higher critical temperature. The detailed shape in this region remains sensitive to the reconstruction. Away from this interval, the central PD--TFD differences are typically only a few hundredths of a percent, with additional longitudinal dressing producing an even smaller change. Thus, the more complete dressing substantially modifies the shallow potential barrier while leaving the bulk entropy close to the TFD result. This is because the entropy receives contributions from all thermally populated species, whereas the barrier depends on much smaller free-energy differences over a narrow field interval.


\section{Discussion}
\label{sec:6}
We have extended the partial dressing resummation (PD) method of the thermal effective potential, incorporating the effect of vacuum masses for the gauge bosons in spontaneously broken non-Abelian gauge theories. The method relies on using dressed masses for scalar, gauge-boson and ghost excitations, which are determined by solving coupled gap equations with fully mass-dependent thermal self-energies. Mixing effects are included in both the gap equations and the dressed tadpole $V'_{\rm eff}$. The longitudinal and transverse gauge excitations are treated separately, and the dressing is formulated in a general $R_\xi$ gauge.

The method has been applied to the Standard Model at $\xi=1$, with the self-energies in the gap equations restricted to their temperature-dependent contributions. The coupled gap equations converge rapidly at representative points in both the broken and symmetric phases. Retaining the full mass dependence of the thermal self-energies of the gauge bosons produces sizeable corrections to the dressed gauge boson masses beyond the leading $\mathcal O(T^2)$ approximation. Such corrections enhance the longitudinal $W^L$ and $Z^L$ masses for small scalar backgrounds and suppress them at larger fields, cf. Fig.~\ref{fig:thermal_masses}. The effect of the gauge boson resummation on the dressed scalar masses remains on the other hand small. Away from the critical temperature, the PD prescriptions $\mathrm{PD}_{\rm(sc)}$ and $\mathrm{PD}_{\rm(sc+L)}$, the former only accounting for scalar vacuum masses and the latter including those of the longitudinal gauge bosons, give similar potential profiles, with a generally shallower broken minimum than the commonly used truncated full-dressing prescription (TFD). At the respective critical temperatures,  $\mathrm{PD}_{\rm(sc)}$ reduces the barrier by more than an order of magnitude relative to TFD and shifts the broken minimum towards the origin. The additional longitudinal gauge dressing considered in this paper gives a further $22\%$ reduction of the barrier relative to  $\mathrm{PD}_{\rm(sc)}$, despite its small additional effect on the scalar masses. The barrier remains present across the tested small-field reconstructions. 

Including transverse modes in the gap equations ($\mathrm{PD}_{\rm(sc+g)}$) removes the potential barrier altogether. The broken minimum moves continuously towards the origin as the temperature increases, with a critical temperature $T_c=148.9$ GeV, in qualitative agreement with the SM crossover found on the lattice~\cite{DOnofrio:2015gop}. A test of the sensitivity to momentum-dependent corrections leaves the barrier suppression intact, providing further support for the robustness of the qualitative result. However, the non-Abelian transverse modes are strongly coupled in this region, so that the PD resummation alone is not sufficient to determine the physical nature of the transition. The problematic nature of the resummation is evidenced by the fact that the gap equations result in negative transverse masses for $W$ bosons. The calculation shows nevertheless that the transverse modes and their associated strong, long-range interactions play a crucial role in determining the nature of the electroweak transition.

The changes in the potential are more pronounced than those in the effective number of relativistic degrees of freedom $g_{\star s}$. It is found that $\mathrm{PD}_{\rm(sc)}$ and $\mathrm{PD}_{\rm(sc+L)}$ remain close to TFD, with all three prescriptions lowering $g_{\star s}$ by approximately one percent near the transition relative to the undressed result.

In the SM benchmark, longitudinal gauge dressing produces moderate changes to the overall potential. In extensions beyond the SM where gauge interactions control symmetry breaking or scalar thermal masses, accurately resumming their contributions becomes especially relevant, particularly when masses become of the order of the temperature. Our framework retains the full mass dependence of thermal self-energies and the coupled feedback between sectors, and can be readily adapted to such theories. This also opens applications to calculations  of thermodynamic properties when the affected sector contributes significantly to the plasma entropy.

A complete treatment through two-loop order, including contributions with overlapping loop momenta, can be achieved by extending the hybrid partial dressing method to non-Abelian gauge theories~\cite{Boyd:1993tz,Berthiaume:2026dkt}. The diagram-by-diagram organisation of the self-energies allows the required contributions to be selected for the hybrid gap equations, while the self-consistent diagonalisation can accommodate mixing. The appropriate diagrammatic counting requires particular care, since derivative gauge interactions generate bubble self-energies already at leading $\mathcal O(g^2T^2)$. We leave this extension, including the explicit two-loop diagrams and the subtractions needed to avoid double counting, for future work.

The general $R_\xi$ formulation provides a starting point for assessing the residual gauge dependence of the results, although it does not by itself guarantee gauge-independent predictions~\cite{Patel:2011th,Athron:2022jyi}. A systematic study of renormalisation-scale dependence should accompany these checks, with consistent renormalisation-group improvement and the higher-order thermal terms required for cancellations of the scale dependence.


\section*{Note added}

During the completion of this paper, Ref.~\cite{Qiu:2026bje} appeared on the arXiv, featuring gap equations for longitudinal gauge bosons beyond the high-temperature regime in a modified PD implementation. Our treatment follows a standard PD procedure, includes results in which the gap equations are also solved for transverse gauge bosons, and seems to differ in the handling of mixing effects in the neutral gauge boson sector.

\section*{Acknowledgements}

\begingroup
\setlength{\emergencystretch}{1em}
The authors acknowledge support from the Cluster of Excellence “Precision Physics, Fundamental Interactions, and Structure of Matter” (PRISMA++ EXC 2118/1) funded by the Deutsche Forschungsgemeinschaft (DFG, German Research Foundation) within the German Excellence Strategy (Project No. 390831469).
\par
\endgroup

\newpage


\appendix
\label{appendix}


\section{Tensor decomposition in the static infrared limit}
\label{appendixA}
At finite temperature, the presence of the thermal medium singles out the four-velocity $u^\mu$, thereby reducing Lorentz invariance to spatial rotational invariance in the plasma rest frame. As a consequence, the most general form of the self-energy tensor in covariant gauges consistent with these symmetries can be constructed as a linear combination of the tensor structures $\delta^{\mu\nu}$, $k^\mu k^\nu$, $u^\mu u^\nu$, and $k^\mu u^\nu + u^\mu k^\nu$. It is convenient to reorganise this general decomposition in terms of a standard tensor basis adapted to the thermal medium,
\begin{equation}
\Pi_A^{\mu \nu}(k)_{ab} = P_T^{\mu \nu} \Pi_{A,T}(k)_{ab}  + P_L^{\mu \nu} \Pi_{A,L}(k)_{ab} + P_C^{\mu \nu} \Pi_{A,C}(k)_{ab} + P_D^{\mu \nu} \Pi_{A,D}(k)_{ab},
\label{eq:Pimunugeneral}
\end{equation}
where the four tensor structures are defined as follows~\cite{Weldon:1996kb}:

\begin{align}
P_T^{\mu \nu} &= \delta^{\mu \nu} -\frac{{\widetilde{k}}^\mu {\widetilde{k}}^\nu}{k^2} - \frac{{k}^\mu {k}^\nu}{k^2}, \label{eq:PT} \\[0.2em]
P_L^{\mu \nu} &= \frac{{\widetilde{k}}^\mu {\widetilde{k}}^\nu}{k^2},  \label{eq:PL}\\[0.2em]
P_C^{\mu \nu} &= \frac{{k}^\mu {\widetilde{k}}^\nu + {\widetilde{k}}^\mu {k}^\nu}{k^2}, \label{eq:PC} \\[0.2em]
P_D^{\mu \nu} &= \frac{{k}^\mu {k}^\nu}{k^2} .  \label{eq:PD}
\end{align}
Here we introduce the vector ${\widetilde{k}}^\mu = ((k \cdot u) k^\mu-k^2 u^\mu) / |\mathbf{k}|$, which is orthogonal to $k^\mu$ with ${\widetilde{k}}^2 = k^2$. Among these tensors, $P_T^{\mu\nu}$, $P_L^{\mu\nu}$, and $P_D^{\mu\nu}$ are genuine projection operators satisfying
\begin{equation}
P_X^{\mu\rho}P_Y^{\rho\nu}
=\delta_{XY}P_X^{\mu\nu},
\qquad X,Y=T,L,D,
\end{equation}
while $P_C^{\mu\nu}$ encodes mixing between the longitudinal spatial and time-like directions and is therefore not a projector. The tensor $P_T^{\mu\nu}$ projects onto the spatial subspace transverse to $\mathbf{k}$, while $P_L^{\mu\nu}$, although commonly referred to as longitudinal, does not in general project onto the purely spatial longitudinal direction, but rather onto the corresponding direction within the time--longitudinal subspace defined by $k^\mu$ and $u^\mu$. In the plasma rest frame $u^\mu = (1,0,0,0)$, the expressions simplify considerably. In this frame, ${\widetilde{k}}^\mu = (-|\mathbf{k}|,k_0 \,\hat{\mathbf{k}})$ and the tensor structures in Eqs.~\eqref{eq:PT}–\eqref{eq:PD} can be written in matrix form as
\begin{align}
P_T^{\mu \nu} &= \left(
\begin{array}{c c}
0 & 0 \\
0 &\delta^{mn} - \hat{k}^m \hat{k}^n \\
\end{array}
\label{eq:PT_rest}
\right), \\
P_L^{\mu \nu} &=  \frac{1}{k^2} \left(
\begin{array}{c c}
|\mathbf{k}|^2 & -k_0k^n \\
-k^m k_0 & k_0^2 \hat{k}^m \hat{k}^n \\
\end{array}
\label{eq:PL_rest}
\right), \\
P_C^{\mu \nu} &=  \frac{1}{k^2} \left(
\begin{array}{c c}
-2k_0 |\mathbf{k}| & (k_0^2 -  |\mathbf{k}|^2)\hat{k}^n \\
\hat{k}^m(k_0^2 -  |\mathbf{k}|^2) & 2 k_0 |\mathbf{k}| \hat{k}^m \hat{k}^n \\
\end{array}
\label{eq:PC_rest}
\right), \\
P_D^{\mu \nu} &=  \frac{1}{k^2} \left(
\begin{array}{c c}
k_0^2& k_0k^n \\
k^m k_0 & k^m k^n \\
\end{array}
\label{eq:PD_rest}
\right).
\end{align}
For the purpose of this paper we are interested in the static infrared limit, which is defined by setting the temporal component of the external momentum to zero and sending the spatial momentum to zero: $\Pi^{\mu\nu}(k^0=0,|\mathbf{k}|\to0)\equiv\Pi^{\mu\nu}(0)$\cite{Carrington:1991hz}. In this limit, the purely temporal component of the self-energy tensor is directly connected to the coefficient of the longitudinal projection operator
\begin{align}
\Pi_A^{0 0}(0)_{ab} = \Pi_{A,L}(0)_{ab}.
\label{eq:Pi00simpl}
\end{align}
The spatial components yield
\begin{align}
\Pi_A^{m n}(0)_{ab} = (\delta^{mn} - \hat{k}^m \hat{k}^n ) \Pi_{A,T}(0)_{ab} + \hat{k}^m \hat{k}^n \, \Pi_{A,D}(0)_{ab}.
\label{eq:Pimnraw}
\end{align}
At non-zero spatial momentum, rotational symmetry permits a term proportional to $\hat{k}^m\hat{k}^n$. We assume that the static self-energy tensor admits a finite, direction-independent limit as $|\mathbf{k}|\to0$. Under this assumption, the coefficient of $\hat{k}^m\hat{k}^n$ must vanish in the limit, which implies
\begin{equation}
\Pi_{A,T}(0)_{ab} = \Pi_{A,D}(0)_{ab}.
\end{equation}
As a result, Eq.~\eqref{eq:Pimnraw} reduces to
\begin{equation}
\Pi_A^{mn}(0)_{ab} = \delta^{mn}\,\Pi_{A,T}(0)_{ab}.
\label{eq:Pimnsimpl}
\end{equation}
Under the same assumption, the mixed temporal--spatial
components must vanish in the static limit. Since $\Pi_A^{m0}(0)_{ab}=-\hat{k}^m\Pi_{A,C}(0)_{ab}$, a direction-independent limit requires
\[
    \Pi_A^{m0}(0)_{ab}=\Pi_A^{0m}(0)_{ab}=0,
\]
and hence
\begin{equation}
    \Pi_{A,C}(0)_{ab}=0.
    \label{eq:PiCzero}
\end{equation}
Finally, this gives a simplified expression for Eq.~\eqref{eq:Pimunugeneral} in the infrared limit,
\begin{equation}
\Pi^{\mu \nu}_A(0)_{ab} = (P_T^{\mu \nu} + P_D^{\mu \nu}) \Pi_{A,T}(0)_{ab}  + P_{L}^{\mu \nu} \,\Pi_{A,L}(0)_{ab},
\end{equation}
where the scalar functions $\Pi_{A,L}(0)_{ab}$ and $\Pi_{A,T}(0)_{ab}$ can be extracted from Eq.~\eqref{eq:Pi00simpl} and Eq.~\eqref{eq:Pimnsimpl},
respectively.

The purely kinematic analysis presented above is consistent with the known structure of thermal corrections in perturbation theory. In particular, at leading order in the high-temperature expansion, only the longitudinal (electric) component receives a non-vanishing static correction of order $T^2$. This is the static hard-thermal-loop (HTL) result, corresponding to the Debye screening mass,
\begin{equation}
\Pi_{A,L}^{\text{HTL}}(0)_{ab} = (m_D^2)_{ab} \sim g^2 T^2 .
\end{equation}
In contrast, the transverse (magnetic) self-energy does not acquire a constant $\mathcal{O}(T^2)$ contribution at this order and vanishes,
\begin{equation}
\Pi_{A,T}^{\text{HTL}}(0)_{ab} = 0
\qquad
\text{(leading order in $T$)}.
\end{equation}
Beyond the leading high-temperature approximation, mass-dependent thermal corrections appear, including
bosonic contributions proportional to $mT$. In this regime, non-vanishing corrections to the transverse self-energy are generated.


\section{Thermal sum-integrals}
\label{sec:appendix_B}
At finite temperature, the one-loop self-energies involve sums over Matsubara frequencies and integrals over spatial momentum. Since only the thermal parts enter the gap equations, we subtract the corresponding zero-temperature contributions,
\begin{equation}
 \mathcal S_{\mathrm{B/F}}[F]\equiv
 T\sum_n\int\frac{\mathrm d^3\mathbf p}{(2\pi)^3}
 F(\omega_{n,\mathrm{B/F}},\mathbf p)
 -\int\frac{\mathrm d^4p}{(2\pi)^4}F(p),
 \label{eq:thermal_sum_measure}
\end{equation}
where $\omega_{n,\mathrm B}=2\pi nT$ and $\omega_{n,\mathrm F}=(2n+1)\pi T$. The vacuum integral uses the same masses and regulator as the sum-integral; dimensional regularisation is understood before subtraction. The expressions below are given after removing the regulator. For Euclidean momentum $p=(p_0,\mathbf p)$, we denote the scalar propagator factors by
\begin{equation}
 \mathcal D_a(p)\equiv\frac1{p^2+a},\qquad
 \mathcal D_b(p)\equiv\frac1{p^2+b},\qquad
 a=m_1^2,\quad b=m_2^2.
 \label{eq:thermal_propagator_factors}
\end{equation}
Their common momentum argument and the temperature dependence of the sum-integrals are suppressed below. All mass derivatives are taken at fixed $T$.

The Matsubara sums are evaluated by contour integration, separating the vacuum term from contributions weighted by the Bose--Einstein or Fermi--Dirac distributions~\cite{Laine:2016hma}. The remaining momentum integrals are expressed through the thermal functions defined in Eq.~\eqref{eq:JB}. The single-mass integrals are
\begin{align}
 I_1^{\mathrm{B/F}}(a)&=\mathcal S_{\mathrm{B/F}}[\mathcal D_a]
 =\frac{T^2}{\pi^2}J_{\mathrm{B/F}}'\!\left(\frac a{T^2}\right),
 \label{eq:I1_as_dJB}\\
 I_2^{\mathrm{B/F}}(a)&=\mathcal S_{\mathrm{B/F}}[p_0^2\mathcal D_a^2]
 =a\,\partial_a I_1^{\mathrm{B/F}}(a)-\frac12 I_1^{\mathrm{B/F}}(a),
 \label{eq:I2_as_dJB}\\
 I_3^{\mathrm{B/F}}(a)&=\mathcal S_{\mathrm{B/F}}[\mathcal D_a^2]
 =-\partial_a I_1^{\mathrm{B/F}}(a),
 \label{eq:I3_as_dJB}\\
 I_4^{\mathrm{B/F}}(a)&=\mathcal S_{\mathrm{B/F}}[p_i^2\mathcal D_a^2]
 =\frac12 I_1^{\mathrm{B/F}}(a).
 \label{eq:I4_as_dJB}
\end{align}
Here and below, no sum over $i$ is implied. Primes on $J_{\mathrm{B/F}}$ denote derivatives with respect to its dimensionless argument. The negative Fermi--Dirac thermal weight is already contained in $J_{\mathrm F}'$; signs from closed fermion loops and multiplicities remain in the diagrammatic prefactors.

Partial fractions and spatial integration by parts give the mixed-mass integrals
\begin{align}
 I_5^{\mathrm{B/F}}(a,b)&=\mathcal S_{\mathrm{B/F}}[p^2\mathcal D_a\mathcal D_b]
 =\frac{aI_1^{\mathrm{B/F}}(a)-bI_1^{\mathrm{B/F}}(b)}{a-b},
 \label{eq:I5_as_dJB}\\
 I_6^{\mathrm{B/F}}(a,b)&=\mathcal S_{\mathrm{B/F}}[p_0^2\mathcal D_a\mathcal D_b]
 =I_5^{\mathrm{B/F}}(a,b)-3I_8^{\mathrm{B/F}}(a,b),
 \label{eq:I6_as_dJB}\\
 I_7^{\mathrm{B/F}}(a,b)&=\mathcal S_{\mathrm{B/F}}[\mathcal D_a\mathcal D_b]
 =\frac{I_1^{\mathrm{B/F}}(b)-I_1^{\mathrm{B/F}}(a)}{a-b},
 \label{eq:I7_as_dJB}\\
 I_8^{\mathrm{B/F}}(a,b)&=\mathcal S_{\mathrm{B/F}}[p_i^2\mathcal D_a\mathcal D_b]
 =\frac{T^4}{2\pi^2}\,
 \frac{J_{\mathrm{B/F}}(a/T^2)-J_{\mathrm{B/F}}(b/T^2)}{a-b}.
 \label{eq:I8_as_dJB}
\end{align}
The gauge projectors introduce two further integrals:
\begin{align}
 \mathcal I_6^{\mathrm{B/F}}(a,b)&=\mathcal S_{\mathrm{B/F}}\!\left[
 \frac{p_0^2}{p^2}\mathcal D_a\mathcal D_b\right]
 =\frac{I_6^{\mathrm{B/F}}(a,0)-I_6^{\mathrm{B/F}}(b,0)}{b-a},
 \label{eq:J6_as_dJB}\\
 \mathcal I_8^{\mathrm{B/F}}(a,b)&=\mathcal S_{\mathrm{B/F}}\!\left[
 \frac{p_i^2}{p^2}\mathcal D_a\mathcal D_b\right]
 =\frac{I_8^{\mathrm{B/F}}(a,0)-I_8^{\mathrm{B/F}}(b,0)}{b-a}.
 \label{eq:projector_integral_reduction}
\end{align}
These obey
\begin{equation}
 \mathcal I_6^{\mathrm{B/F}}(a,b)+3\mathcal I_8^{\mathrm{B/F}}(a,b)=I_7^{\mathrm{B/F}}(a,b).
 \label{eq:projector_integral_identity}
\end{equation}
All equal-mass limits follow from
\begin{equation}
 \lim_{b\to a}\frac{f(a)-f(b)}{a-b}=\partial_a f(a),
 \label{eq:thermal_integral_equal_mass}
\end{equation}
provided $f$ is differentiable at $a$. Higher propagator powers follow by mass differentiation. Equal-mass limits must be distinguished from genuine infrared singularities: for example, $I_3^{\mathrm B}(m^2)\sim T/(8\pi m)$ as $m/T\to0^+$. For negative squared masses, the analytic continuation and real-part prescription of the numerical setup are used.


\section{Neutral mixing}
\label{app:neutral_mixing_replacements}

The self-energies in Sec.~\ref{sec:4} are expressed in the tree-level $(W^\pm,Z,\gamma)$ basis. We collect the replacement rules whenever an internal neutral gauge boson appears in the self-energies. The following rules are written with the Lorentz projectors factored out. In each Lorentz sector $X=T,L,D$, the real symmetric neutral mass matrix is diagonalised by an orthogonal rotation,
\begin{equation}
U_{\hat Z,X}^T M_{\hat Z,X}^2 U_{\hat Z,X}
=
\operatorname{diag}
\left(M_{\hat Z,X,1}^2,M_{\hat Z,X,2}^2\right).
\end{equation}
Here $M_{\hat Z,D}^2=M_{\hat Z,T}^2$ and $U_{\hat Z,D}=U_{\hat Z,T}$, as in Sec.~\ref{subsec:gap_equations}. For each Lorentz sector $X$, the neutral mass eigenmodes $B_k^X$ are defined by $(Z,\gamma)_a=(U_{\hat Z,X})_{ak}B_k^X$. For a tree-level coupling vector $q$, define $q_{X,k}=q_a(U_{\hat Z,X})_{ak}$. The propagator contraction is then
\begin{equation}
q_a\,\mathcal D_X(p)_{ab}\,q_b
=
\sum_{k=1}^2
\frac{q_{X,k}^2}{p^2+M_{\hat Z,X,k}^2}.
\label{eq:neutral_replacement_basic}
\end{equation}

Below, $I$ denotes the relevant bosonic thermal function $I_r^{\mathrm B}$ or $\mathcal I_r^{\mathrm B}$ of Appendix~\ref{sec:appendix_B}, with squared-mass arguments. The replacements act on each internal propagator before any equal-mass reduction. Other internal masses are dressed according to the gap equations.

For vertices selecting only the tree-level $Z$ component, the single-propagator and mixed scalar--$Z$ replacements are given by
\begin{align}
    I(m_{Z,X}^2)
    &\longrightarrow
    \sum_{k=1}^2
    (U_{\hat Z,X})_{Zk}^2 \, I(M_{\hat Z, X, k}^2),
    \label{eq:neutral_replacement_Z_single}
    \\
    I(m_{Z,X}^2,m_S^2)
    &\longrightarrow
    \sum_{k=1}^2
    (U_{\hat Z,X})_{Zk}^2
    I(M_{\hat Z,X,k}^2,M_S^2),
    \label{eq:replacement_Z_scalar}
\end{align}
with $S\in\{h,G^0\}$. For two internal neutral propagators, both are projected:
\begin{equation}
\begin{aligned}
    I(m_{Z,X}^2,m_{Z,Y}^2)
    \longrightarrow{}&
    \sum_{k,k'=1}^2
    (U_{\hat Z, X})_{Zk}^2 (U_{\hat Z, Y})_{Zk'}^2
    I(M_{\hat Z, X, k}^2,M_{\hat Z, Y, k'}^2).
\end{aligned}
\label{eq:replacement_ZZ}
\end{equation}
In particular, terms containing two identical propagators use $I_3^{\mathrm B}(a)=I_7^{\mathrm B}(a,a)$ before this replacement, and therefore retain the double sum over the eigenmodes.

For the charged sectors, the neutral coupling vectors in the $(Z,\gamma)$ basis are
\begin{equation}
    q_G=
    \begin{pmatrix}
        \dfrac{g(c_w^2-s_w^2)}{2c_w}\\
        e
    \end{pmatrix},
    \qquad
    q_W=
    \begin{pmatrix}
        gc_w\\
        e
    \end{pmatrix}.
    \label{eq:neutral_vectors_q}
\end{equation}
The vector $q_G$ describes the charged-Goldstone neutral gauge interactions, while $q_W$ describes the pure gauge interactions involving $W^\pm$. For the $W^\pm G^\mp V$ vertex, $V=Z,\gamma$, extracting the common factor $g^2h$ gives
\begin{equation}
    r_{GW}=
    \begin{pmatrix}
        -\dfrac{s_w^2}{2c_w}\\
        \dfrac{s_w}{2}
    \end{pmatrix}.
    \label{eq:neutral_vector_rGW}
\end{equation}
Its relative sign is retained when projecting onto the dressed eigenmodes.

For any of these vectors, the replacements take the form
\begin{align}
    q_Z^2 I(m_{Z,X}^2)
    +q_\gamma^2 I(m_{\gamma,X}^2)
    &\longrightarrow
    \sum_{k=1}^2 q_{X,k}^2 \, I(M_{\hat Z, X, k}^2),
    \label{eq:neutral_replacement_single}
    \\
    q_Z^2 I(m_{Z,X}^2,\mathcal{M}^2)
    +q_\gamma^2 I(m_{\gamma,X}^2,\mathcal{M}^2)
    &\longrightarrow
    \sum_{k=1}^2 q_{X,k}^2 \, I(M_{\hat Z, X, k}^2,\mathcal{M}^2).
    \label{eq:neutral_replacement_mixed}
\end{align}
Here $\mathcal{M}^2$ denotes the other internal squared mass, dressed according to the gap equations. The quartic $G^+G^-VV$ tadpoles and mixed $G^\pm$--$V$ bubbles in $\Pi_{G^\pm}(0)$, with $V=Z,\gamma$, use $q=q_G$. Only the $D$ sector contributes to the mixed bubble. The charged-$W$ gauge tadpoles and gauge bubbles use $q=q_W$. Contributions containing two $W^\pm G^\mp V$ vertices use $q=r_{GW}$ and the common factor $g^4h^2$ from the vertices. All Lorentz and diagrammatic prefactors multiply these replacements as in Sec.~\ref{sec:gbselfenergies}. For $U_{\hat Z,X}=\mathbf1_2$, the sums recover the separate $Z$ and photon contributions evaluated with their corresponding dressed masses.

The same projection principle applies to internal neutral ghosts. Restricting the similarity transformation in Eq.~\eqref{eq:ghost_mass_diagonalisation_general} to the neutral subspace, the two vertex projections are
\begin{equation}
    q_{\eta,k}^{R}=(q_W)_a(U_\eta)_{ak},
    \qquad
    q_{\eta,k}^{L}=(U_\eta^{-1})_{ka}(q_W)_a.
\end{equation}
Their product replaces the squared projected coupling of the symmetric gauge-boson sector. For the upper-triangular SM ghost matrix in Eq.~\eqref{eq:ghost_neutral_mass_matrix}, with a massless photon ghost, the charged-$W$ ghost-loop replacements become
\begin{align}
    g^2c_w^2 I_r^{\mathrm B}
        (m_{\eta^Z}^2,m_{\eta^\pm}^2)
    +e^2 I_r^{\mathrm B}(0,m_{\eta^\pm}^2)
    & \longrightarrow
    \sum_{k=1}^2 q_{\eta,k}^{R}q_{\eta,k}^{L}
    I_r^{\mathrm B}(M_{\eta,k}^2,M_{\eta^\pm}^2)
    \nonumber\\
    &\quad=
    g^2c_w^2 I_r^{\mathrm B}
        (M_{\eta^Z}^2,M_{\eta^\pm}^2)
    +e^2 I_r^{\mathrm B}(0,M_{\eta^\pm}^2)
        \nonumber\\
    &\qquad\quad\quad\quad\quad\quad
    -gc_we\,\Pi_{\bar\eta^Z\eta^\gamma}(0)\,
    \mathcal I_r^{\mathrm B}
        (M_{\eta^Z}^2,M_{\eta^\pm}^2),
    \label{eq:neutral_ghost_replacement}
\end{align}
with $r=6,8$.

At each gap iteration, these replacements incorporate the updated masses and neutral gauge-boson and ghost mixing through the internal propagators, with the interaction vertices retained at tree level.


\section{Momentum dependence}
\label{app:momentum_dependence}

As discussed in Sec.~\ref{sec:pd_limitations}, PD evaluates the self-energies entering the gap equations at vanishing external momentum. Momentum-dependent insertions are consequently neglected, so diagrams involving overlapping loop momenta are not generically reproduced~\cite{Boyd:1993tz,Bahl:2024bcg}. We test the sensitivity of the reconstructed potential to this approximation by retaining the quadratic momentum dependence of selected scalar self-energies, including thermal and renormalised vacuum contributions. The zero-momentum gap solutions are kept fixed, and the correction is introduced through the scalar propagators entering the tadpole. This isolates the effect of the retained momentum dependence on the reconstructed PD potential.

\begin{figure}[t]
    \centering
    \includegraphics[width=0.71\textwidth]{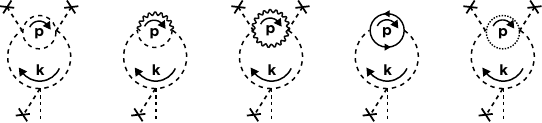}
    \caption{Representative one-loop momentum-dependent self-energy insertions in the electroweak scalar sector. From left to right, the insertions arise from scalar--scalar, mixed scalar--vector, vector--vector, fermion, and ghost loops. The scalar--scalar, vector--vector and ghost bubbles contain two background-dependent cubic vertices and therefore carry an explicit factor of $h^2$. Momentum-independent tadpole and seagull contributions are not shown.}
    \label{fig:scalar_momentum_diagrams}
\end{figure}

The relevant one-loop bubble topologies are illustrated in Fig.~\ref{fig:scalar_momentum_diagrams}. The scalar--scalar, vector--vector and ghost bubbles contain two background-dependent cubic vertices and therefore carry an explicit factor of $h^2$. We focus on the mixed gauge--scalar and fermion bubbles, which have no such vertex suppression and provide a direct probe of momentum dependence in the small-field region. The size of the individual contributions also depends on the infrared behaviour of the loop integrals. Momentum-independent tadpole and seagull diagrams contribute to $\Pi(0)$ but not to the momentum coefficients considered below.

We therefore retain terms through quadratic order in the external momentum and write the approximation formally as
\begin{equation}
    \Pi(k_0,\mathbf{k})
    =
    \Pi(0)
    +\delta Z_0\,k_0^2
    +\delta Z_s\,\mathbf{k}^2
    +\mathcal O(k^4).
    \label{eq:self_energy_derivative_expansion}
\end{equation}
Terms of fourth and higher order in the external momentum are neglected at the level of the loop integrands. For this diagnostic, we include these coefficients through the inverse scalar propagator
\begin{equation}
    \mathcal D^{-1}(k_0,\mathbf{k})
    =
    Z_0 k_0^2
    +Z_s\mathbf{k}^2
    +M^2,
    \label{eq:Z_dressed_propagator}
\end{equation}
with
\begin{align}
    Z_0=1+\delta Z_0,
    \qquad
    Z_s=1+\delta Z_s.
\end{align}
Here $M^2=m^2+\Pi^{\mathrm{th}}(0)$ is obtained from the thermal gap equations. We determine $\delta Z_0$ and $\delta Z_s$ by expanding the Euclidean loop integrands to quadratic order in the external momentum before performing the Matsubara sum, treating the external frequency as continuous during this expansion. In mixed gauge--scalar loops, the scalar carries momentum $p$ and the vector $p-k$. At finite temperature, thermal self-energies can be non-analytic at vanishing external momentum. Taking the frequency and spatial momentum to zero in different orders can then give different results~\cite{Das:1987yb,Arnold:1992qy}. We use the coefficients obtained from the expansion above to test how the retained local momentum terms affect the PD potential.

For each mixed gauge--scalar bubble, let $M_S^2$ denote the internal scalar squared mass and $M_X^2$ the vector squared mass in sector $X=T,L,D$. The factors $C_X$ include the couplings and internal-state multiplicities. The thermal contributions at quadratic order in the external momentum are
\begin{align}
    \delta Z_0^{(T),\mathrm{th}}=0,
    \qquad
    \delta Z_s^{(T),\mathrm{th}}
    =
    -\frac{8C_T}{3}\,
    I_7^{\rm B}(M_S^2,M_T^2),
    \label{eq:ZT}
\end{align}
for the transverse contribution, and
\begin{align}
    \delta Z_0^{(L),\mathrm{th}}
    =
    -12C_L\,\mathcal{I}_8^{\rm B}(M_S^2,M_L^2),
    \qquad
    \delta Z_s^{(L),\mathrm{th}}
    =
    -\frac{4C_L}{3}\,\mathcal{I}_6^{\rm B}(M_S^2,M_L^2),
    \label{eq:ZL}
\end{align}
for the longitudinal contribution. The gauge-direction contribution is
\begin{align}
    \delta Z_0^{(D),\mathrm{th}}
    =-C_D\Big[
        -4 I_7^{\rm B}(M_S^2,M_D^2)
        -M_D^2\,
        &\partial_{M_D^2}I_7^{\rm B}(M_S^2,M_D^2)
        +4\mathcal{I}_6^{\rm B}(M_S^2,M_D^2)
        \nonumber\\[-0.2em]
        &-8\,
        \partial_{M_D^2}I_6^{\rm B}(M_S^2,M_D^2)
        -2M_D^2\,
        \partial_{M_D^2}^2 I_6^{\rm B}(M_S^2,M_D^2)
    \Big],
    \label{eq:ZD0}
\end{align}
and
\begin{equation}
    \delta Z_s^{(D),\mathrm{th}}
    =
    \left.\delta Z_0^{(D),\mathrm{th}}\right|_{
    I_6^{\rm B}\to I_8^{\rm B},\;\mathcal{I}_6^{\rm B}\to \mathcal{I}_8^{\rm B}}.
    \label{eq:ZDs}
\end{equation}
Here the derivatives act only on the second mass-squared argument of the integrals, at fixed couplings and temperature. The replacements are understood to apply also to the mass derivatives of
the corresponding loop functions. The thermal contribution of a given mixed gauge--scalar loop is then
\begin{equation}
    \delta Z_\alpha^{(VS),\mathrm{th}}
    =
    \delta Z_\alpha^{(T),\mathrm{th}}
    +\delta Z_\alpha^{(L),\mathrm{th}}
    +\delta Z_\alpha^{(D),\mathrm{th}},
\end{equation}

with $\alpha=0,s$. The factors $C_X$ for all retained mixed loops are listed below.
\begin{equation}
\renewcommand{\arraystretch}{1.4}
\begin{array}{c|c|c|c}
\text{external scalar} & \text{internal scalar}
& \text{internal vector} & C_X \\ \hline
h & G^\pm & W^\pm & g^2/2 \\
h & G^0 & B_k^X &
\dfrac{g^2}{4c_w^2}(U_{\hat Z,X})_{Zk}^2 \\
G^0 & G^\pm & W^\pm & g^2/2 \\
G^0 & h & B_k^X &
\dfrac{g^2}{4c_w^2}(U_{\hat Z,X})_{Zk}^2 \\
G^\pm & h & W^\pm & g^2/4 \\
G^\pm & G^0 & W^\pm & g^2/4 \\
G^\pm & G^\pm & B_k^X &
\left[ \dfrac{g(c_w^2-s_w^2)}{2c_w}(U_{\hat Z,X})_{Zk}
+e(U_{\hat Z,X})_{\gamma k} \right]^2
\end{array}
\label{eq:momentum_VS_couplings}
\end{equation}
For each row, $M_S^2$ is the dressed mass squared of the listed internal scalar. For charged-vector loops, $M_X^2=M_{W,X}^2$. The charged-vector entries include the charge multiplicities of the internal states. No additional polarisation multiplicity is to be applied to Eqs.~\eqref{eq:ZT}--\eqref{eq:ZDs}. For each neutral entry, the loop functions use $M_X^2=M_{\hat Z,X,k}^2$ and are summed over $k=1,2$, following Appendix~\ref{app:neutral_mixing_replacements}.

For a fermion bubble with internal masses $m_1$ and $m_2$, the thermal temporal and spatial contributions are
\begin{align}
    \delta Z_0^{(F),\mathrm{th}}
    ={}&2N_cC_F\left[
    2\partial_{m_2^2}I_6^{\rm F}(m_1^2,m_2^2)
    -\partial_{m_2^2}I_5^{\rm F}(m_1^2,m_2^2)
    +2m_2^2\partial_{m_2^2}^2I_6^{\rm F}(m_1^2,m_2^2)\right]
    \nonumber\\
    &-2N_cD_Fm_1m_2\left[
    \partial_{m_2^2}I_7^{\rm F}(m_1^2,m_2^2)
    +2\partial_{m_2^2}^2I_6^{\rm F}(m_1^2,m_2^2)\right]
    \label{eq:fermion_Z_0_general}
\end{align}
and
\begin{equation}
    \delta Z_s^{(F),\mathrm{th}}
    =
    \left.\delta Z_0^{(F),\mathrm{th}}\right|_{
    I_6^{\rm F}\to I_8^{\rm F}}.
    \label{eq:fermion_Z_s_general}
\end{equation}
The fermion with mass $m_1$ carries momentum $p$, while the fermion with mass $m_2$ carries momentum $p-k$. The derivatives act only on the integrals, at fixed temperature and vertex factors; equal masses are inserted after differentiation. For the SM with $V_{\rm CKM}=\mathbf{1}$, the coefficients are
\begin{equation}
\renewcommand{\arraystretch}{1.35}
\begin{array}{c|c|c|c|c}
\text{external scalar} & \text{internal quark 1} & \text{internal quark 2} & C_F & D_F \\ \hline
h & m_q & m_q & y_q^2 & -y_q^2 \\
G^0 & m_q & m_q & y_q^2 & y_q^2 \\
G^\pm & m_t & m_b & y_t^2+y_b^2 & 2y_ty_b
\end{array}
\label{eq:momentum_F_couplings}
\end{equation}
We use $N_c=3$ and the tree-level fermion masses $m_q=y_qh/\sqrt{2}$. The neutral entries are summed over $q=t,b$; the charged entry applies to either $G^+$ or $G^-$ without an additional multiplicity factor.

We also include the renormalised vacuum parts of the retained loops. These are obtained from the unsubtracted loop integrands by replacing the Matsubara sum with the continuous frequency integral and subtracting the ultraviolet poles in the $\overline{\mathrm{MS}}$ scheme at scale $\mu_R$. The projector factors are kept in $d=4-2\epsilon$ dimensions until subtraction, including the resulting finite terms. The internal masses and projected couplings are the same as in the thermal contributions; here ``vacuum'' denotes the part without thermal occupation factors.

For each mixed gauge--scalar loop, define
\begin{equation}
    \Delta_X(u)= uM_S^2+(1-u)M_X^2,
    \qquad
    \mathcal L_X=\int_0^1\mathrm{d}u\,
    \ln\frac{\Delta_X(u)}{\mu_R^2},
    \qquad X=T,L,D.
    \label{eq:vacuum_Z_parameters}
\end{equation}
The vacuum contributions are
\begin{align}
    \delta Z_0^{(T),\mathrm{vac}}&=0,
    &\delta Z_s^{(T),\mathrm{vac}}
    &=\frac{C_T}{16\pi^2}
    \left(\frac83\mathcal L_T+\frac89\right),
    \label{eq:vacuum_ZT}\\
    \delta Z_0^{(L),\mathrm{vac}}
    &=\frac{C_L}{16\pi^2}
    \left(3\mathcal L_L+\frac12\right),
    &\delta Z_s^{(L),\mathrm{vac}}
    &=\frac{C_L}{16\pi^2}
    \left(\frac13\mathcal L_L-\frac7{18}\right).
    \label{eq:vacuum_ZL}
\end{align}
For the gauge-direction contribution,
\begin{equation}
    \delta Z_0^{(D),\mathrm{vac}}
    =\delta Z_s^{(D),\mathrm{vac}}
    =-\frac{C_D}{16\pi^2}\int_0^1\mathrm du\,
    \left[
        \ln\frac{\Delta_D(u)}{\mu_R^2}
        +\frac{u M_S^2}{\Delta_D(u)}
        +\frac{u^2M_S^4}{2\Delta_D(u)^2}
    \right].
    \label{eq:vacuum_ZD}
\end{equation}
The factors $C_X$ are given in Eq.~\eqref{eq:momentum_VS_couplings}, and the three sectors are summed for each mixed loop. Again, in total we have
\begin{equation}
    \delta Z_\alpha^{(VS),\mathrm{vac}}
    =
    \delta Z_\alpha^{(T),\mathrm{vac}}
    +\delta Z_\alpha^{(L),\mathrm{vac}}
    +\delta Z_\alpha^{(D),\mathrm{vac}}.
\end{equation}

For the fermion bubbles, with
$\Delta_F(u)=um_1^2+(1-u)m_2^2$, the common result is
\begin{equation}
\begin{aligned}
    \delta Z_0^{(F),\mathrm{vac}}
    &=\delta Z_s^{(F),\mathrm{vac}}=\frac{N_c}{8\pi^2}\int_0^1\mathrm du\,u(1-u)
    \left[
        -3C_F\ln\frac{\Delta_F(u)}{\mu_R^2}
        -C_F+\frac{D_Fm_1m_2}{\Delta_F(u)}
    \right],
\end{aligned}
    \label{eq:vacuum_ZF}
\end{equation}
using the masses and coefficients in Eq.~\eqref{eq:momentum_F_couplings}.

Summing the mixed gauge--scalar and fermion loops specified above,
we obtain
\begin{equation}
    \delta Z_{\alpha,i}
    =\delta Z_{\alpha,i}^{\mathrm{th}}
    +\delta Z_{\alpha,i}^{\mathrm{vac}},
    \qquad \alpha=0,s,
    \label{eq:scalar_Z_total}
\end{equation}
for each scalar mode $i=h,G^0,G^\pm$.
To estimate the impact of the momentum dependence on partial dressing, we implement the correction at the level of the scalar tadpole, following the conventional PD prescription discussed in Sec.~\ref{sec:3}. We first determine the masses from the zero-momentum gap equations of Sec.~\ref{subsec:gap_equations}, which continue to resum only thermal self-energies. We then set $Z_{\alpha,i}=1+\delta Z_{\alpha,i}$, using the coefficients in Eq.~\eqref{eq:scalar_Z_total}, evaluated with the dressed bosonic masses and tree-level fermion masses. These factors are not fed back into the gap equations. For each scalar mode $i=h,G^0,G^\pm$, we schematically implement
\begin{equation}
T\sum_n\int\frac{\mathrm d^3p}{(2\pi)^3}
\frac{1}{\omega_n^2+\mathbf p^2+M_i^2}
\quad\longrightarrow\quad
T\sum_n\int\frac{\mathrm d^3p}{(2\pi)^3}
\frac{1}{Z_{0,i}\omega_n^2+Z_{s,i}\mathbf p^2+M_i^2}.
\label{eq:Z_tadpole_replacement}
\end{equation}
The coefficients are determined at vanishing external momentum; the resulting local propagator is used throughout the momentum integral and Matsubara sum to test its effect on the reconstructed potential. This replacement is applied to both the thermal and Coleman--Weinberg scalar tadpoles, keeping the tree-level vertices and multiplicities of the PD prescription. The two charged Goldstone modes share identical $Z$ factors. The corrected potential is then obtained by integrating the resulting tadpole over the background field. We denote the result including these momentum-dependent corrections by $\mathrm{PD}_{\rm(sc+g)}+\mathrm{MO}$.

\begin{figure}[t!]
    \centering
    \includegraphics[]{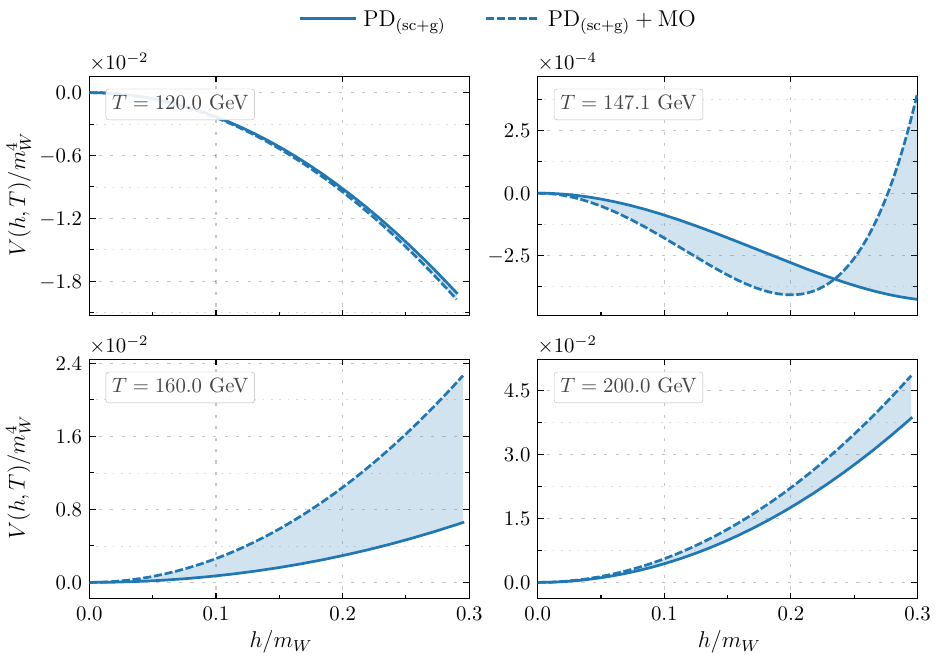}
    \caption{Comparison of the $\mathrm{PD}_{\rm(sc+g)}$ effective potential (blue solid) with the result including the retained quadratic momentum dependence of the scalar self-energies ($\mathrm{PD}_{\rm(sc+g)}+\mathrm{MO}$, blue dashed), at $T=120.0,147.1,160.0$ and $200.0\,\mathrm{GeV}$. The scalar $Z$ factors include thermal and renormalised vacuum contributions and are applied to both the thermal and Coleman--Weinberg scalar tadpoles, with the zero-momentum gap solutions kept fixed. Explicit calculations show that the corrections do not produce a potential barrier at the transition temperature, preserving the continuous character of the electroweak transition found in zero-momentum $\rm{PD}_{\rm{(sc+g)}}$.}
    \label{fig:MO_Test}
\end{figure}

Fig.~\ref{fig:MO_Test} compares the resulting potential with the zero-momentum PD result at four benchmark temperatures. At $T=120.0\,\mathrm{GeV}$, the profiles remain close. Near the transition, at $T=147.1\,\mathrm{GeV}$, the retained local momentum terms appreciably change the small-field profile. At $T=160.0\,\mathrm{GeV}$ and $200.0\,\mathrm{GeV}$, the origin remains the minimum within the displayed interval, although the magnitude and curvature of the potential change. A temperature scan confirms that the broken minimum of the examined resummed branch still approaches the origin continuously, with only a small shift in the transition temperature.

\newpage
\bibliographystyle{osajnl}  
\bibliography{lib}  

@article{Kajantie:1996mn,
    author = "Kajantie, K. and Laine, M. and Rummukainen, K. and Shaposhnikov, Mikhail E.",
    title = "{Is there a~ hot electroweak phase transition at $m_H \gtrsim m_W$?}",
    eprint = "hep-ph/9605288",
    archivePrefix = "arXiv",
    reportNumber = "CERN-TH-96-126, HD-THEP-96-15, IUHET-333",
    doi = "10.1103/PhysRevLett.77.2887",
    journal = "Phys. Rev. Lett.",
    volume = "77",
    pages = "2887--2890",
    year = "1996"
}

@article{Kajantie:1996qd,
    author = "Kajantie, K. and Laine, M. and Rummukainen, K. and Shaposhnikov, Mikhail E.",
    title = "{A Nonperturbative analysis of the finite T phase transition in SU(2) x U(1) electroweak theory}",
    eprint = "hep-lat/9612006",
    archivePrefix = "arXiv",
    reportNumber = "BI-TP-96-54, CERN-TH-96-334A, HD-THEP-96-48",
    doi = "10.1016/S0550-3213(97)00164-8",
    journal = "Nucl. Phys. B",
    volume = "493",
    pages = "413--438",
    year = "1997"
}

@article{DOnofrio:2014rug,
    author        = "D'Onofrio, Michela and Rummukainen, Kari and Tranberg, Anders",
    title         = "{Sphaleron Rate in the Minimal Standard Model}",
    eprint        = "1404.3565",
    archivePrefix = "arXiv",
    primaryClass  = "hep-ph",
    doi           = "10.1103/PhysRevLett.113.141602",
    journal       = "Phys. Rev. Lett.",
    volume        = "113",
    pages         = "141602",
    year          = "2014"
}

@article{DOnofrio:2015gop,
    author = "D'Onofrio, Michela and Rummukainen, Kari",
    title = "{Standard model cross-over on the lattice}",
    eprint = "1508.07161",
    archivePrefix = "arXiv",
    primaryClass = "hep-ph",
    reportNumber = "HIP-2015-30-TH",
    doi = "10.1103/PhysRevD.93.025003",
    journal = "Phys. Rev. D",
    volume = "93",
    number = "2",
    pages = "025003",
    year = "2016"
}

@article{Anderson:1991zb,
    author = "Anderson, Greg W. and Hall, Lawrence J.",
    title = "{The Electroweak phase transition and baryogenesis}",
    reportNumber = "LBL-31169, UCB-PTH-91-41",
    doi = "10.1103/PhysRevD.45.2685",
    journal = "Phys. Rev. D",
    volume = "45",
    pages = "2685--2698",
    year = "1992"
}

@article{Espinosa:1993bs,
    author = "Espinosa, J. R. and Quiros, M.",
    title = "{The Electroweak phase transition with a singlet}",
    eprint = "hep-ph/9301285",
    archivePrefix = "arXiv",
    reportNumber = "IEM-FT-67-93",
    doi = "10.1016/0370-2693(93)91111-Y",
    journal = "Phys. Lett. B",
    volume = "305",
    pages = "98--105",
    year = "1993"
}

@article{Profumo:2007wc,
    author        = "Profumo, Stefano and Ramsey-Musolf, Michael J. and Shaughnessy, Gabe",
    title         = "{Singlet Higgs Phenomenology and the Electroweak Phase Transition}",
    eprint        = "0705.2425",
    archivePrefix = "arXiv",
    primaryClass  = "hep-ph",
    doi           = "10.1088/1126-6708/2007/08/010",
    journal       = "JHEP",
    volume        = "08",
    pages         = "010",
    year          = "2007"
}

@article{Espinosa:2011ax,
    author        = "Espinosa, Jose R. and Konstandin, Thomas and Riva, Francesco",
    title         = "{Strong Electroweak Phase Transitions in the Standard Model with a Singlet}",
    eprint        = "1107.5441",
    archivePrefix = "arXiv",
    primaryClass  = "hep-ph",
    doi           = "10.1016/j.nuclphysb.2011.09.010",
    journal       = "Nucl. Phys. B",
    volume        = "854",
    pages         = "592--630",
    year          = "2012"
}

@article{Curtin:2014jma,
    author = "Curtin, David and Meade, Patrick and Yu, Chiu-Tien",
    title = "{Testing Electroweak Baryogenesis with Future Colliders}",
    eprint = "1409.0005",
    archivePrefix = "arXiv",
    primaryClass = "hep-ph",
    reportNumber = "YITP-SB-14-33",
    doi = "10.1007/JHEP11(2014)127",
    journal = "JHEP",
    volume = "11",
    pages = "127",
    year = "2014"
}

@article{Turok:1990zg,
    author = "Turok, Neil and Zadrozny, John",
    title = "{Electroweak baryogenesis in the two doublet model}",
    reportNumber = "PUPT-90-1216",
    doi = "10.1016/0550-3213(91)90356-3",
    journal = "Nucl. Phys. B",
    volume = "358",
    pages = "471--493",
    year = "1991"
}

@article{Cline:1996mga,
    author        = "Cline, James M. and Lemieux, Pierre-Anthony",
    title         = "{Electroweak Phase Transition in Two Higgs Doublet Models}",
    eprint        = "hep-ph/9609240",
    archivePrefix = "arXiv",
    doi           = "10.1103/PhysRevD.55.3873",
    journal       = "Phys. Rev. D",
    volume        = "55",
    pages         = "3873--3881",
    year          = "1997"
}

@article{Fromme:2006cm,
    author = "Fromme, Lars and Huber, Stephan J. and Seniuch, Michael",
    title = "{Baryogenesis in the two-Higgs doublet model}",
    eprint = "hep-ph/0605242",
    archivePrefix = "arXiv",
    reportNumber = "CERN-PH-TH-2006-094, BI-TP-2006-18",
    doi = "10.1088/1126-6708/2006/11/038",
    journal = "JHEP",
    volume = "11",
    pages = "038",
    year = "2006"
}

@article{Cline:2011mm,
    author = "Cline, James M. and Kainulainen, Kimmo and Trott, Michael",
    title = "{Electroweak Baryogenesis in Two Higgs Doublet Models and B meson anomalies}",
    eprint = "1107.3559",
    archivePrefix = "arXiv",
    primaryClass = "hep-ph",
    doi = "10.1007/JHEP11(2011)089",
    journal = "JHEP",
    volume = "11",
    pages = "089",
    year = "2011"
}

@article{Dorsch:2013wja,
    author        = "Dorsch, G. C. and Huber, S. J. and No, J. M.",
    title         = "{A strong electroweak phase transition in the 2HDM after LHC8}",
    eprint        = "1305.6610",
    archivePrefix = "arXiv",
    primaryClass  = "hep-ph",
    doi           = "10.1007/JHEP10(2013)029",
    journal       = "JHEP",
    volume        = "10",
    pages         = "029",
    year          = "2013"
}

@article{Laine:2017hdk,
    author = "Laine, M. and Meyer, M. and Nardini, G.",
    title = "{Thermal phase transition with full 2-loop effective potential}",
    eprint = "1702.07479",
    archivePrefix = "arXiv",
    primaryClass = "hep-ph",
    doi = "10.1016/j.nuclphysb.2017.04.023",
    journal = "Nucl. Phys. B",
    volume = "920",
    pages = "565--600",
    year = "2017"
}

@article{Patel:2012pi,
    author        = "Patel, Hiren H. and Ramsey-Musolf, Michael J.",
    title         = "{Stepping Into Electroweak Symmetry Breaking: Phase Transitions and Higgs Phenomenology}",
    eprint        = "1212.5652",
    archivePrefix = "arXiv",
    primaryClass  = "hep-ph",
    doi           = "10.1103/PhysRevD.88.035013",
    journal       = "Phys. Rev. D",
    volume        = "88",
    pages         = "035013",
    year          = "2013"
}

@article{Chiang:2014hia,
  author        = {Chiang, Cheng-Wei and Yamada, Toshifumi},
  title         = {Electroweak phase transition in
                   {Georgi--Machacek} model},
  journal       = {Phys. Lett. B},
  volume        = {735},
  pages         = {295--300},
  year          = {2014},
  doi           = {10.1016/j.physletb.2014.06.048},
  eprint        = {1404.5182},
  archivePrefix = {arXiv},
  primaryClass  = {hep-ph}
}

@article{Niemi:2018asa,
  author        = {Niemi, Lauri and Patel, Hiren H.
                   and Ramsey-Musolf, Michael J.
                   and Tenkanen, Tuomas V. I. and Weir, David J.},
  title         = {Electroweak phase transition in the real triplet
                   extension of the {SM}: Dimensional reduction},
  journal       = {Phys. Rev. D},
  volume        = {100},
  number        = {3},
  pages         = {035002},
  year          = {2019},
  doi           = {10.1103/PhysRevD.100.035002},
  eprint        = {1802.10500},
  archivePrefix = {arXiv},
  primaryClass  = {hep-ph}
}

@article{Niemi:2020hto,
    author = "Niemi, Lauri and Ramsey-Musolf, Michael J. and Tenkanen, Tuomas V. I. and Weir, David J.",
    title = "{Thermodynamics of a Two-Step Electroweak Phase Transition}",
    eprint = "2005.11332",
    archivePrefix = "arXiv",
    primaryClass = "hep-ph",
    reportNumber = "HIP-2020-11/TH, ACFI-T20-05",
    doi = "10.1103/PhysRevLett.126.171802",
    journal = "Phys. Rev. Lett.",
    volume = "126",
    number = "17",
    pages = "171802",
    year = "2021"
}

@article{Chung:2012vg,
    author = "Chung, Daniel J. H. and Long, Andrew J. and Wang, Lian-Tao",
    title = "{125 GeV Higgs boson and electroweak phase transition model classes}",
    eprint = "1209.1819",
    archivePrefix = "arXiv",
    primaryClass = "hep-ph",
    doi = "10.1103/PhysRevD.87.023509",
    journal = "Phys. Rev. D",
    volume = "87",
    number = "2",
    pages = "023509",
    year = "2013"
}

@article{Ramsey-Musolf:2019lsf,
    author = "Ramsey-Musolf, Michael J.",
    title = "{The electroweak phase transition: a collider target}",
    eprint = "1912.07189",
    archivePrefix = "arXiv",
    primaryClass = "hep-ph",
    reportNumber = "ACFI-T19-14",
    doi = "10.1007/JHEP09(2020)179",
    journal = "JHEP",
    volume = "09",
    pages = "179",
    year = "2020"
}

@article{Kuzmin:1985mm,
    author        = "Kuzmin, V. A. and Rubakov, V. A. and Shaposhnikov, M. E.",
    title         = "{On the Anomalous Electroweak Baryon Number Nonconservation in the Early Universe}",
    doi           = "10.1016/0370-2693(85)91028-7",
    journal       = "Phys. Lett. B",
    volume        = "155",
    pages         = "36--42",
    year          = "1985"
}

@article{Cohen:1993nk,
    author        = "Cohen, Andrew G. and Kaplan, David B. and Nelson, Ann E.",
    title         = "{Progress in Electroweak Baryogenesis}",
    eprint        = "hep-ph/9302210",
    archivePrefix = "arXiv",
    doi           = "10.1146/annurev.ns.43.120193.000331",
    journal       = "Ann. Rev. Nucl. Part. Sci.",
    volume        = "43",
    pages         = "27--70",
    year          = "1993"
}

@article{Morrissey:2012db,
    author = "Morrissey, David E. and Ramsey-Musolf, Michael J.",
    title = "{Electroweak baryogenesis}",
    eprint = "1206.2942",
    archivePrefix = "arXiv",
    primaryClass = "hep-ph",
    reportNumber = "NPAC-12-08",
    doi = "10.1088/1367-2630/14/12/125003",
    journal = "New J. Phys.",
    volume = "14",
    pages = "125003",
    year = "2012"
}

@article{Sakharov:1967dj,
    author = "Sakharov, A. D.",
    title = "{Violation of CP Invariance, C asymmetry, and baryon asymmetry of the universe}",
    doi = "10.1070/PU1991v034n05ABEH002497",
    journal = "Pisma Zh. Eksp. Teor. Fiz.",
    volume = "5",
    pages = "32--35",
    year = "1967"
}

@article{Witten:1984rs,
    author        = "Witten, Edward",
    title         = "{Cosmic Separation of Phases}",
    doi           = "10.1103/PhysRevD.30.272",
    journal       = "Phys. Rev. D",
    volume        = "30",
    pages         = "272--285",
    year          = "1984"
}

@article{Kosowsky:1991ua,
    author        = "Kosowsky, Arthur and Turner, Michael S. and Watkins, Richard",
    title         = "{Gravitational Radiation from Colliding Vacuum Bubbles}",
    doi           = "10.1103/PhysRevD.45.4514",
    journal       = "Phys. Rev. D",
    volume        = "45",
    pages         = "4514--4535",
    year          = "1992"
}

@article{Kamionkowski:1993fg,
    author        = "Kamionkowski, Marc and Kosowsky, Arthur and Turner, Michael S.",
    title         = "{Gravitational Radiation from First-Order Phase Transitions}",
    eprint        = "astro-ph/9310044",
    archivePrefix = "arXiv",
    doi           = "10.1103/PhysRevD.49.2837",
    journal       = "Phys. Rev. D",
    volume        = "49",
    pages         = "2837--2851",
    year          = "1994"
}

@article{Hindmarsh:2013xza,
    author = "Hindmarsh, Mark and Huber, Stephan J. and Rummukainen, Kari and Weir, David J.",
    title = "{Gravitational waves from the sound of a first order phase transition}",
    eprint = "1304.2433",
    archivePrefix = "arXiv",
    primaryClass = "hep-ph",
    reportNumber = "HIP-2013-07-TH",
    doi = "10.1103/PhysRevLett.112.041301",
    journal = "Phys. Rev. Lett.",
    volume = "112",
    pages = "041301",
    year = "2014"
}

@article{Hindmarsh:2015qta,
    author = "Hindmarsh, Mark and Huber, Stephan J. and Rummukainen, Kari and Weir, David J.",
    title = "{Numerical simulations of acoustically generated gravitational waves at a first order phase transition}",
    eprint = "1504.03291",
    archivePrefix = "arXiv",
    primaryClass = "astro-ph.CO",
    reportNumber = "HIP-2015-13-TH",
    doi = "10.1103/PhysRevD.92.123009",
    journal = "Phys. Rev. D",
    volume = "92",
    number = "12",
    pages = "123009",
    year = "2015"
}

@article{Caprini:2015zlo,
    author = "Caprini, Chiara and others",
    title = "{Science with the space-based interferometer eLISA. II: Gravitational waves from cosmological phase transitions}",
    eprint = "1512.06239",
    archivePrefix = "arXiv",
    primaryClass = "astro-ph.CO",
    reportNumber = "DESY-15-246",
    doi = "10.1088/1475-7516/2016/04/001",
    journal = "JCAP",
    volume = "04",
    pages = "001",
    year = "2016"
}

@article{Caprini:2019egz,
    author = "Caprini, Chiara and others",
    title = "{Detecting gravitational waves from cosmological phase transitions with LISA: an update}",
    eprint = "1910.13125",
    archivePrefix = "arXiv",
    primaryClass = "astro-ph.CO",
    reportNumber = "DESY-19-159, IPPP/19/27, HIP-2019-14/TH, MITP/19-066, IFT-UAM/CSIC-19-139",
    doi = "10.1088/1475-7516/2020/03/024",
    journal = "JCAP",
    volume = "03",
    pages = "024",
    year = "2020"
}

@article{Dolan:1973qd,
    author        = "Dolan, L. and Jackiw, R.",
    title         = "{Symmetry Behavior at Finite Temperature}",
    doi           = "10.1103/PhysRevD.9.3320",
    journal       = "Phys. Rev. D",
    volume        = "9",
    pages         = "3320--3341",
    year          = "1974"
}

@article{Weinberg:1974hy,
    author        = "Weinberg, Steven",
    title         = "{Gauge and Global Symmetries at High Temperature}",
    doi           = "10.1103/PhysRevD.9.3357",
    journal       = "Phys. Rev. D",
    volume        = "9",
    pages         = "3357--3378",
    year          = "1974"
}

@article{Carrington:1991hz,
    author        = "Carrington, M. E.",
    title         = "{The Effective Potential at Finite Temperature in the Standard Model}",
    doi           = "10.1103/PhysRevD.45.2933",
    journal       = "Phys. Rev. D",
    volume        = "45",
    pages         = "2933--2944",
    year          = "1992"
}

@article{Espinosa:1992gq,
    author        = "Espinosa, J. R. and Quiros, M. and Zwirner, F.",
    title         = "{On the Phase Transition in the Scalar Theory}",
    eprint        = "hep-ph/9206227",
    archivePrefix = "arXiv",
    doi           = "10.1016/0370-2693(92)90129-R",
    journal       = "Phys. Lett. B",
    volume        = "291",
    pages         = "115--124",
    year          = "1992"
}

@article{Fendley:1987ef,
    author        = "Fendley, Paul",
    title         = "{The Effective Potential and the Coupling Constant at High Temperature}",
    doi           = "10.1016/0370-2693(87)90599-5",
    journal       = "Phys. Lett. B",
    volume        = "196",
    pages         = "175--180",
    year          = "1987"
}

@article{Parwani:1991gq,
    author        = "Parwani, Rajesh R.",
    title         = "{Resummation in a Hot Scalar Field Theory}",
    eprint        = "hep-ph/9204216",
    archivePrefix = "arXiv",
    doi           = "10.1103/PhysRevD.45.4695",
    journal       = "Phys. Rev. D",
    volume        = "45",
    pages         = "4695--4705",
    year          = "1992",
    note          = "[Erratum: Phys. Rev. D 48, 5965 (1993)]"
}

@article{Arnold:1992rz,
    author        = "Arnold, Peter Brockway and Espinosa, Olivier",
    title         = "{The Effective Potential and First-Order Phase Transitions: Beyond Leading Order}",
    eprint        = "hep-ph/9212235",
    archivePrefix = "arXiv",
    doi           = "10.1103/PhysRevD.47.3546",
    journal       = "Phys. Rev. D",
    volume        = "47",
    pages         = "3546--3579",
    year          = "1993",
    note          = "[Erratum: Phys. Rev. D 50, 6662 (1994)]"
}

@article{Boyd:1992xn,
    author = "Boyd, C. Glenn and Brahm, David E. and Hsu, Stephen D. H.",
    title = "{Corrections to the electroweak effective action at finite temperature}",
    eprint = "hep-ph/9206235",
    archivePrefix = "arXiv",
    reportNumber = "EFI-92-22, CALT-68-1795, HUTP-92-A027",
    doi = "10.1103/PhysRevD.48.4952",
    journal = "Phys. Rev. D",
    volume = "48",
    pages = "4952--4962",
    year = "1993"
}

@article{Espinosa:1992kf,
    author = "Espinosa, J. R. and Quiros, M. and Zwirner, F.",
    title = "{On the nature of the electroweak phase transition}",
    eprint = "hep-ph/9212248",
    archivePrefix = "arXiv",
    reportNumber = "CERN-TH-6577-92, IEM-FT-58-92",
    doi = "10.1016/0370-2693(93)90450-V",
    journal = "Phys. Lett. B",
    volume = "314",
    pages = "206--216",
    year = "1993"
}

@article{Boyd:1993tz,
    author = "Boyd, C. Glenn and Brahm, David E. and Hsu, Stephen D. H.",
    title = "{Resummation methods at finite temperature: The Tadpole way}",
    eprint = "hep-ph/9304254",
    archivePrefix = "arXiv",
    reportNumber = "CALT-68-1858, HUTP-93-A011, EFI-93-22",
    doi = "10.1103/PhysRevD.48.4963",
    journal = "Phys. Rev. D",
    volume = "48",
    pages = "4963--4973",
    year = "1993"
}

@misc{Berthiaume:2026dkt,
    author = {Berthiaume, Rapha{\"e}l and Curtin, David and Luke, Michael and Rasovic, Andrija and Roy, Jyotirmoy},
    title = "{Hybrid Partial Dressing: Correct Effective Potentials at All Temperatures}",
    eprint = "2609.30366",
    archivePrefix = "arXiv",
    primaryClass = "hep-ph",
    month = "9",
    year = "2026"
}

@article{Dine:1992wr,
    author        = "Dine, Michael and Leigh, Robert G. and Huet, Patrick
                     and Linde, Andrei D. and Linde, Dmitri A.",
    title         = "{Towards the Theory of the Electroweak Phase Transition}",
    eprint        = "hep-ph/9203203",
    archivePrefix = "arXiv",
    primaryClass  = "hep-ph",
    doi           = "10.1103/PhysRevD.46.550",
    journal       = "Phys. Rev. D",
    volume        = "46",
    pages         = "550--571",
    year          = "1992"
}

@article{Curtin:2016urg,
    author        = "Curtin, David and Meade, Patrick and Ramani, Harikrishnan",
    title         = "{Thermal Resummation and Phase Transitions}",
    eprint        = "1612.00466",
    archivePrefix = "arXiv",
    primaryClass  = "hep-ph",
    doi           = "10.1140/epjc/s10052-018-6268-0",
    journal       = "Eur. Phys. J. C",
    volume        = "78",
    number        = "9",
    pages         = "787",
    year          = "2018"
}

@article{Lofgren:2023sep,
    author = {L{\"o}fgren, Johan},
    title = "{Stop comparing resummation methods}",
    eprint = "2301.05197",
    archivePrefix = "arXiv",
    primaryClass = "hep-ph",
    doi = "10.1088/1361-6471/ad074b",
    journal = "J. Phys. G",
    volume = "50",
    number = "12",
    pages = "125008",
    year = "2023"
}

@article{Curtin:2022ovx,
    author = "Curtin, David and Roy, Jyotirmoy and White, Graham",
    title = "{Gravitational waves and tadpole resummation: Efficient and easy convergence of finite temperature QFT}",
    eprint = "2211.08218",
    archivePrefix = "arXiv",
    primaryClass = "hep-ph",
    doi = "10.1103/PhysRevD.109.116001",
    journal = "Phys. Rev. D",
    volume = "109",
    number = "11",
    pages = "116001",
    year = "2024"
}

@article{Bahl:2024bcg,
    author        = "Bahl, Henning and Carena, Marcela and Ireland, Aurora
                     and Wagner, Carlos E. M.",
    title         = "{Improved Thermal Resummation for Multi-Field Potentials}",
    eprint        = "2404.12439",
    archivePrefix = "arXiv",
    primaryClass  = "hep-ph",
    doi           = "10.1007/JHEP09(2024)153",
    journal       = "JHEP",
    volume        = "09",
    pages         = "153",
    year          = "2024"
}

@article{Bittar:2025lcr,
    author = "Bittar, Pedro and Roy, Subhojit and Wagner, Carlos E. M.",
    title = "{Self consistent thermal resummation: a case study of the phase transition in 2HDM}",
    eprint = "2504.02024",
    archivePrefix = "arXiv",
    primaryClass = "hep-ph",
    reportNumber = "EFI-25-3",
    doi = "10.1007/JHEP12(2025)021",
    journal = "JHEP",
    volume = "12",
    pages = "021",
    year = "2025"
}

@misc{Zhao:2026tob,
    author = "Zhao, Yu-Bo and Li, Ji-Wei and Wang, Zhi-Wei",
    title = "{Thermal Resummation Effects on Electroweak Phase Transitions and Gravitational Waves in Conformal and Non-Conformal 2HDMs}",
    eprint = "2611.26417",
    archivePrefix = "arXiv",
    primaryClass = "hep-ph",
    month = "11",
    year = "2026"
}

@article{Kajantie:1995dw,
    author = "Kajantie, K. and Laine, M. and Rummukainen, K. and Shaposhnikov, Mikhail E.",
    title = "{Generic rules for high temperature dimensional reduction and their application to the standard model}",
    eprint = "hep-ph/9508379",
    archivePrefix = "arXiv",
    reportNumber = "CERN-TH-95-226, HU-TFT-95-50, IUHET-312",
    doi = "10.1016/0550-3213(95)00549-8",
    journal = "Nucl. Phys. B",
    volume = "458",
    pages = "90--136",
    year = "1996"
}

@article{Laine:2015kra,
    author = "Laine, M. and Meyer, M.",
    title = "{Standard Model thermodynamics across the electroweak crossover}",
    eprint = "1503.04935",
    archivePrefix = "arXiv",
    primaryClass = "hep-ph",
    doi = "10.1088/1475-7516/2015/07/035",
    journal = "JCAP",
    volume = "07",
    pages = "035",
    year = "2015"
}

@article{Saikawa:2018rcs,
    author = "Saikawa, Ken'ichi and Shirai, Satoshi",
    title = "{Primordial gravitational waves, precisely: The role of thermodynamics in the Standard Model}",
    eprint = "1803.01038",
    archivePrefix = "arXiv",
    primaryClass = "hep-ph",
    doi = "10.1088/1475-7516/2018/05/035",
    journal = "JCAP",
    volume = "05",
    pages = "035",
    year = "2018"
}

@article{Braaten:1995cm,
    author = "Braaten, Eric and Nieto, Agustin",
    title = "{Effective field theory approach to high temperature thermodynamics}",
    eprint = "hep-ph/9501375",
    archivePrefix = "arXiv",
    reportNumber = "NUHEP-TH-95-2",
    doi = "10.1103/PhysRevD.51.6990",
    journal = "Phys. Rev. D",
    volume = "51",
    pages = "6990--7006",
    year = "1995"
}

@article{Farakos:1994kx,
    author = "Farakos, K. and Kajantie, K. and Rummukainen, K. and Shaposhnikov, Mikhail E.",
    title = "{3-D physics and the electroweak phase transition: Perturbation theory}",
    eprint = "hep-ph/9404201",
    archivePrefix = "arXiv",
    reportNumber = "CERN-TH-6973-94, IUHET-273",
    doi = "10.1016/0550-3213(94)90173-2",
    journal = "Nucl. Phys. B",
    volume = "425",
    pages = "67--109",
    year = "1994"
}

@article{Croon:2020cgk,
    author = "Croon, Djuna and Gould, Oliver and Schicho, Philipp and Tenkanen, Tuomas V. I. and White, Graham",
    title = "{Theoretical uncertainties for cosmological first-order phase transitions}",
    eprint = "2009.10080",
    archivePrefix = "arXiv",
    primaryClass = "hep-ph",
    reportNumber = "HIP-2020-26/TH",
    doi = "10.1007/JHEP04(2021)055",
    journal = "JHEP",
    volume = "04",
    pages = "055",
    year = "2021"
}

@article{Ekstedt:2022bff,
    author = "Ekstedt, Andreas and Schicho, Philipp and Tenkanen, Tuomas V. I.",
    title = "{DRalgo: A package for effective field theory approach for thermal phase transitions}",
    eprint = "2205.08815",
    archivePrefix = "arXiv",
    primaryClass = "hep-ph",
    reportNumber = "HIP-2022-11/TH, NORDITA 2022-030",
    doi = "10.1016/j.cpc.2023.108725",
    journal = "Comput. Phys. Commun.",
    volume = "288",
    pages = "108725",
    year = "2023"
}

@article{Ekstedt:2024etx,
    author = "Ekstedt, Andreas and Schicho, Philipp and Tenkanen, Tuomas V. I.",
    title = "{Cosmological phase transitions at three loops: The final verdict on perturbation theory}",
    eprint = "2405.18349",
    archivePrefix = "arXiv",
    primaryClass = "hep-ph",
    reportNumber = "HIP-2024-15/TH",
    doi = "10.1103/PhysRevD.110.096006",
    journal = "Phys. Rev. D",
    volume = "110",
    number = "9",
    pages = "096006",
    year = "2024"
}

@article{Chala:2024xll,
    author = "Chala, Mikael and Criado, Juan Carlos and Gil, Luis and Miras, Javier L{\'o}pez",
    title = "{Higher-order-operator corrections to phase-transition parameters in dimensional reduction}",
    eprint = "2406.02667",
    archivePrefix = "arXiv",
    primaryClass = "hep-ph",
    doi = "10.1007/JHEP10(2024)025",
    journal = "JHEP",
    volume = "10",
    pages = "025",
    year = "2024"
}

@article{Gould:2021oba,
    author = "Gould, Oliver and Tenkanen, Tuomas V. I.",
    title = "{On the perturbative expansion at high temperature and implications for cosmological phase transitions}",
    eprint = "2104.04399",
    archivePrefix = "arXiv",
    primaryClass = "hep-ph",
    reportNumber = "NORDITA 2021-010",
    doi = "10.1007/JHEP06(2021)069",
    journal = "JHEP",
    volume = "06",
    pages = "069",
    year = "2021"
}

@article{Schicho:2022wty,
    author = "Schicho, Philipp and Tenkanen, Tuomas V. I. and White, Graham",
    title = "{Combining thermal resummation and gauge invariance for electroweak phase transition}",
    eprint = "2203.04284",
    archivePrefix = "arXiv",
    primaryClass = "hep-ph",
    reportNumber = "HIP-2022-2/TH, NORDITA 2022-009",
    doi = "10.1007/JHEP11(2022)047",
    journal = "JHEP",
    volume = "11",
    pages = "047",
    year = "2022"
}

@article{Chala:2025oul,
    author = "Chala, Mikael and Gil, Luis and Ren, Zhe",
    title = "{Phase transitions in dimensional reduction up to three loops}",
    eprint = "2505.14335",
    archivePrefix = "arXiv",
    primaryClass = "hep-ph",
    doi = "10.1088/1674-1137/adf322",
    journal = "Chin. Phys. C",
    volume = "49",
    number = "12",
    pages = "123105",
    year = "2025"
}

@article{Bernardo:2025vkz,
    author = "Bernardo, Fabio and Klose, Philipp and Schicho, Philipp and Tenkanen, Tuomas V. I.",
    title = "{Higher-dimensional operators at finite temperature affect gravitational-wave predictions}",
    eprint = "2503.18904",
    archivePrefix = "arXiv",
    primaryClass = "hep-ph",
    reportNumber = "HIP-2025-6/TH",
    doi = "10.1007/JHEP08(2025)109",
    journal = "JHEP",
    volume = "08",
    pages = "109",
    year = "2025"
}

@article{Navarrete:2025yxy,
    author = {Navarrete, Pablo and Paatelainen, Risto and Sepp{\"a}nen, Kaapo and Tenkanen, Tuomas V. I.},
    title = "{Cosmological phase transitions without high-temperature expansions}",
    eprint = "2507.07014",
    archivePrefix = "arXiv",
    primaryClass = "hep-ph",
    reportNumber = "HIP-2025-20/TH",
    doi = "10.1007/JHEP01(2026)113",
    journal = "JHEP",
    volume = "01",
    pages = "113",
    year = "2026"
}

@misc{Gould:2026rqz,
    author = "Gould, Oliver and Rassouli, Farbod-Sayyed and Saffin, Paul M.",
    title = "{Effective field theory for strong phase transitions}",
    eprint = "2610.10692",
    archivePrefix = "arXiv",
    primaryClass = "hep-ph",
    month = "10",
    year = "2026"
}

@article{Cornwall:1974vz,
    author = "Cornwall, John M. and Jackiw, R. and Tomboulis, E.",
    title = "{Effective Action for Composite Operators}",
    reportNumber = "MIT-CTP-419",
    doi = "10.1103/PhysRevD.10.2428",
    journal = "Phys. Rev. D",
    volume = "10",
    pages = "2428--2445",
    year = "1974"
}

@article{Amelino-Camelia:1993rvt,
    author = "Amelino-Camelia, G.",
    title = "{Selfconsistently improved finite temperature effective potential for gauge theories}",
    eprint = "hep-ph/9305222",
    archivePrefix = "arXiv",
    reportNumber = "BUHEP-93-12",
    doi = "10.1103/PhysRevD.49.2740",
    journal = "Phys. Rev. D",
    volume = "49",
    pages = "2740--2751",
    year = "1994"
}

@misc{Klose:2026pux,
    author = "Klose, Philipp",
    title = "{Thermal Resummation for Very Strong First-order Phase Transitions}",
    eprint = "2607.21252",
    archivePrefix = "arXiv",
    primaryClass = "hep-ph",
    month = "7",
    year = "2026"
}

@misc{Banik:2026ljf,
    author = "Banik, Amitayus and Kainulainen, Kimmo",
    title = "{Consistent Thermal Resummation and Phase Transitions with 2PI Methods}",
    eprint = "2608.04102",
    archivePrefix = "arXiv",
    primaryClass = "hep-ph",
    month = "8",
    year = "2026"
}

@article{Jackiw:1974cv,
    author = "Jackiw, R.",
    title = "{Functional evaluation of the effective potential}",
    doi = "10.1103/PhysRevD.9.1686",
    journal = "Phys. Rev. D",
    volume = "9",
    pages = "1686",
    year = "1974"
}

@article{Coleman:1973jx,
    author = "Coleman, Sidney R. and Weinberg, Erick J.",
    title = "{Radiative Corrections as the Origin of Spontaneous Symmetry Breaking}",
    doi = "10.1103/PhysRevD.7.1888",
    journal = "Phys. Rev. D",
    volume = "7",
    pages = "1888--1910",
    year = "1973"
}

@inproceedings{Quiros:1999jp,
    author = "Quiros, Mariano",
    title = "{Finite temperature field theory and phase transitions}",
    booktitle = "{ICTP Summer School in High-Energy Physics and Cosmology}",
    eprint = "hep-ph/9901312",
    archivePrefix = "arXiv",
    reportNumber = "IEM-FT-187-99",
    pages = "187--259",
    month = "1",
    year = "1999"
}

@book{Laine:2016hma,
    author = "Laine, Mikko and Vuorinen, Aleksi",
    title = "{Basics of Thermal Field Theory}",
    eprint = "1701.01554",
    archivePrefix = "arXiv",
    primaryClass = "hep-ph",
    doi = "10.1007/978-3-319-31933-9",
    publisher = "Springer",
    volume = "925",
    year = "2016"
}

@book{Kapusta:2006pm,
    author = "Kapusta, J. I. and Gale, Charles",
    title = "{Finite-temperature field theory: Principles and applications}",
    doi = "10.1017/CBO9780511535130",
    isbn = "978-0-521-17322-3, 978-0-521-82082-0, 978-0-511-22280-1",
    publisher = "Cambridge University Press",
    series = "Cambridge Monographs on Mathematical Physics",
    year = "2011"
}

@article{Matsubara:1955ws,
    author = "Matsubara, Takeo",
    title = "{A New Approach to Quantum-Statistical Mechanics}",
    journal = "Prog. Theor. Phys.",
    volume = "14",
    pages = "351--378",
    year = "1955",
    doi = "10.1143/PTP.14.351"
}

@article{Linde:1980ts,
    author = "Linde, Andrei D.",
    title = "{Infrared Problem in Thermodynamics of the Yang-Mills Gas}",
    reportNumber = "LEBEDEV-80-106",
    doi = "10.1016/0370-2693(80)90769-8",
    journal = "Phys. Lett. B",
    volume = "96",
    pages = "289--292",
    year = "1980"
}

@article{Patel:2011th,
    author = "Patel, Hiren H. and Ramsey-Musolf, Michael J.",
    title = "{Baryon Washout, Electroweak Phase Transition, and Perturbation Theory}",
    eprint = "1101.4665",
    archivePrefix = "arXiv",
    primaryClass = "hep-ph",
    doi = "10.1007/JHEP07(2011)029",
    journal = "JHEP",
    volume = "07",
    pages = "029",
    year = "2011"
}

@article{Martin:2018emo,
    author = "Martin, Stephen P. and Patel, Hiren H.",
    title = "{Two-loop effective potential for generalized gauge fixing}",
    eprint = "1808.07615",
    archivePrefix = "arXiv",
    primaryClass = "hep-ph",
    doi = "10.1103/PhysRevD.98.076008",
    journal = "Phys. Rev. D",
    volume = "98",
    number = "7",
    pages = "076008",
    year = "2018"
}

@book{Peskin:1995ev,
    author = "Peskin, Michael E. and Schroeder, Daniel V.",
    title = "{An Introduction to quantum field theory}",
    doi = "10.1201/9780429503559",
    isbn = "978-0-201-50397-5, 978-0-429-50355-9, 978-0-429-49417-8",
    publisher = "Addison-Wesley",
    address = "Reading, USA",
    year = "1995"
}

@article{tHooft:1971qjg,
    author = "'t Hooft, Gerard",
    editor = "Taylor, J. C.",
    title = "{Renormalizable Lagrangians for Massive Yang-Mills Fields}",
    doi = "10.1016/0550-3213(71)90139-8",
    journal = "Nucl. Phys. B",
    volume = "35",
    pages = "167--188",
    year = "1971"
}

@article{Fujikawa:1972fe,
    author = "Fujikawa, K. and Lee, B. W. and Sanda, A. I.",
    title = "{Generalized Renormalizable Gauge Formulation of Spontaneously Broken Gauge Theories}",
    reportNumber = "FERMILAB-PUB-72-055-T, NAL-THY-55, EFI-72-22",
    doi = "10.1103/PhysRevD.6.2923",
    journal = "Phys. Rev. D",
    volume = "6",
    pages = "2923--2943",
    year = "1972"
}

@article{Faddeev:1967fc,
    author = "Faddeev, L. D. and Popov, V. N.",
    editor = "Hsu, Jong-Ping and Fine, D.",
    title = "{Feynman Diagrams for the Yang-Mills Field}",
    doi = "10.1016/0370-2693(67)90067-6",
    journal = "Phys. Lett. B",
    volume = "25",
    pages = "29--30",
    year = "1967"
}

@article{Weldon:1996kb,
    author = "Weldon, H. Arthur",
    title = "{Structure of the gluon propagator at finite temperature}",
    eprint = "hep-ph/9701279",
    archivePrefix = "arXiv",
    doi = "10.1006/aphy.1998.5873",
    journal = "Annals Phys.",
    volume = "271",
    pages = "141--156",
    year = "1999"
}

@article{Nielsen:1975fs,
    author = "Nielsen, N. K.",
    title = "{On the Gauge Dependence of Spontaneous Symmetry Breaking in Gauge Theories}",
    reportNumber = "Print-75-0792 (AARHUS)",
    doi = "10.1016/0550-3213(75)90301-6",
    journal = "Nucl. Phys. B",
    volume = "101",
    pages = "173--188",
    year = "1975"
}

@article{Fukuda:1975di,
    author = "Fukuda, Reijiro and Kugo, Taichiro",
    title = "{Gauge Invariance in the Effective Action and Potential}",
    reportNumber = "RIFP-237",
    doi = "10.1103/PhysRevD.13.3469",
    journal = "Phys. Rev. D",
    volume = "13",
    pages = "3469",
    year = "1976"
}

@article{Weinberg:1987vp,
    title = {Understanding complex perturbative effective potentials},
    author = {Weinberg, Erick J. and Wu, Aiqun},
    journal = {Phys. Rev. D},
    volume = {36},
    issue = {8},
    pages = {2474--2480},
    numpages = {0},
    year = {1987},
    month = {Oct},
    publisher = {American Physical Society},
    doi = {10.1103/PhysRevD.36.2474}
}

@article{ParticleDataGroup:2026mpi,
    author = "Takahashi, F. and others",
    collaboration = "Particle Data Group",
    title = "{Review of Particle Physics}",
    doi = "10.1142/s0217751x26300115",
    journal = "Int. J. Mod. Phys. A",
    volume = "41",
    number = "22",
    pages = "2630011",
    year = "2026"
}

@article{Hempfling:1994ar,
    author = "Hempfling, Ralf and Kniehl, Bernd A.",
    title = "{On the relation between the fermion pole mass and MS Yukawa coupling in the standard model}",
    eprint = "hep-ph/9408313",
    archivePrefix = "arXiv",
    reportNumber = "KEK-TH-408, KEK-PREPRINT-94-74",
    doi = "10.1103/PhysRevD.51.1386",
    journal = "Phys. Rev. D",
    volume = "51",
    pages = "1386--1394",
    year = "1995"
}

@article{Chetyrkin:1999qi,
    author = "Chetyrkin, K. G. and Steinhauser, M.",
    title = "{The Relation between the MS-bar and the on-shell quark mass at order alpha(s)**3}",
    eprint = "hep-ph/9911434",
    archivePrefix = "arXiv",
    reportNumber = "DESY-99-174, TTP-99-47",
    doi = "10.1016/S0550-3213(99)00784-1",
    journal = "Nucl. Phys. B",
    volume = "573",
    pages = "617--651",
    year = "2000"
}

@article{Melnikov:2000qh,
    author = "Melnikov, Kirill and Ritbergen, Timo van",
    title = "{The Three loop relation between the MS-bar and the pole quark masses}",
    eprint = "hep-ph/9912391",
    archivePrefix = "arXiv",
    reportNumber = "SLAC-PUB-8321, TTP-99-51",
    doi = "10.1016/S0370-2693(00)00507-4",
    journal = "Phys. Lett. B",
    volume = "482",
    pages = "99--108",
    year = "2000"
}

@article{Athron:2022jyi,
    author = "Athron, Peter and Balazs, Csaba and Fowlie, Andrew and Morris, Lachlan and White, Graham and Zhang, Yang",
    title = "{How arbitrary are perturbative calculations of the electroweak phase transition?}",
    eprint = "2208.01319",
    archivePrefix = "arXiv",
    primaryClass = "hep-ph",
    doi = "10.1007/JHEP01(2023)050",
    journal = "JHEP",
    volume = "01",
    pages = "050",
    year = "2023"
}

@article{Das:1987yb,
    author = "Das, Ashok K. and Karev, A.",
    title = "{The Derivative Expansion and the Anomaly at Finite Temperature}",
    reportNumber = "UR-982, ER13065-473",
    doi = "10.1103/PhysRevD.36.623",
    journal = "Phys. Rev. D",
    volume = "36",
    pages = "623",
    year = "1987"
}

@article{Arnold:1992qy,
    author = "Arnold, Peter Brockway and Vokos, Stamatis and Bedaque, Paulo F. and Das, Ashok K.",
    title = "{On the analytic structure of the selfenergy for massive gauge bosons at finite temperature}",
    eprint = "hep-ph/9211334",
    archivePrefix = "arXiv",
    reportNumber = "UW-PT-92-19, ANL-HEP-PR-92-51, UR-1287, ER-40685-736",
    doi = "10.1103/PhysRevD.47.4698",
    journal = "Phys. Rev. D",
    volume = "47",
    pages = "4698--4704",
    year = "1993"
}

@misc{Qiu:2026bje,
    author = "Qiu, Dayun and Huang, Fa Peng",
    title = "{Towards Consistent Thermal Resummation: Improved Partial Dressing for Precise Electroweak Phase Transition Predictions}",
    eprint = "2611.26303",
    archivePrefix = "arXiv",
    primaryClass = "hep-ph",
    month = "11",
    year = "2026"
}

@article{Buchmuller:1993bq,
    author = "Buchmuller, W. and Fodor, Z. and Helbig, T. and Walliser, D.",
    title = "{The weak electroweak phase transition}",
    eprint = "hep-ph/9303251",
    archivePrefix = "arXiv",
    reportNumber = "DESY-93-021",
    doi = "10.1006/aphy.1994.1082",
    journal = "Annals Phys.",
    volume = "234",
    pages = "260--299",
    year = "1994"
}

@article{Buchmuller:1994qy,
    author = "Buchmuller, W. and Philipsen, O.",
    title = "{Phase structure and phase transition of the SU(2) Higgs model in three-dimensions}",
    eprint = "hep-ph/9411334",
    archivePrefix = "arXiv",
    reportNumber = "DESY-94-202",
    doi = "10.1016/0550-3213(95)00124-B",
    journal = "Nucl. Phys. B",
    volume = "443",
    pages = "47--69",
    year = "1995"
}

\end{document}